\documentclass[twoside,twocolumn,9pt]{article}
\usepackage{extsizes}
\usepackage[super,sort&compress,comma]{natbib} 
\usepackage[version=3]{mhchem}
\usepackage[left=1.5cm, right=1.5cm, top=1.785cm, bottom=2.0cm]{geometry}
\usepackage{balance}
\usepackage{times,newtxmath}
\usepackage{sectsty}
\usepackage{graphicx}
\usepackage{lastpage}
\usepackage[format=plain,justification=justified,singlelinecheck=false,font={stretch=1.125,small,sf},labelfont=bf,labelsep=space]{caption}
\usepackage{float}
\usepackage{fancyhdr}
\usepackage{fnpos}
\usepackage[english]{babel}
\usepackage{array}
\usepackage{droidsans}
\usepackage{charter}
\usepackage[T1]{fontenc}
\usepackage[usenames,dvipsnames]{xcolor}
\usepackage{setspace}
\usepackage[compact]{titlesec}
\usepackage{amssymb}
\usepackage{amstext}
\usepackage{amsmath}
\usepackage{amsfonts}
\usepackage[T1]{fontenc}
\usepackage[latin9]{inputenc}
\usepackage{wasysym}

\def\be{\begin{equation}}
\def\ee{\end{equation}}
\def\pas{\,\,\mathrm{Pa}}
\def\kpas{\,\,\mathrm{kPa}}
\def\mum{\,\,\mathrm{\mu m^2}}
\def\secs{\,\,\mathrm{s}}
\def\mm{\,\,\mathrm{mm}}
\def\mgml{\,\,\mathrm{mg/mL}}

\definecolor{cream}{RGB}{222,217,201}

\begin{document}

\pagestyle{fancy}
\thispagestyle{plain}
\fancypagestyle{plain}{

\fancyhead[C]{\includegraphics[width=18.5cm]{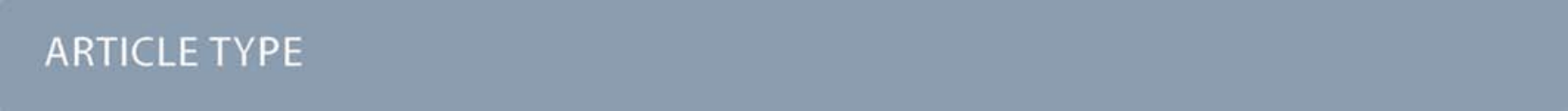}}
\fancyhead[L]{\hspace{0cm}\vspace{1.5cm}\includegraphics[height=30pt]{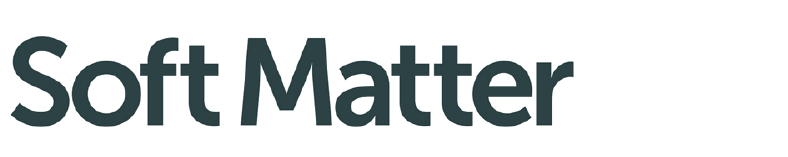}}
\fancyhead[R]{\hspace{0cm}\vspace{1.7cm}\includegraphics[height=55pt]{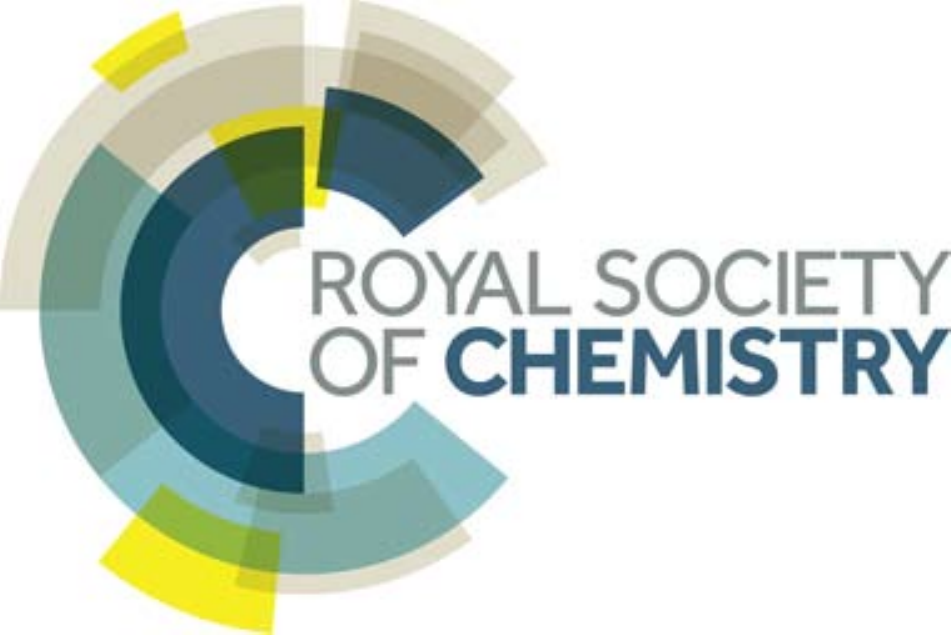}}
\renewcommand{\headrulewidth}{0pt}
}

\makeFNbottom
\makeatletter
\renewcommand\LARGE{\@setfontsize\LARGE{15pt}{17}}
\renewcommand\Large{\@setfontsize\Large{12pt}{14}}
\renewcommand\large{\@setfontsize\large{10pt}{12}}
\renewcommand\footnotesize{\@setfontsize\footnotesize{7pt}{10}}
\makeatother

\renewcommand{\thefootnote}{\fnsymbol{footnote}}
\renewcommand\footnoterule{\vspace*{1pt}%
\color{cream}\hrule width 3.5in height 0.4pt \color{black}\vspace*{5pt}} 
\setcounter{secnumdepth}{5}

\makeatletter 
\renewcommand\@biblabel[1]{#1}            
\renewcommand\@makefntext[1]%
{\noindent\makebox[0pt][r]{\@thefnmark\,}#1}
\makeatother 
\renewcommand{\figurename}{\small{Fig.}~}
\sectionfont{\sffamily\Large}
\subsectionfont{\normalsize}
\subsubsectionfont{\bf}
\setstretch{1.125} 
\setlength{\skip\footins}{0.8cm}
\setlength{\footnotesep}{0.25cm}
\setlength{\jot}{10pt}
\titlespacing*{\section}{0pt}{4pt}{4pt}
\titlespacing*{\subsection}{0pt}{15pt}{1pt}

\fancyfoot{}
\fancyfoot[LO,RE]{\vspace{-7.1pt}\includegraphics[height=9pt]{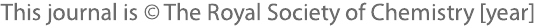}}
\fancyfoot[CO]{\vspace{-7.1pt}\hspace{13.2cm}\includegraphics{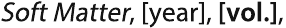}}
\fancyfoot[CE]{\vspace{-7.2pt}\hspace{-14.2cm}\includegraphics{head_foot/RF}}
\fancyfoot[RO]{\footnotesize{\sffamily{1--\pageref{LastPage} ~\textbar  \hspace{2pt}\thepage}}}
\fancyfoot[LE]{\footnotesize{\sffamily{\thepage~\textbar\hspace{3.45cm} 1--\pageref{LastPage}}}}
\fancyhead{}
\renewcommand{\headrulewidth}{0pt} 
\renewcommand{\footrulewidth}{0pt}
\setlength{\arrayrulewidth}{1pt}
\setlength{\columnsep}{6.5mm}
\setlength\bibsep{1pt}

\makeatletter 
\newlength{\figrulesep} 
\setlength{\figrulesep}{0.5\textfloatsep} 

\newcommand{\topfigrule}{\vspace*{-1pt}%
\noindent{\color{cream}\rule[-\figrulesep]{\columnwidth}{1.5pt}} }

\newcommand{\botfigrule}{\vspace*{-2pt}%
\noindent{\color{cream}\rule[\figrulesep]{\columnwidth}{1.5pt}} }

\newcommand{\dblfigrule}{\vspace*{-1pt}%
\noindent{\color{cream}\rule[-\figrulesep]{\textwidth}{1.5pt}} }

\makeatother

\twocolumn[
  \begin{@twocolumnfalse}
\vspace{3cm}
\sffamily
\begin{tabular}{m{4.5cm} p{13.5cm} }

\includegraphics{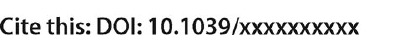} & \noindent\LARGE{\textbf{Poroelasticity of (bio)polymer networks during compression: theory and experiment$^\dag$}} \\
\vspace{0.3cm} & \vspace{0.3cm} \\

& \noindent\large{Melle T.J.J.M. Punter,\textit{$^{\ast a}$} Bart E. Vos,\textit{$^{\ast b c}$} Bela M. Mulder,\textit{$^{a}$} and Gijsje H. Koenderink\textit{$^{b d}$}$^{\ddag}$} \\

\includegraphics{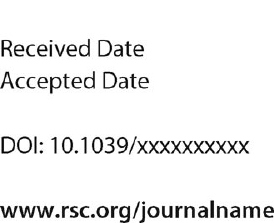} & \noindent\normalsize
{Soft living tissues like cartilage can be considered as biphasic materials comprised of a fibrous complex biopolymer network and a viscous background liquid. 
Here, we show by a combination of experiment and theoretical analysis that both the hydraulic permeability and the elastic properties of (bio)polymer networks can be determined with simple ramp compression experiments in a commercial rheometer. 
In our approximate closed-form solution of the poroelastic equations of motion, we find the normal force response during compression as a combination of network stress and fluid pressure. 
Choosing fibrin as a biopolymer model system with controllable pore size, measurements of the full time-dependent normal force during compression are found to be in excellent agreement with the theoretical calculations. 
The inferred elastic response of large-pore  ($\mathrm{\mu m}$) fibrin networks depends on the strain rate, suggesting a strong interplay between network elasticity and fluid flow. 
Phenomenologically extending the calculated normal force into the regime of nonlinear elasticity, we find strain-stiffening of small-pore (sub-$\mathrm{\mu m}$) fibrin networks to occur at an onset average tangential stress at the gel-plate interface that depends on the polymer concentration in a power-law fashion. 
The inferred permeability of small-pore fibrin networks scales approximately inverse squared with the fibrin concentration, implying with a microscopic cubic lattice model that the thickness of the fibrin fibers decreases with protein concentration.
Our theoretical model provides a new method to obtain the hydraulic permeability and the elastic properties of biopolymer networks and hydrogels with simple compression experiments, and paves the way to study the relation between fluid flow and elasticity in biopolymer networks during dynamical compression. 
} \\

\end{tabular}

 \end{@twocolumnfalse} \vspace{0.6cm}

  ]

\renewcommand*\rmdefault{bch}\normalfont\upshape
\rmfamily
\section*{}
\vspace{-1cm}


\footnotetext{$\ast$~These authors contributed equally to this work.}
\footnotetext{\textit{$^{a}$~AMOLF, Theory of Biomolecular Matter, Science Park 104, 1098XG Amsterdam, the Netherlands}}
\footnotetext{\textit{$^{b}$~AMOLF, Biological Soft Matter, Science Park 104, 1098XG Amsterdam, the Netherlands}}
\footnotetext{\textit{$^{c}$}~Current address: \textit{ZMBE, Mechanics of cellular systems Group,
Institute of Cell Biology, Westf{\"a}lische Wilhelms-Universit{\"a}t, Von-Esmarch-Stra{\ss}e 56, 48149 M{\"u}nster, Germany.}}
\footnotetext{\textit{$^{d}$}~Current address: \textit{Department of Bionanoscience, Kavli Institute of Nanoscience Delft, Delft University of Technology, 2629HZ Delft, The Netherlands}}
\footnotetext{$\ddag$~Corresponding Author, email: G.Koenderink@amolf.nl}
\footnotetext{\dag~Electronic Supplementary Information (ESI) available: construction of the approximate solution of the poroelastic equations of motion (S1), description of and results from the cubic lattice model (S2), overview and discussion of the results from fits of the approximate solution to the measured normal force in all experiments (S3). See DOI: 10.1039/cXsm00000x/}


Soft biopolymer networks have essential functions 
in living cells\cite{Pritchard2014MechanicsTissue,Huber2015CytoskeletalUp}, the extracellular matrix\cite{Mouw2014ExtracellularDeconstruction,Vogel2018UnravelingMatrix} and the process of blood coagulation\cite{Weisel2017FibrinProperties,Bagoly2017ClotHemostasis}. 
Their mechanical properties are determined by the network's hydraulic permeability and (visco)elastic properties. 

The permeability of biopolymer networks determines mass transport in soft tissues\cite{Swartz2007InterstitialTissues,Wilson2005ATissues,Avendano2019ApplicationMatrix}, the dynamic behaviour of cells\cite{Moeendarbary2013TheMaterial,Charras2009AnimalHydraulics} and the (dis)functioning of blood clots in hemostasis and thrombosis\cite{Voronov2013SimulationResolution,Chen2018ThrombusTherapy,Santos2016ThrombusStroke}. 
Conventionally, the permeability of porous materials is inferred from the measured flow rate of a liquid through the material\cite{Tokita2000,Pieters2012AnValues}. 
There are alternative approaches to measure the permeability of porous materials, such as microfluidic devices\cite{Maity2019ResponseHydrogels}, but sticky biopolymer gels are prone to block such devices. 
Another complicating factor is that separate measurements on biopolymer gels are required for a characterization of their elastic properties. The elastic properties of biopolymer networks are essential for the physiological function of tissues and in wound healing\cite{Storm2005NonlinearGels}. 
For instance, arteries need to be extendable to provide blood pressure capacitance and pulse smoothing in the blood circulation\cite{Shadwick1999MechanicalArteries}, and blood clots are required to be resilient structural scaffolds in wound healing\cite{Schultz2010ExtracellularWounds}. 
The elastic properties of biopolymer networks have been studied extensively in shear\cite{Storm2005NonlinearGels,Burla2019FromNetworks}, extension\cite{Brown2009MultiscaleWater,Roeder2009FibrilMatrices} and static compression\cite{vanOosten2016UncouplingStretch-stiffening}. 
The dynamic response during compression, however, remains largely unexplored\cite{Kim2014StructuralCompression,Kim2016Foam-likeNetworks}.

\begin{figure}[t!]  
  \centering
  \begin{tabular}{
  c 
  c 
  }
    \includegraphics[scale=0.55]{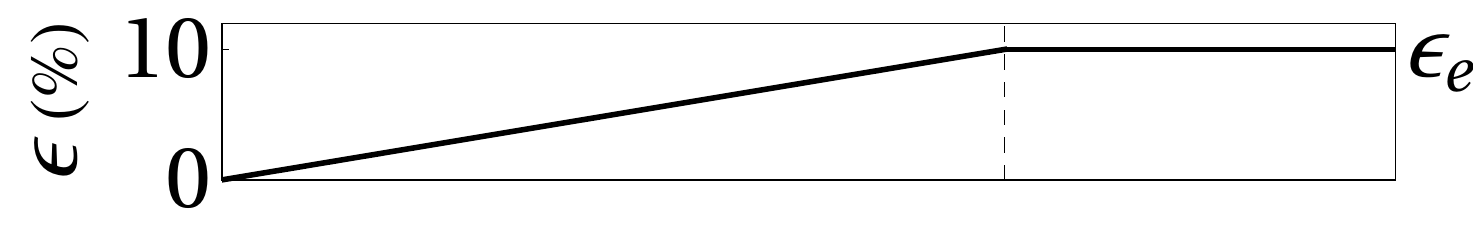}
    \\
    \setbox1=\hbox{\includegraphics[scale=0.55]{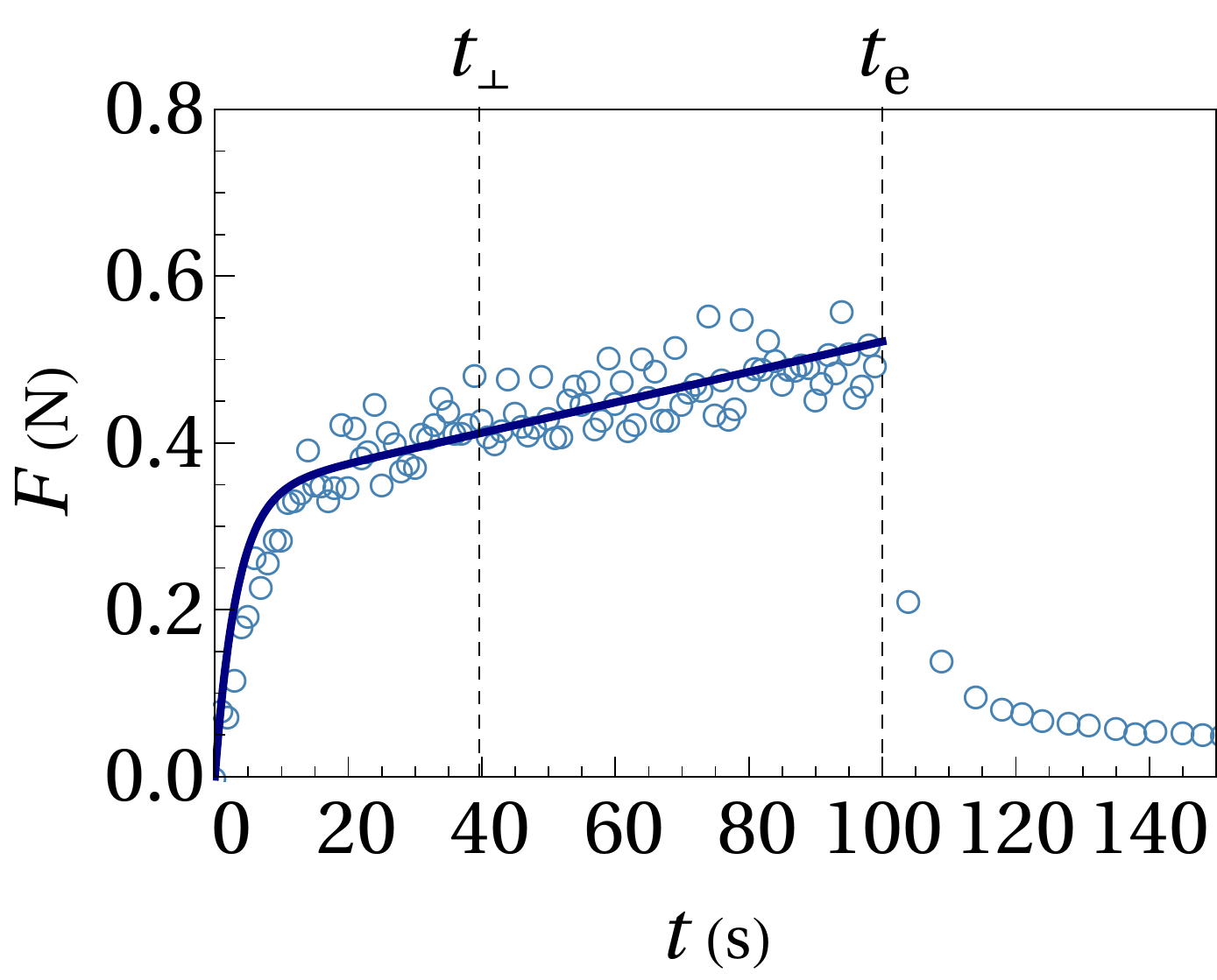}}
    \includegraphics[scale=0.55]{figures/ExampleCoarse.pdf}\llap{\makebox[0.265\wd1][l]{\raisebox{3.5cm}{\includegraphics[height=2cm, width=2cm]{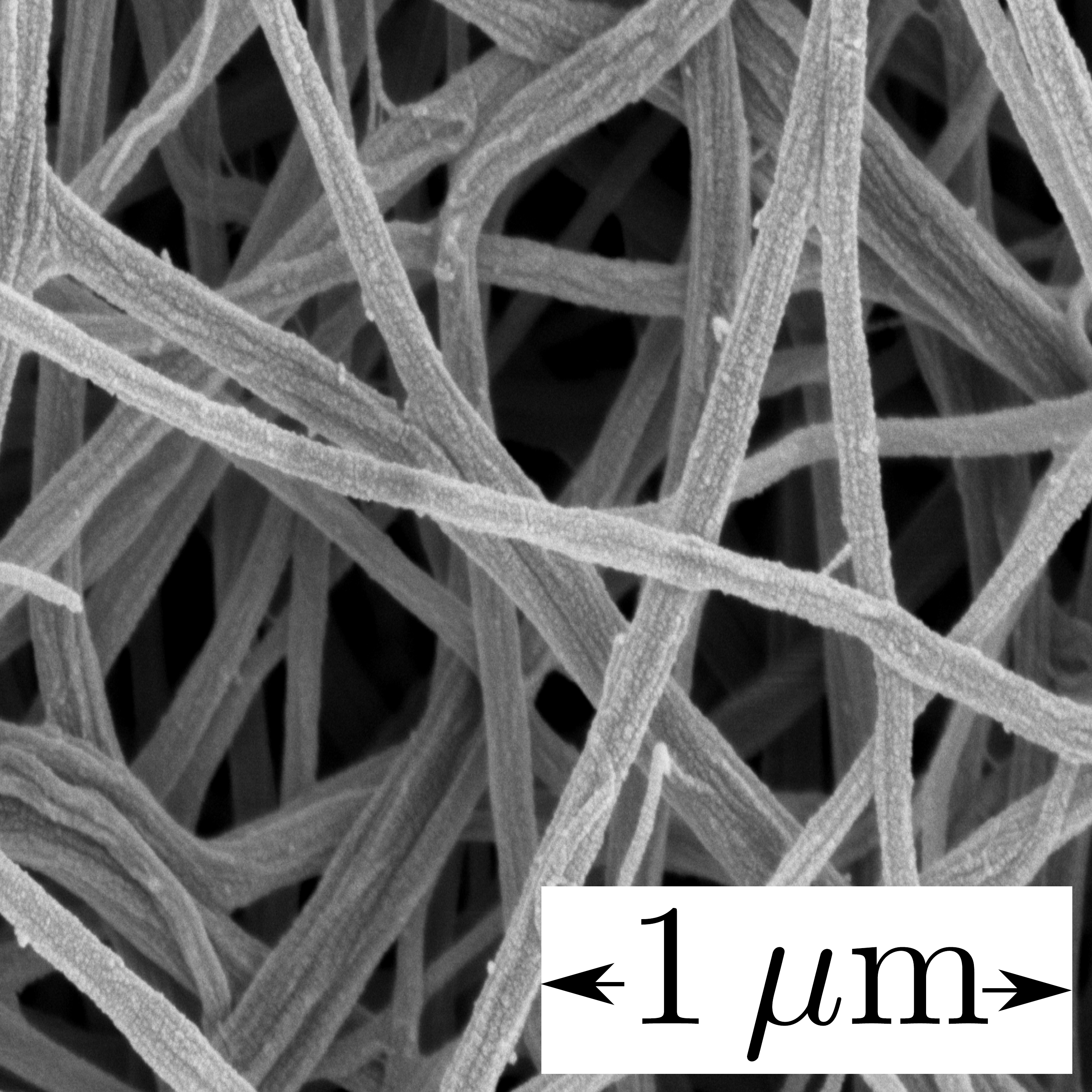}}}}
  \end{tabular}
\caption{
The measured normal force $F$ (blue circles) of a large-pore fibrin gel (mesh size $\xi \sim 1 \,\,\mathrm{\mu m}$ and fibrinogen concentration $c=2\mgml$) with initial radius $a=20\mm$ and height $h=1\mm$ in response to ramp compression in  $t_\mathrm{e} = 100\secs$ up to 10\% engineering strain $\epsilon_\mathrm{e}$. 
During the pressurizing time $t_\perp = 40\secs$ the fluid pressure builds up to its maximal value. 
Subsequently, the compression of the fibrin network keeps increasing the normal force. 
After compression, the fluid pressure contribution decreases to zero; the residual normal force consists only of a static network response. 
Assuming the independently measured shear modulus $G_0 = 139\pas$ to be constant during compression, a fit of the calculated normal force in equation \eqref{eq:Fn_simplified} (blue curve) gives the permeability of the fibrin network as $k=(1.26\pm0.03)\cdot 10^{-1}\mum$ and its oedometric modulus as $M=K+4G_0/3=1.5\pm0.1\kpas$, with $K$ the bulk modulus and $\pm$ denotes the estimation uncertainty. Inset: scanning electron microscopy image of a large-pore fibrin gel with $c=2\mgml$. 
}
\label{fig:example_large}
\end{figure}

Seeking to establish both the permeability and the elastic properties of dynamically compressed biopolymer networks in a single test, we consider slow ramp compression tests with a commercial rheometer. 
For interpretation of the measured normal force, we use the theory of poroelasticity\cite{Macminn2016LargeMaterial,coussy2004poromechanics,deBoer2000TheoryMedia}, as applied to polymer gels\cite{Doi2009GelDynamics}. 
Previously, this theory has been applied successfully to other systems, e.g. interstitial fluid flow through (mineralized) bone tissue\cite{Brynk2011ExperimentalFluctuations,Cardoso2013AdvancesFlow}. 
In short, this theory constructs a stress field in a poroelastic material whose physical origin is twofold: elastic stresses from the deformed network, and pressure from the fluid. 
Network stress and fluid pressure are tightly coupled: when a fast compressive deformation is applied, for example, stresses are generated in the network because it is forced to deform in a volume-conserving manner. 
By equilibrium conditions on the overall stress field, pressure is induced in the fluid, prompting fluid flow through the porous material by Darcy's law\cite{coussy2004poromechanics}. 

We propose an approximate closed-form solution to the poroelastic equations of motion from which we calculate the normal force, during ramp compression, of a disk-like cylindrical gel bonded to the plates of a parallel-plate rheometer. 
The theoretical calculation of the normal force allows us to infer the permeability and elastic properties of a biopolymer network, or any other gel bonded to the plates. 
The calculated normal force separates the contribution of the fluid and the gel network to the measured normal force, describing the full temporal evolution during ramp compression, which, to the best of our knowledge, is lacking in literature, see for example Kim \textit{et al.}\cite{Kim2016Foam-likeNetworks}. 
To test the approximate solution, we use covalently cross-linked fibrin gels as a model system. Fibrin is a fibrous protein structure that is the main structural component of blood clots. The formation of a fibrin gel starts with thrombin cleaving fibrinopeptides from dissolved fibrinogen molecules to obtain fibrin monomers. 
The fibrin monomers then assemble in a half-staggered manner, forming elongated protofibrils of two molecules thick. These protofibrils interconnect to form the relatively thick fibrin fibers that constitute the fibrin network\cite{Weisel2017FibrinProperties}. 
The fibers themselves are immersed in fluid, making a hydrogel with a solid volume fraction of typically less than 1\%. 

\begin{figure}[t!]  
  \centering
    \setbox1=\hbox{\includegraphics[width=0.45\textwidth]{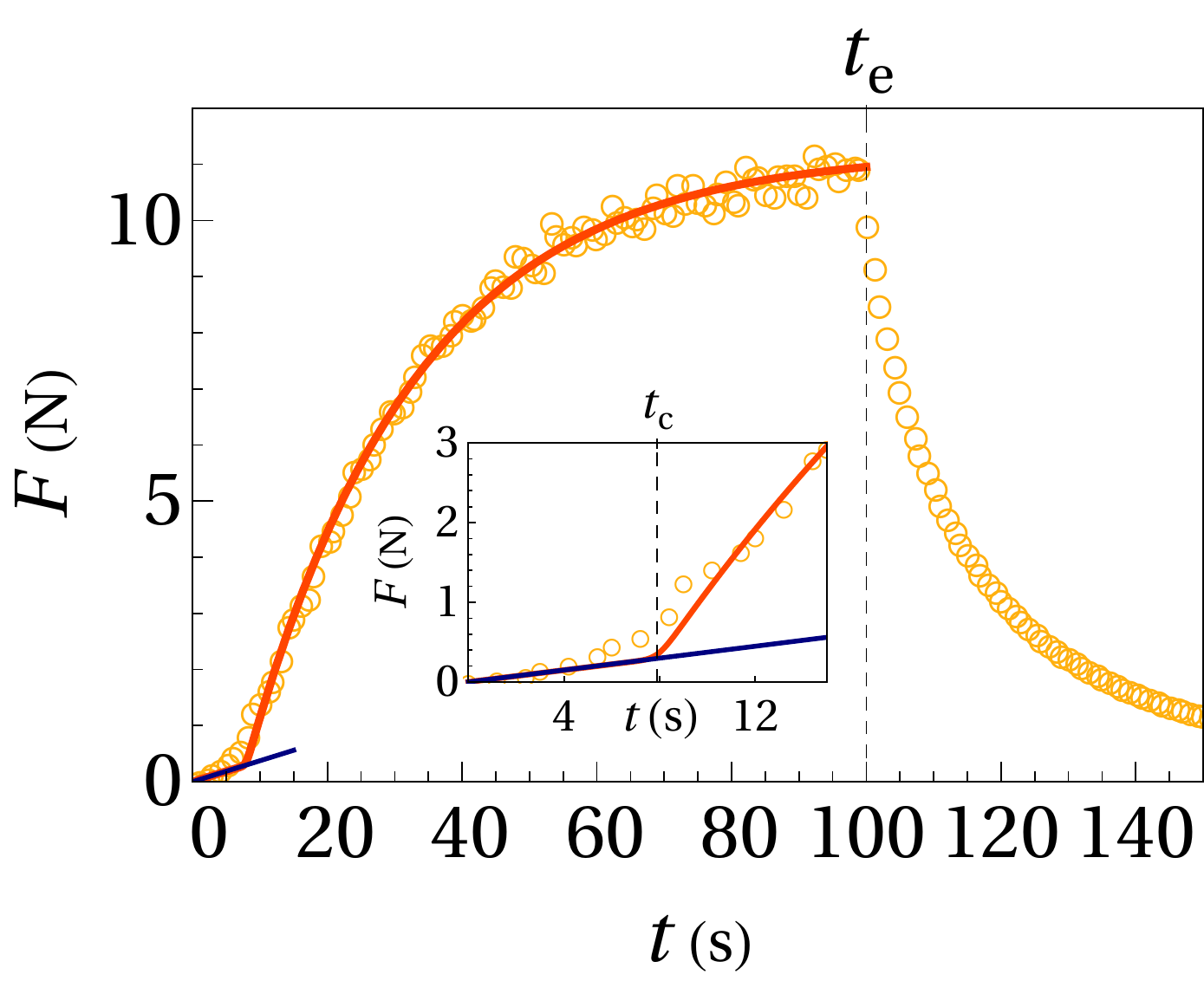}}
    \includegraphics[width=0.45\textwidth]{figures/ExampleFine.pdf}\llap{\makebox[0.23\wd1][l]{\raisebox{4.15cm}{\includegraphics[height=1.85cm]{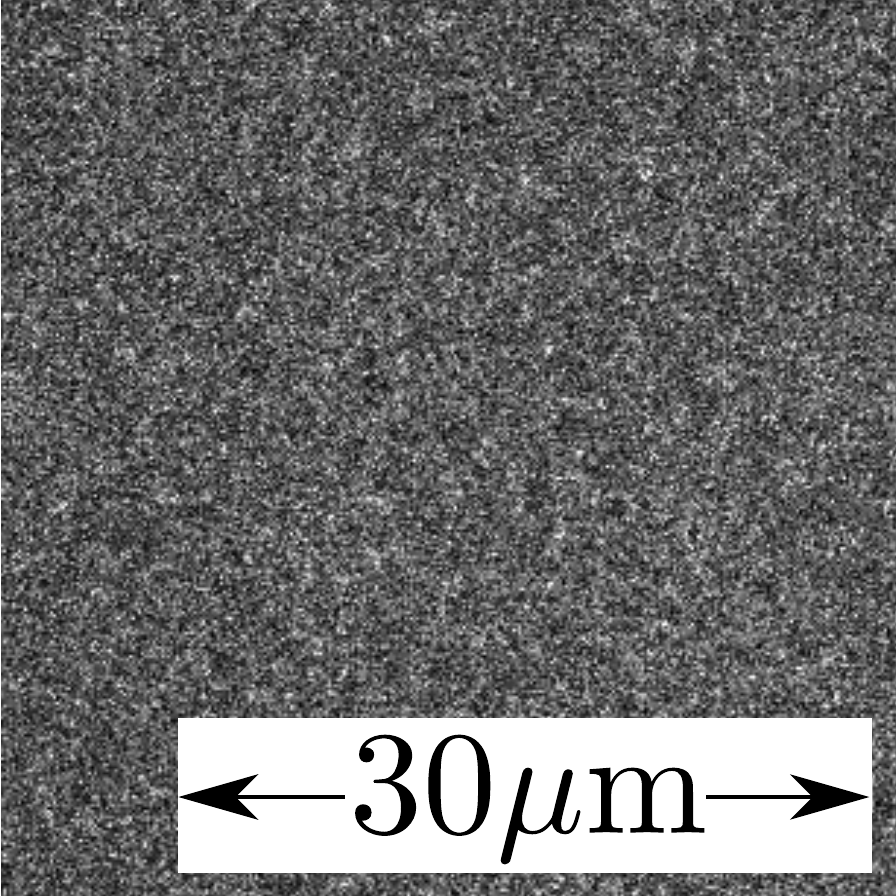}}}}
\caption{
The measured normal force $F$ (orange circles) of a small-pore fibrin gel (mesh size $\xi \sim 0.1\,\,\mathrm{\mu m}$) probed under identical conditions as the large-pore gel of Figure \ref{fig:example_large}. The relatively small pores cause a large pressurizing time $t_\perp = 300\secs$ and fluid pressure. 
Before the onset time of strain-stiffening $t_\mathrm{c} = 7.9\pm0.3\secs$, with $\pm$ denoting the estimation uncertainty, the normal force follows the time dependence expected for a volume-conserving linear elastic solid with the measured initial shear modulus $G_0 = 49\pas$ (blue line in the inset). 
Fitting equation \eqref{eq:Fn_simplified}, combined with \eqref{eq:G_variable}, we find the permeability as $k = (3.86\pm0.15)\cdot 10^{-3}\mum$, the augmented shear modulus after stiffening as $G_\mathrm{c} = 589\pm19\pas$ and the oedometric modulus as $M = 0\pm5\kpas$ (red curve). 
The latter could not be estimated due to the small contribution of network elasticity to the normal force. 
Top right inset: single 2D fluorescence microscopy image of a fine fibrin network. As the mesh size of this network is smaller than the diffraction limit of light\cite{DeCagny2016PorosityGels}, the network cannot be resolved with fluorescence microscopy.
}
\label{fig:example_small}
\end{figure}

\begin{figure*}[t!]
	\centering
	\includegraphics[width=1\textwidth]{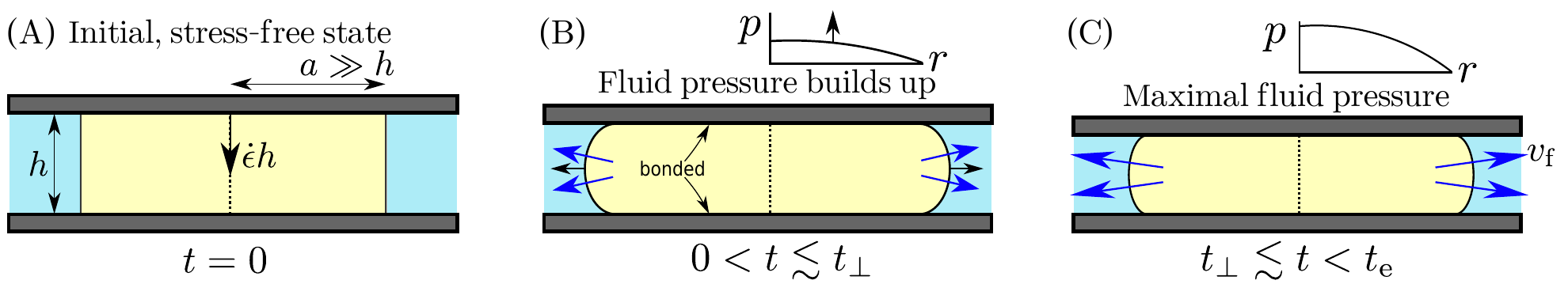}
	\caption{Two compression phases of (\textbf{A}) an initially stress-free cylindrical fibrin gel (yellow) of radius $a$ and height $h$ with high aspect ratio $S\equiv a/h \gg 1$ (Figure not on scale). 
	The gel is ramp compressed in a parallel-plate rheometer with the upper plate (gray) having a constant velocity $\dot{\epsilon} h$, where $\dot{\epsilon}$ is the strain rate. 
	\textbf{B}) As compression commences, the fibrin network starts to bulge out (black arrows) because the gel is bonded to the plates, causing the fluid pressure $p$ to build up in a pressurizing time $t_\perp$. 
	The build-up proceeds until the fluid outflow velocity $v_\mathrm{f}$ (blue arrows) due to the fluid pressure gradient, $v_\mathrm{f} \propto - \partial p / \partial r$ with $r$ the radial coordinate in the fibrin gel, is such that the outward bulging of the gel network, which induces the fluid pressure, stabilizes. 
	\textbf{C}) Afterwards, the gel is compressed further at maximal fluid pressure until at time $t_\mathrm{e}$ the compression stops. 
	}
	\label{fig:CompressionPhases}
\end{figure*}

Fibrin gels are convenient experimental model systems because their permeability can be manipulated through the polymerization conditions, with pore sizes that can be either several tens of nanometers or a few microns\cite{DeCagny2016PorosityGels}. 
We perform ramp compression tests on disk-like cylindrical fibrin gels in which we vary the fibrinogen concentration, the amount of strain, the strain rate and the dimensions of the gel. 
The normal force response of fibrin gels during compression can be explained with our theoretical solution, accounting for the time-dependent build-up of fluid pressure, see Figure \ref{fig:CompressionPhases}--\ref{fig:example_large}. 
Since for large-pore fibrin gels the vast majority of fluid pressure builds up in only a few seconds, we also considered small-pore fibrin gels having a prolonged phase of pressure build-up, see Figure \ref{fig:example_small}. 

We find that the flow of fluid through large-pore fibrin networks has a strong influence on its mechanical response under both small (5--10\% compressive strain) and large ($\leq$80\%) compression. 
Phenomenologically extending the theoretical solution to include strain-stiffening during compression, we show that strain-stiffening of small-pore fibrin networks occurs during pressure build-up around an onset average tangential stress at the gel-plate interface, similar as in shear rheology\cite{Storm2005NonlinearGels}. 
The onset stress depends on the fibrinogen concentration in a power-law fashion.  
Fluid flow through the fibrin network inhibits strain-stiffening, implying a nontrivial interplay between network elasticity and fluid flow. 
The permeability of small-pore fibrin networks is found to scale approximately with the inverse squared of the fibrinogen concentration, implying with a microscopic cubic lattice model that the mass density per unit axial length of the fibrin fibers decreases with the overall fibrinogen concentration in the gel. 

After introducing the compression experiments on fibrin gels, we develop the theoretical background, the theoretical calculation for the normal force, and its phenomenological extensions. 
The strain rate dependence of the elastic properties of large-pore fibrin networks is discussed, as well as the onset stress for strain-stiffening and the inferred permeability of small-pore fibrin networks. 

\section{Materials and Methods}
\subsection{Materials}
Human plasma fibrinogen (contains naturally occuring Factor XIIIa; plasminogen, von Willebrand Factor and Fibronectin depleted) and human $\alpha$-thrombin were obtained in lyophilized form from Enzyme Research Laboratories (Swansea, United Kingdom). 
All chemicals were obtained from Sigma Aldrich (Zwijndrecht, The Netherlands). 
Fibrinogen was dissolved in water at 37\textdegree C for 15~min to its original concentration (approximately 13~mg/ml) and dialysed against fibrin buffer containing 20~mM 4-(2-hydroxyethyl)-1-piperazineethanesulfonic acid (HEPES) and 150~mM NaCl at a pH of 7.4, and stored at -80\textdegree C. 
Prior to use, the fibrinogen was quickly thawed at 37\textdegree C, and then diluted in a final assembly buffer containing 20~mM HEPES, 150~mM NaCl and 5~mM CaCl$_2$ (large-pore gels). 
Dense networks (small-pore gels) with an average pore size of 0.08 $\mu$m, as determined by light scattering measurements \cite{DeCagny2016PorosityGels,Yeromonahos2010NanostructureClot}, were obtained in small-pore-gel assembly buffer (400 mM $\textrm{NaCl}$, 3.2 mM CaCl$_2$ and 50 mM Tris-HCl) at a pH of 8.5 \cite{Bale1985EffectsCreep/peptides}. \\

Fibrin polymerization was initiated by the addition and quick mixing of 0.5~U/ml of thrombin from a 20~U/ml thrombin stock, kept on ice for a maximum of 24 hours. 
After addition of thrombin, the mixture was quickly transferred to the rheometer to allow \textit{in situ} polymerization. 
During polymerization, we measured the linear elastic shear modulus $G'$ of the fibrin gels by measuring the stress response to a small oscillatory shear strain with an amplitude of $0.5\%$ and a frequency of 0.5~Hz. 
In this way the shear modulus just before compression $G_0$ was determined. 

\subsection{Compression experiments}

To measure the normal force produced by fibrin gels under ramp uniaxial compression, we use an Anton Paar rheometer (Physica MCR 501, Graz, Austria) to compress disk-like fibrin gels, confined between two impermeable surfaces: a stainless steel bottom plate and a steel top plate, separated by an initial gap $h$ of 1 or 0.5~mm. We used top plates with diameters of either 20 or 40~mm. 
The plates of the rheometer were held at 37\textdegree{C} throughout the experiment. 
To provide external hydrostatic pressure and to prevent the formation of a fibrin membrane at the free boundary of the gel, we immersed the gel in mineral oil\cite{Macrae2018AInvasion}.

In the analysis of the time-dependent normal force, the liquid can safely be assumed to be incompressible at the pressures we apply: the force transducer in the rheometer can apply normal forces up to 50~N, implying the maximum engineering stress to be of the order of 10 kPa. 
To verify that in the compression experiments only liquid is expelled while the network remains intact, we collected the expelled liquid and confirmed by spectrophotometric measurements of the absorbance at 280 and 320~nm that no protein was present.

\section{Theoretical framework}\label{sec:modeling}

To calculate the normal force response of a fibrin gel under compression we use the theory of linear poroelasticity which rests on the following three principles\cite{coussy2004poromechanics,deBoer2000TheoryMedia,Macminn2016LargeMaterial}. 
\textbf{1)}. Assuming fluid flow through the fibrin network to be in the regime of low Reynolds number, and because the gel is immersed in fluid, we can neglect, respectively, inertia and gravity, implying the overall \textit{force balance} of the fibrin gel to read \cite{Macminn2016LargeMaterial}
\begin{equation}
\nabla\cdot\left(\boldsymbol{\sigma'} - p\boldsymbol{{1}}\right)=\textbf{0},\label{eq:fb}
\end{equation}
where $\boldsymbol{{\sigma'}}$ is the Terzaghi effective stress of the fibrin network: the network stress relative to the pressure $p$ of the permeating fluid. 
We take the effective stress to be that of a linear elastic isotropic homogeneous solid with the bulk modulus $K$ and the shear modulus $G$ as elastic constants. 
\textbf{2)} Assuming the fibrin fibers and the fluid both to be individually incompressible, we find from \textit{mass conservation} the incompressibility condition for a fibrin gel as\cite{Doi2009GelDynamics}
\begin{equation}
\nabla\cdot\boldsymbol{V}=0,\label{eq:incompressibility}
\end{equation}
where $\boldsymbol{V}\equiv\phi_{\mathrm{{f}}}\boldsymbol{v}_{\mathrm{{f}}}+\phi_{\mathrm{{n}}}\boldsymbol{v}_{\mathrm{{n}}}$ is the gel velocity: a sum of the local volume-averaged velocity $\boldsymbol{v}_i$ of the fluid and the fibrin network weighted with their respective volume fractions $\phi_i$.  
\textbf{3)} Finally, in the low Reynolds number regime, \textit{Darcy's law} governs the flow of fluid through the fibrin network \cite{Macminn2016LargeMaterial}
\begin{equation}
\phi_{\mathrm{{f}}}\left(\boldsymbol{v}_{\mathrm{{f}}}-\boldsymbol{v}_{\mathrm{{n}}}\right)=-\frac{k}{\eta}\nabla p,\label{eq:Darcy}
\end{equation}
where $k$ is the permeability of the fibrin network and $\eta$ the viscosity of the fluid, which we take to be equal to that of pure water at 37\textdegree{C}. 

For a compressed fibrin gel with initial radius $a$ and height $h$, experiencing no friction with the rheometer plates, equation \eqref{eq:fb}-\eqref{eq:Darcy} can be solved exactly\cite{Armstrong1984AnCartilage}. 
During slow ramp compression, the fluid pressure in a frictionless gel becomes maximum after some pressurizing time $t_\parallel$. 
In our high aspect ratio $S\equiv a/h\gg1$ compression tests on fibrin gels, however, where the network binds to the plates, this binding strongly influences its mechanical response, see Figure \ref{fig:CompressionPhases}. 
Here, we propose an approximate solution to equation \eqref{eq:fb}-\eqref{eq:Darcy} for a disk-like bonded gel during compression, see section S1 of the the Supplementary Information for a full derivation$^\dag$. 
This solution assumes that the shear stress induced by the friction between the gel and the rheometer plate dominates the network stress in the gel. 
For a given bulk modulus $K$ and shear modulus $G$, the local increase in the radial force on the gel network per unit of volume due to inhomogeneous radial strain is given by $M\partial_r (1/r) \partial_r r U$, with $M = K + 4 G / 3$ the oedometric modulus, $U$ the radial displacement and $r$ the radial coordinate. 
The shear stress dominates the network stress when the ratio of this radial force to the local increase in radial force due to bending of the gel network $G \partial_z^2 U$, with $z$ the vertical coordinate, is small, i.e., $M / G S^2 \ll 1$. 
The solution interpolates between initial volume-conserving (VC) compression, during which no significant outflow of fluid occurs and of which the network displacement field and fluid pressure are well known \cite{Qiao2015AnalyticalLayers}, and pressurized compression in which the fluid pressure is maximal, see Figure \ref{fig:CompressionPhases}C. 
The dominant part of the normal force $F$ is found as
\begin{equation}
\frac{F}{\pi a^{2}}=T(t)\left(\frac{\eta a^{2}}{8k}\dot{{\epsilon}}+M\epsilon(t)\right) + \left( 1 - T(t) \right) 2 G \epsilon,\label{eq:Fn_simplified}
\end{equation}
where $\dot{\epsilon} \equiv v/h$ is the strain rate with $v$ the velocity of the upper plate and $h$ the initial height of the gel, and $\epsilon \equiv \dot{\epsilon} t$ is the engineering strain. 
Equation \eqref{eq:Fn_simplified} shows that when $T(t)\approx 1$, $F$ is composed of two contributions: the first stems from the compression-induced fluid pressure and the second from the normal force response of the fibrin network. 
The first term is proportional to the strain rate $\dot{{\epsilon}}$ and increases with decreasing permeability $k$ of the network, whilst the second term is proportional to the engineering strain $\epsilon$, and grows proportionally to the oedometric modulus $M$ of the network. 
Equation (\ref{eq:Fn_simplified}) gives a quantitative prediction for the evolution of the normal force from the outset of compression into the pressurized phase, needed to describe the normal force during the full range of ramp compression, see the blue curve in Figure \ref{fig:example_large}. 

The transition function $T(t)$ is given by
\begin{align}
\label{eq:T_function}T(t) &= 1-\exp\left(-12\frac{t}{t_{\perp}}\right),\\
\label{eq:pressurizing_time}t_\perp &= \frac{h^{2}\eta}{k G},
\end{align}
where $t_{\perp}$ is the pressurizing time. 
The rate of fluid outflow increases until the fluid pressure distribution in the gel is maximal, i.e., the gel is pressurized when $T(t)\approx 1$. 
Once pressurized, equation \eqref{eq:Fn_simplified} agrees with the normal force for load-controlled compression\cite{Doi2009GelDynamics}. 
The time scale of relaxation $t_{\perp}$ does not depend on the initial radius $a$ of the gel, because the main contribution to the pressure of the fluid is induced by bending of the fibrin network, i.e., from vertical, $h$-dependent curvature in the radial displacement field. 
For comparison, radial gradients in the radial strain field relax on a time scale $t_{\parallel} = a^2 \eta/kM = t_{\perp}S^{2}G/M \gg t_{\perp}$, implying a geometry-induced separation of time scales for relaxations of vertical and radial strain gradients. 

The compression of small-pore fibrin networks suggests the network to strain-stiffen, see Figure \ref{fig:example_small}, but only for small fluid pressures where the gel deforms approximately volume-conserving. 
We accommodate this phenomenologically by replacing $Gt\rightarrow\int_{0}^{t}dt'G(t')$ in the approximate solution wherever $G$ enters, with the shear modulus $G(t)$ increasing instantaneously at an onset time $t_{\mathrm{{c}}}$. 
Notwithstanding that strain-stiffening is a continuous process, this instantaneous increase is, in the absence of knowledge of the details, a minimal form to incorporate strain-stiffening, and gives
\begin{equation}
G(t)\equiv G_{0}+\left(G_\mathrm{c}-G_{0}\right)H\left(t-t_\mathrm{c}\right),\label{eq:G_variable}
\end{equation}
with $G_{0}$ the measured shear modulus of the undeformed gel, $G_{\mathrm{c}}$ the augmented shear modulus and $H(t)$ the Heaviside step function.

In simple shear experiments \cite{Piechocka2010StructuralMechanics}, strain-stiffening starts to occur at an onset shear stress $\sigma_{\mathrm{{c}}}$. 
In our compression experiments, we assume the onset stress to be proportional to the average tangential network stress $\bar{{\sigma}}'_{rz}$ at the gel-plate interface at time $t=t_{\mathrm{{c}}}$
\begin{equation}
\bar{{\sigma}}'_{rz}\equiv\frac{1}{\pi a^{2}}\int_{0}^{a}dr2\pi r\sigma'_{rz}(z=h,t=t_{\mathrm{{c}}}),
\end{equation}
which can be calculated using the solution for the network displacement field presented in section S1 of the Supplementary Information$^\dag$, giving 
\begin{equation}
\sigma_\mathrm{c}\propto T(t_\mathrm{c})\frac{\eta h a }{6k} \dot{\epsilon}. \label{eq:critical_stress}
\end{equation}
The onset stress is a property of the fibrin network, implying its magnitude to be independent of the aspect ratio $S = a / h$ of the gel, contrary to what equation (\ref{eq:critical_stress}) suggests at a first glance. 
Below, we show, however, that $\sigma_{\mathrm{{c}}}$ is indeed independent of the aspect ratio.

To study the mechanical response of the fibrin networks outside of the linear regime, we performed compression tests on large-pore fibrin gels up to 80\% compressive strain, $\epsilon \leq 0.8$. 
For these experiments, we assume a phenomenological form for the normal force in the pressurized phase, based on equation (\ref{eq:Fn_simplified}), by retaining the form of the fluid pressure term, but with a strain-dependent permeability, and by replacing the elastic contribution with the Toll model normal force response of a fibrous network under large compression \cite{Toll1998PackingReinforcements,Kim2016Foam-likeNetworks}, giving
\begin{equation}
\frac{F}{\pi a^{2}}=\frac{\eta a^{2}}{8k(\epsilon)}\dot{{\epsilon}} + b E_{\mathrm{{f}}}\left(\phi^{3}(\epsilon)-\phi_{0}^{3}\right),\label{eq:Fn_large_compression}
\end{equation}
where $E_{\mathrm{{f}}}$ is the Young's modulus of a single fibrin fiber, $\phi(\epsilon)=\phi_{0}/\left(1-\epsilon\right)$ is (approximately) the strain dependent volume fraction of the fibrin network with $\phi_{0}$ the volume fraction in the initial state, and $k(\epsilon)=k_{0}\left(1-\epsilon\right)$ is (approximately) the permeability of the fibrin network, with $k_{0}$ its initial value, see section S2 of the Supplementary Information$^\dag$ for more information. Finally, $b$ is a proportionality constant. 

\section{Results}

\begin{figure}[b!]
	\centering
	\includegraphics[width=0.45\textwidth]{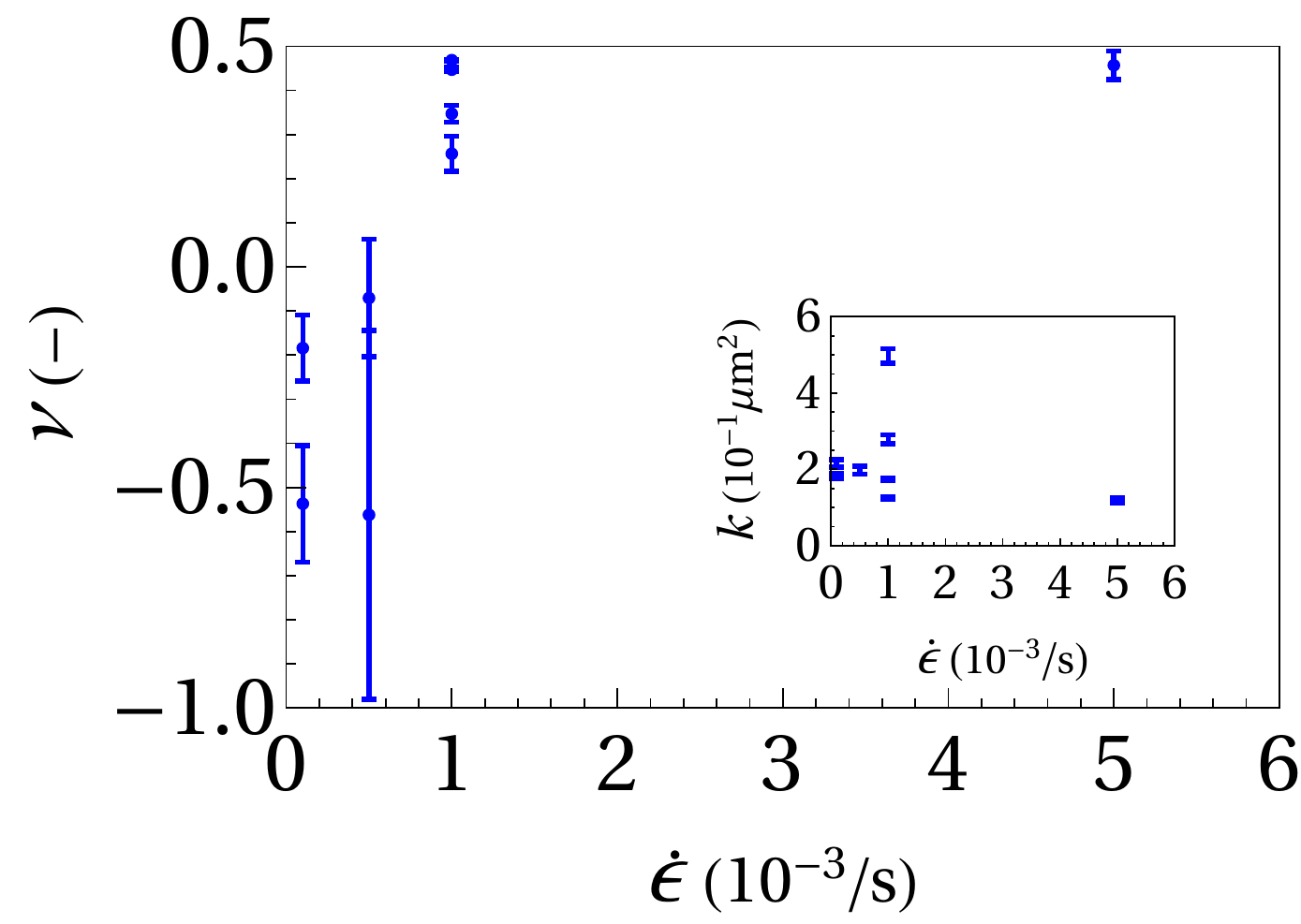}
	\caption{Poisson's ratio $\nu$ of large-pore fibrin gels as a function of the strain rate $\dot{{\epsilon}}$, inferred from ramp compression measurements. All gels were prepared at a fibrinogen concentration of 2 mg/mL, were compressed up to an engineering strain of 5\% or 10\% and have an aspect ratio of either $S\equiv a/h=20$ or $S=10$, with $a$ the radius and $h$ the height of the gel before compression. Poisson's ratio grows with strain rate, suggesting that the fluid velocity has a marked influence on the elastic response of the fibrin network. Inset: the fitted permeability $k$ is independent of the strain rate, as expected, though it shows a large sample-to-sample variation. 
	}
	\label{fig:small_compression_Poisson_ratio_vs_StrainRate}
\end{figure}

To probe the capability of our theoretical framework to infer both the permeability and the elastic properties of a biopolymer network from simple compression tests, we performed compression experiments on both large-pore and small-pore fibrin fiber networks. 
Large-pore fibrin networks have a mesh size of about $\xi \sim 1\,\mathrm{\mu m}$, whereas  small-pore networks have $\xi \sim 0.1\, \mathrm{\mu m}$\cite{DeCagny2016PorosityGels}. 
Therefore, we expect the latter to have a much smaller permeability $k \propto \xi^2$ and, from equation \eqref{eq:Fn_simplified} and \eqref{eq:pressurizing_time}, a larger normal force in the pressurized phase and a larger pressurizing time $t_\perp$, which we indeed observe by comparing the maximum normal force during compression and the pressurizing time between the large-pore and small-pore experiment in, respectively, Figure \ref{fig:example_large} and \ref{fig:example_small}. 
For details of the fitting procedure and all fit results, we refer to section S3 of the Supplementary Information$^\dag$. 

\subsection{Large-pore fibrin gels}\label{sec:large-pore}

Comparing the measured normal force of large-pore and small-pore fibrin networks, see respectively Figure \ref{fig:example_large} and the inset of Figure \ref{fig:example_small}, we observe no supralinear initial increase for large-pore fibrin, see section \ref{sec:small-pore} for further elaboration on this point. 
Therefore, we assume the shear modulus of the large-pore samples to remain equal to $G_0$ throughout compression, i.e., the independently measured shear modulus just before compression, while we fit the permeability $k$ and the oedometric modulus $M$ as free parameters. 
The different large-pore gel samples show a large sample-to-sample variability for the fitted permeability $k$ under equal conditions, see the inset of Figure \ref{fig:small_compression_Poisson_ratio_vs_StrainRate}, and do not suggest dependence of the permeability on the strain rate. 
The compressibility of the fibrin network seems to decrease with strain rate, however, as evidenced by an increasing Poisson's ratio $\nu = ( M - 2 G_0 ) / ( 2 M - 2 G_0 )$, see Figure \ref{fig:small_compression_Poisson_ratio_vs_StrainRate}. 
As fibrin and other biopolymer networks are known to exhibit (strong) nonlinear effects, even for small strains\cite{Brown2009MultiscaleWater,Ban2019StrongNetworks,Steinwachs2016Three-dimensionalNetworks,Shivers2019NonlinearTransition}, we consider this Poisson's ratio to be an effective value over the range of applied compressive strain. 
For strain rates close to zero, i.e., $\dot{\epsilon} = 0.1 \cdot 10^{-3}\,\mathrm{/s}$, we find negative values for Poisson's ratio with a large estimation uncertainty. 
In earlier work \cite{vanOosten2016UncouplingStretch-stiffening}, it was found that in the static limit fibrin networks seem to have a Poisson's ratio of zero, although no uncertainty estimation was given. 
If one calculates the radial extension of a static linear elastic solid bound to the plates \cite{Qiao2015AnalyticalLayers}, however, it is found that for negative Poisson's ratios, $-1\leq \nu \leq 0$, the maximum radial extension is very small: it is less than $30\,\mathrm{\mu m}$ for a gel with radius $a=20\,\mathrm{mm}$ and $h=1\,\mathrm{mm}$  under 10\% compressive strain. 
As the mesh size of a large-pore fibrin network is about $\xi \sim 1\,\mathrm{\mu m}$, this radial extension is on the boundary of being meaningful in the poroelastic continuum approach we use. 
Therefore, we deem our finding of negative Poisson's ratio near the static limit to be consistent with literature. 

\begin{figure}[t!]
	\centering
	\includegraphics[width=0.45\textwidth]{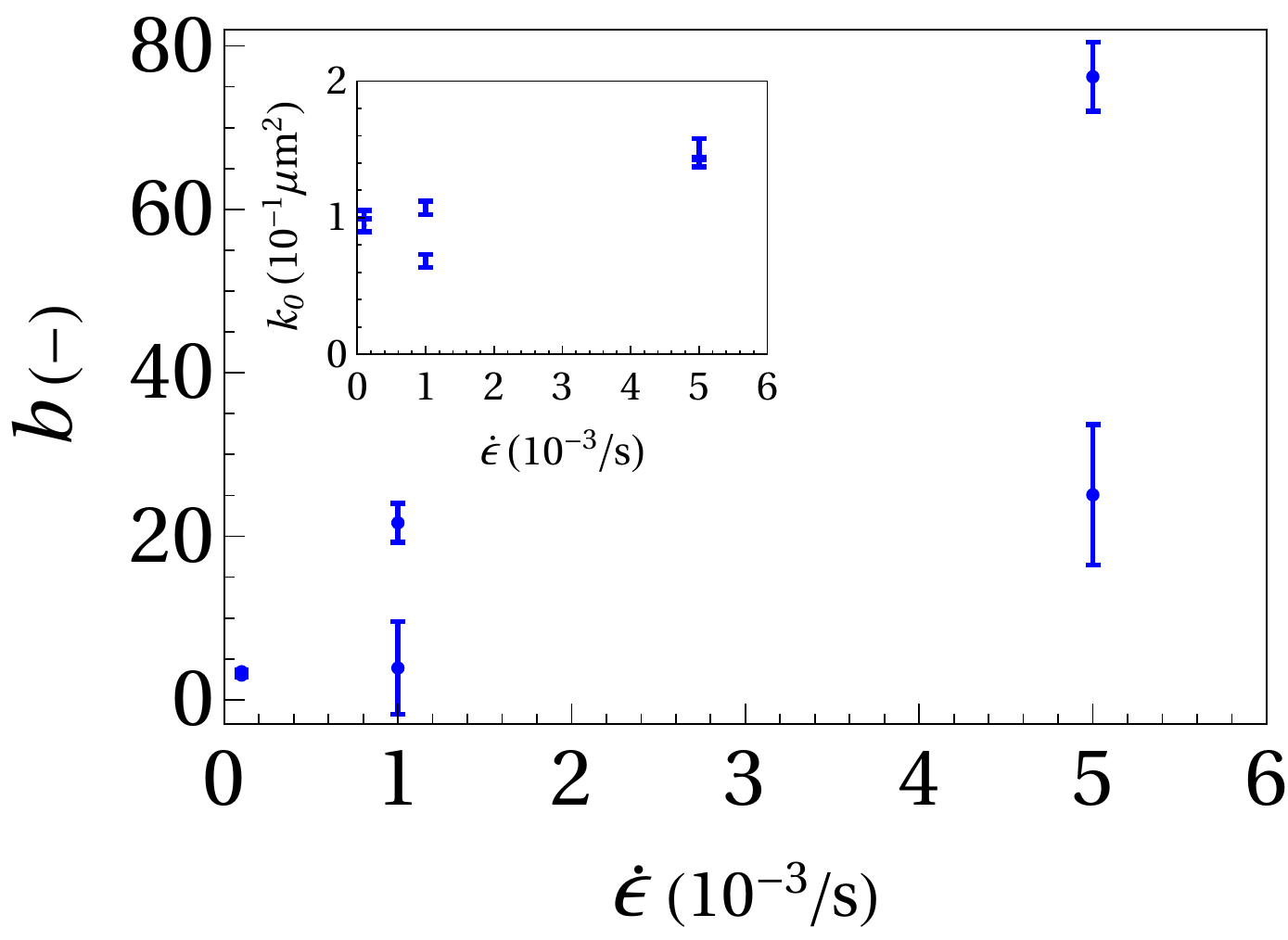}
	\caption{The proportionality constant $b$ of the Toll model obtained by fitting the normal force response, see equation \eqref{eq:Fn_large_compression}, of a large-pore fibrin fiber network in the pressurized phase during compression experiments up to 80\% strain as a function of strain rate $\dot{{\epsilon}}$, at a fibrinogen concentration of 2 mg/mL. For increasing strain rate, $b$ grows, suggesting a higher fluid velocity to induce a stronger mechanical response of the fibrin network, similar to the increase in Poisson's ratio $\nu$ for the 5--10\% compression experiments shown in Figure \ref{fig:small_compression_Poisson_ratio_vs_StrainRate}. Inset: although there is a large sample-to-sample variation, see also the inset of Figure \ref{fig:small_compression_Poisson_ratio_vs_StrainRate}, the fitted initial permeability $k_0$ does not vary appreciably with strain rate, as expected. 
	}
	\label{fig:b_versus_strain_rate}
\end{figure}

To further investigate the dependence of the mechanical response of the fibrin network on the applied strain rate, we performed compression experiments of eight consecutive compression ramps of 10\% engineering train, while we let the gel relax fully in between the ramps. 
Using equation (\ref{eq:Fn_large_compression}) we find the permeability at zero strain $k_0$ and the proportionality constant of the Toll model\cite{Toll1998PackingReinforcements} $b$ by fitting the maximum value of the normal force during each compression step, where the fluid pressure is assumed to be maximal, see Figure \ref{fig:b_versus_strain_rate}. 
Again, given a large sample-to-sample variability, the estimated initial permeability $k_0$ seems to be independent of the strain rate, as expected, see the inset of Figure \ref{fig:b_versus_strain_rate}. 
The proportionality constant $b$, however, depends significantly on the strain rate. For vanishing strain rate $\dot{\epsilon}$ it approaches a value of order unity, which agrees with literature \cite{Kim2016Foam-likeNetworks}.

\subsection{Small-pore fibrin gels}\label{sec:small-pore}

\begin{figure}[b!]
\centering
\includegraphics[width=0.45\textwidth]{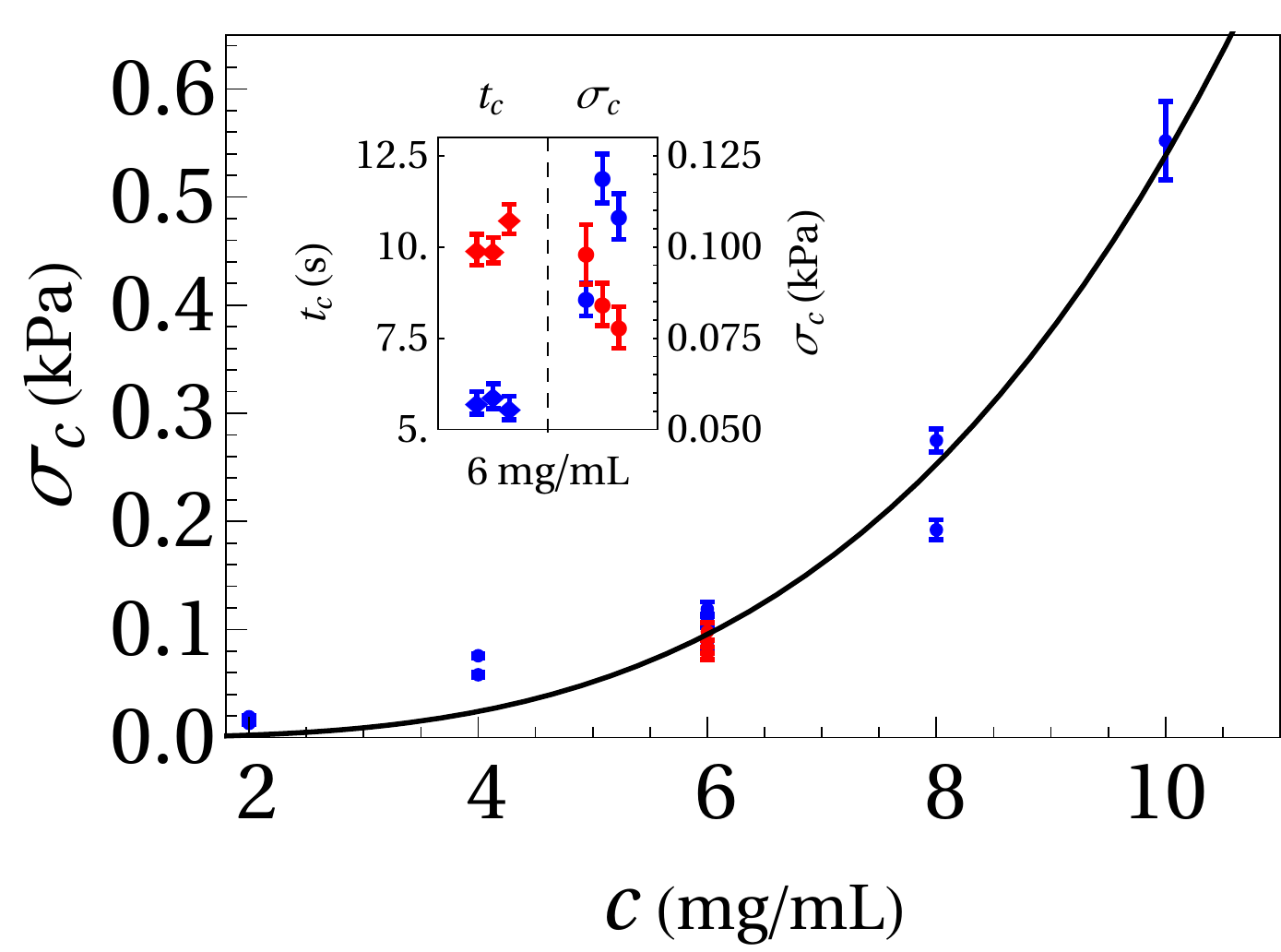}
\caption{The onset stress $\sigma_{\mathrm{{c}}}$ at which the shear modulus of small-pore fibrin gels increases as a function of fibrinogen concentration $c$. 
The onset stress can be fitted with a power law $\sigma_{\mathrm{{c}}}=\sigma_{\mathrm{{c,ref}}}\left(c/c_{\mathrm{{ref}}}\right)^{n}$ where we choose as a reference concentration $c_{\mathrm{{ref}}}=2\mathrm{{\,mg/mL}}$, and we fit $\sigma_{\mathrm{{c},ref}}=2.3\pm0.8\mathrm{{\,Pa}}$ and $n=3.4\pm0.2$, with the estimation uncertainty in brackets. 
The inset shows that the stiffening onset time $t_{\mathrm{c}}$ ($\blacklozenge$) depends on aspect ratio, blue symbols correspond to $S\equiv a / h = 20$ and red to $S=10$, whereas the onset stress $\sigma_\mathrm{c}$ ($\CIRCLE$) is geometry independent.
}
\label{fig:critical_stress}
\end{figure}

Small-pore fibrin gels exhibit a qualitatively different increase in normal force during compression, see the inset of Figure \ref{fig:example_small}, as compared to a large-pore gel, see Figure \ref{fig:example_large}. 
Initially, the normal force increases as one would expect when the volume of the gel is conserved, based on the normal force of a linear elastic volume-conserving  solid\cite{Qiao2015AnalyticalLayers} with a shear modulus equal to that of the uncompressed fibrin network $G_0$, see the blue line in the inset of Figure \ref{fig:example_small}. 
Afterwards, it increases supralinearly before it starts to relax due to fluid outflow, similar as with a large-pore gel. 
The initial supralinear increase suggests that the gel network stiffens while the fluid pressure is still low, which we take into account, as introduced in section \ref{sec:modeling}, by assuming a stepwise increase in the shear modulus at some onset time $t_{\mathrm{c}}$, providing the red curve fit in Figure \ref{fig:example_small}. 
Strain-stiffening in small-pore fibrin gels is to be expected, as their small permeability provides a relatively long volume-conserving compression, which is a shearing deformation, and fibrin networks are known to stiffen under shear\cite{Shah1997StrainClots}. 
Given the fitted values for the onset time $t_{\mathrm{c}}$ at which the shear modulus increases, we can calculate the onset stress $\sigma_{\mathrm{c}}$ at which stiffening occurs for the different fibrinogen concentrations we experimentally realized, see Figure \ref{fig:critical_stress}. 
The black curve is a power law fit to the calculated onset stresses, suggesting a sharp dependence on the fibrinogen concentration of the gel.  

As mentioned in section \ref{sec:modeling}, the calculated onset stress should not depend on the aspect ratio $S$ of the gel, since it is a microscopic property of the fibrin network. 
In the inset of Figure \ref{fig:critical_stress}, the calculated onset stress ($\CIRCLE$) and the fitted onset times ($\blacklozenge$) are shown for the experiments at $c=6\,\mathrm{mg/mL}$, where three experiments have been performed at an aspect ratio of $S=20$ (blue) and three at $S=10$ (red). 
It shows that the onset time at which stiffening occurs depends on the aspect ratio of the gel, but the onset stress does not. Therefore, the measure we defined for the onset stress in equation (\ref{eq:critical_stress}) seems appropriate. The geometry dependence of the stiffening onset time is expected, because for higher aspect ratio less compression is needed to establish a given average shear strain $\gamma_\parallel \propto \dot{\epsilon} t_\mathrm{c} a / h$ in the gel. 

As the gel is compressed further after the onset time $t_\mathrm{c}$, the fluid pressure increases due to increased bending of the fibrin network, thereby increasing the average tangential stress at the sample-plate interface: a measure for the magnitude of the shear stress in the sample. 
Therefore, due to increased shear stress, one would expect the shear modulus to increase further during compression. 
In our model, however, we assume the shear modulus to remain constant at the augmented value, which was attained at the onset time $t_\mathrm{c}$. 
The red curve fit closely matches the measurements throughout compression, see Figure \ref{fig:example_small}, suggesting further strain-stiffening of the fibrin network to be somehow suppressed.  

\begin{figure}[t!]
\centering	\includegraphics[width=0.45\textwidth]{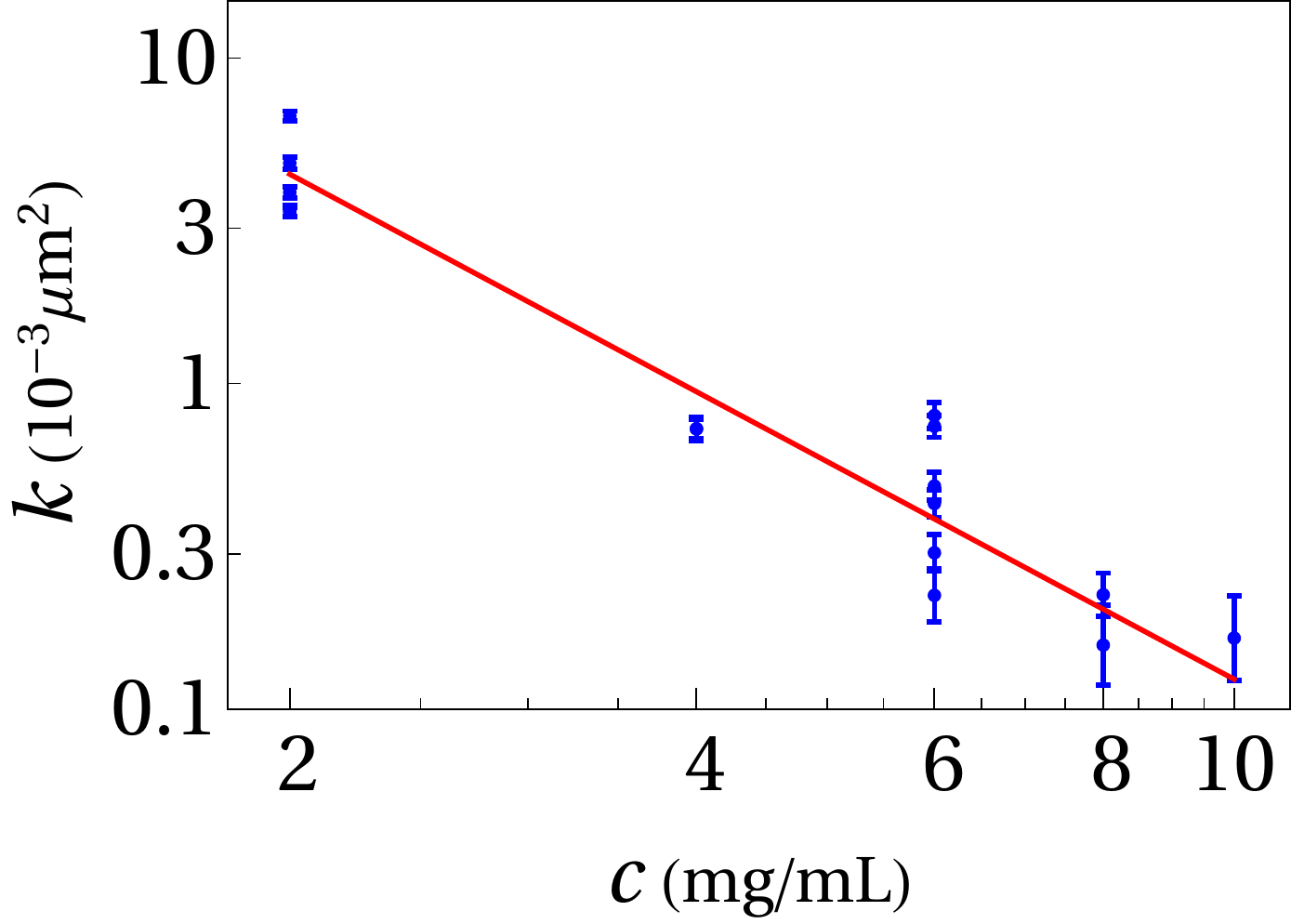}
\caption{The permeability of small-pore fibrin gels as a function of the overall fibrinogen concentration $c$. 
The red line is a fit of $k=k_\mathrm{ref}(c/c_\mathrm{ref})^n$, where we choose $c_\mathrm{ref} = 2 \mgml$ as a reference concentration and fit $k_\mathrm{ref} = (4.4\pm0.3)\cdot 10^{-3} \mum$ and $n=-2.2\pm0.5$, with the estimation uncertainty in brackets. 
This result implies that the mass density per unit length of fibrin fiber decreases with the fibrinogen concentration, see section S2 of the Supplementary Information$^\dag$ for more information. 
}
\label{fig:PermeabilityScaling}
\end{figure}

The fitted permeabilities $k$ of the small-pore fibrin networks scale as $k \propto c^{n}$, with $n=-2.2\pm0.5$, see Figure \ref{fig:PermeabilityScaling}. 
On the basis of a simple cubic lattice model, this result suggests that the mass density per unit fiber length for small-pore fibrin decreases with fibrinogen concentration, see section S2 of the Supplementary Information for more information$^\dag$. 
Finally, due to small pores the equilibrium time $t_\perp$ is generally larger than the compression time $t_\mathrm{e}$ and the fluid pressure contribution is relatively large compared to the network elasticity contribution. 
Therefore, and because of the uncertainty in the normal force measurements, we were unable to estimate the oedometric modulus $M$ from normal force measurements on all small-pore gels, and consequently we cannot calculate Poisson's ratio. 

\section{Discussion and conclusion}

We formulate a closed-form approximate solution to the poroelastic equations of motion which allows, with appropriate phenomenological extensions, to obtain the permeability and the elastic properties of a soft bonded biopolymer network from the measured time-dependent normal force in a ramp compression test. 
This approximate solution, appropriate for disk-like gels bonded to the rheometer plates, differs strongly from that for frictionless gels, it distinguishes the fluid and network contribution to the normal force, and it holds for \textit{all times} during ramp compression, which allows for the quantification of strain stiffening during compression, in contrast to previous approximate approaches\cite{Yamaue2004TheoryConstraint,Doi2009GelDynamics}.  

The normal force contribution of the large-pore fibrin networks is found to increase with increasing strain rate, suggesting fluid flow through the network to make it more resistant against volume change, both for small (5--10\%) and large ($\leq$80\%) compressive strains. 
This strain rate dependence suggests that the network-fluid interactions cause, apart from the fluid pressure, a change in the elastic response of the fibrin fiber network. The microscopic details giving rise to this change in the elastic response are a subject for further research. 

In earlier work the normal force response of fibrin gels has been found to be dependent on the strain rate of compression\cite{Kim2014StructuralCompression,Kim2016Foam-likeNetworks}. 
In these works the contributions of fluid pressure and the fibrin network were not separated in a biphasic model, however, and the network elasticity under compression could not be extracted, see for example Figure 5 in Kim \textit{et al.}\cite{Kim2016Foam-likeNetworks}. 
Our work enables the separation of fluid pressure and network elasticity and shows that, additionally to the fluid pressure contribution, the network elasticity also depends on the applied strain rate. 

The small-pore fibrin networks are found to strain-stiffen close to the start of compression, and the shear modulus is found to remain at the magnitude attained at the onset time although the shear stress on the network still increases. 
Possibly, the flow of fluid through the fiber network suppresses the irreversible changes in the hierarchical structure of the fibrin network which are needed for reversible strain-stiffening\cite{Kurniawan2016FibrinScales}. 

From compression experiments on small-pore fibrin gels we find the permeability to scale inverse squared with fibrinogen concentration. 
With the cubic lattice model this suggests that the fibrin fiber mass density per unit of fiber length decreases with fibrinogen concentration. 
Considering the complex polymerization kinetics of fibrin from its soluble precursor fibrinogen\cite{Weisel2017FibrinProperties}, this dependence could be caused by the enhanced local depletion of protofibrils with increasing fibrinogen concentration, thereby decreasing the mass density per unit length\cite{Weisel1992ComputerControlled}. 

This work quantitatively describes how the complex mechanical behaviour of biopolymer systems can be decomposed into simple physical principles. It provides an alternative method to determine the hydraulic permeability of biopolymer systems based on simple compression measurements, rather than flow-through assays, with the added benefit that their elastic properties are probed at the same time. Therefore, we expect our findings to prove fruitful in, for example, mechanobiological investigations of the relation between fluid flow and the elasticity of biopolymer networks and soft tissues. 

\section*{Conflicts of interest}
There are no conflicts of interest to declare. 

\section*{Acknowledgements}

This work is part of the Industrial Partnership Programme Hybrid Soft Materials that is carried out under an agreement between Unilever R\&D B.V.\ and the Netherlands Organisation for Scientific Research (NWO), and of FOM Program grant no.\ 143, also funded by NWO. 
We thank Kristina Ganzinger for her helpful comments. 

\balance

\bibliography{references}

\providecommand*{\mcitethebibliography}{\thebibliography}
\csname @ifundefined\endcsname{endmcitethebibliography}
{\let\endmcitethebibliography\endthebibliography}{}
\begin{mcitethebibliography}{47}
\providecommand*{\natexlab}[1]{#1}
\providecommand*{\mciteSetBstSublistMode}[1]{}
\providecommand*{\mciteSetBstMaxWidthForm}[2]{}
\providecommand*{\mciteBstWouldAddEndPuncttrue}
  {\def\EndOfBibitem{\unskip.}}
\providecommand*{\mciteBstWouldAddEndPunctfalse}
  {\let\EndOfBibitem\relax}
\providecommand*{\mciteSetBstMidEndSepPunct}[3]{}
\providecommand*{\mciteSetBstSublistLabelBeginEnd}[3]{}
\providecommand*{\EndOfBibitem}{}
\mciteSetBstSublistMode{f}
\mciteSetBstMaxWidthForm{subitem}
{(\emph{\alph{mcitesubitemcount}})}
\mciteSetBstSublistLabelBeginEnd{\mcitemaxwidthsubitemform\space}
{\relax}{\relax}

\bibitem[Pritchard \emph{et~al.}(2014)Pritchard, Shery~Huang, and
  Terentjev]{Pritchard2014MechanicsTissue}
R.~H. Pritchard, Y.~Y. Shery~Huang and E.~M. Terentjev, \emph{Soft Matter},
  2014, \textbf{10}, 1864--1884\relax
\mciteBstWouldAddEndPuncttrue
\mciteSetBstMidEndSepPunct{\mcitedefaultmidpunct}
{\mcitedefaultendpunct}{\mcitedefaultseppunct}\relax
\EndOfBibitem
\bibitem[Huber \emph{et~al.}(2015)Huber, Boire, L{\'{o}}pez, and
  Koenderink]{Huber2015CytoskeletalUp}
F.~Huber, A.~Boire, M.~P. L{\'{o}}pez and G.~H. Koenderink, \emph{Current
  Opinion in Cell Biology}, 2015, \textbf{32}, 39--47\relax
\mciteBstWouldAddEndPuncttrue
\mciteSetBstMidEndSepPunct{\mcitedefaultmidpunct}
{\mcitedefaultendpunct}{\mcitedefaultseppunct}\relax
\EndOfBibitem
\bibitem[Mouw \emph{et~al.}(2014)Mouw, Ou, and
  Weaver]{Mouw2014ExtracellularDeconstruction}
J.~K. Mouw, G.~Ou and V.~M. Weaver, \emph{Nature Reviews Molecular Cell
  Biology}, 2014, \textbf{15}, 771--785\relax
\mciteBstWouldAddEndPuncttrue
\mciteSetBstMidEndSepPunct{\mcitedefaultmidpunct}
{\mcitedefaultendpunct}{\mcitedefaultseppunct}\relax
\EndOfBibitem
\bibitem[Vogel(2018)]{Vogel2018UnravelingMatrix}
V.~Vogel, \emph{Annual Review of Physiology}, 2018, \textbf{80}, 353--387\relax
\mciteBstWouldAddEndPuncttrue
\mciteSetBstMidEndSepPunct{\mcitedefaultmidpunct}
{\mcitedefaultendpunct}{\mcitedefaultseppunct}\relax
\EndOfBibitem
\bibitem[Weisel and Litvinov(2017)]{Weisel2017FibrinProperties}
J.~W. Weisel and R.~I. Litvinov, \emph{Fibrous Proteins: Structures and
  Mechanisms}, Springer International Publishing, 2017, vol.~82, pp.
  405--456\relax
\mciteBstWouldAddEndPuncttrue
\mciteSetBstMidEndSepPunct{\mcitedefaultmidpunct}
{\mcitedefaultendpunct}{\mcitedefaultseppunct}\relax
\EndOfBibitem
\bibitem[Bagoly \emph{et~al.}(2017)Bagoly, Ari{\"{e}}ns, Rijken, Pieters, and
  Wolberg]{Bagoly2017ClotHemostasis}
Z.~Bagoly, R.~A. Ari{\"{e}}ns, D.~C. Rijken, M.~Pieters and A.~S. Wolberg,
  \emph{BioMed Research International}, 2017, \textbf{2017}, year\relax
\mciteBstWouldAddEndPuncttrue
\mciteSetBstMidEndSepPunct{\mcitedefaultmidpunct}
{\mcitedefaultendpunct}{\mcitedefaultseppunct}\relax
\EndOfBibitem
\bibitem[Swartz and Fleury(2007)]{Swartz2007InterstitialTissues}
M.~A. Swartz and M.~E. Fleury, \emph{Annual Review of Biomedical Engineering},
  2007, \textbf{9}, 229--256\relax
\mciteBstWouldAddEndPuncttrue
\mciteSetBstMidEndSepPunct{\mcitedefaultmidpunct}
{\mcitedefaultendpunct}{\mcitedefaultseppunct}\relax
\EndOfBibitem
\bibitem[Wilson(2005)]{Wilson2005ATissues}
W.~Wilson, \emph{Journal of Biomechanical Engineering}, 2005, \textbf{127},
  158\relax
\mciteBstWouldAddEndPuncttrue
\mciteSetBstMidEndSepPunct{\mcitedefaultmidpunct}
{\mcitedefaultendpunct}{\mcitedefaultseppunct}\relax
\EndOfBibitem
\bibitem[Avendano \emph{et~al.}(2019)Avendano, Cortes-Medina, and
  Song]{Avendano2019ApplicationMatrix}
A.~Avendano, M.~Cortes-Medina and J.~W. Song, \emph{Frontiers in Bioengineering
  and Biotechnology}, 2019, \textbf{7}, 1--8\relax
\mciteBstWouldAddEndPuncttrue
\mciteSetBstMidEndSepPunct{\mcitedefaultmidpunct}
{\mcitedefaultendpunct}{\mcitedefaultseppunct}\relax
\EndOfBibitem
\bibitem[Moeendarbary \emph{et~al.}(2013)Moeendarbary, Valon, Fritzsche,
  Harris, Moulding, Thrasher, Stride, Mahadevan, and
  Charras]{Moeendarbary2013TheMaterial}
E.~Moeendarbary, L.~Valon, M.~Fritzsche, A.~R. Harris, D.~A. Moulding, A.~J.
  Thrasher, E.~Stride, L.~Mahadevan and G.~T. Charras, \emph{Nature Materials},
  2013, \textbf{12}, 253--261\relax
\mciteBstWouldAddEndPuncttrue
\mciteSetBstMidEndSepPunct{\mcitedefaultmidpunct}
{\mcitedefaultendpunct}{\mcitedefaultseppunct}\relax
\EndOfBibitem
\bibitem[Charras \emph{et~al.}(2009)Charras, Mitchison, and
  Mahadevan]{Charras2009AnimalHydraulics}
G.~T. Charras, T.~J. Mitchison and L.~Mahadevan, \emph{Journal of Cell
  Science}, 2009, \textbf{122}, 3233--3241\relax
\mciteBstWouldAddEndPuncttrue
\mciteSetBstMidEndSepPunct{\mcitedefaultmidpunct}
{\mcitedefaultendpunct}{\mcitedefaultseppunct}\relax
\EndOfBibitem
\bibitem[Voronov \emph{et~al.}(2013)Voronov, Stalker, Brass, and
  Diamond]{Voronov2013SimulationResolution}
R.~S. Voronov, T.~J. Stalker, L.~F. Brass and S.~L. Diamond, \emph{Annals of
  Biomedical Engineering}, 2013, \textbf{41}, 1297--1307\relax
\mciteBstWouldAddEndPuncttrue
\mciteSetBstMidEndSepPunct{\mcitedefaultmidpunct}
{\mcitedefaultendpunct}{\mcitedefaultseppunct}\relax
\EndOfBibitem
\bibitem[Chen \emph{et~al.}(2018)Chen, Shi, Gong, Zhang, Zhong, Zhang, Zhou,
  and Lou]{Chen2018ThrombusTherapy}
Z.~Chen, F.~Shi, X.~Gong, R.~Zhang, W.~Zhong, R.~Zhang, Y.~Zhou and M.~Lou,
  \emph{American Journal of Neuroradiology}, 2018, \textbf{39},
  1854--1859\relax
\mciteBstWouldAddEndPuncttrue
\mciteSetBstMidEndSepPunct{\mcitedefaultmidpunct}
{\mcitedefaultendpunct}{\mcitedefaultseppunct}\relax
\EndOfBibitem
\bibitem[Santos \emph{et~al.}(2016)Santos, Marquering, Den~Blanken, Berkhemer,
  Boers, Yoo, Beenen, Treurniet, Wismans, Van~Noort, Lingsma, Dippel, Van
  Der~Lugt, Van~Zwam, Roos, Van~Oostenbrugge, Niessen, and
  Majoie]{Santos2016ThrombusStroke}
E.~M. Santos, H.~A. Marquering, M.~D. Den~Blanken, O.~A. Berkhemer, A.~M.
  Boers, A.~J. Yoo, L.~F. Beenen, K.~M. Treurniet, C.~Wismans, K.~Van~Noort,
  H.~F. Lingsma, D.~W. Dippel, A.~Van Der~Lugt, W.~H. Van~Zwam, Y.~B. Roos,
  R.~J. Van~Oostenbrugge, W.~J. Niessen and C.~B. Majoie, \emph{Stroke}, 2016,
  \textbf{47}, 732--741\relax
\mciteBstWouldAddEndPuncttrue
\mciteSetBstMidEndSepPunct{\mcitedefaultmidpunct}
{\mcitedefaultendpunct}{\mcitedefaultseppunct}\relax
\EndOfBibitem
\bibitem[Tokita and Tanaka(1991)]{Tokita2000}
M.~Tokita and T.~Tanaka, \emph{Journal of Chemical Physics}, 1991, \textbf{95},
  4613--4619\relax
\mciteBstWouldAddEndPuncttrue
\mciteSetBstMidEndSepPunct{\mcitedefaultmidpunct}
{\mcitedefaultendpunct}{\mcitedefaultseppunct}\relax
\EndOfBibitem
\bibitem[Pieters \emph{et~al.}(2012)Pieters, Undas, Marchi, De~Maat, Weisel,
  and Ari{\"{e}}ns]{Pieters2012AnValues}
M.~Pieters, A.~Undas, R.~Marchi, M.~P.~M. De~Maat, J.~W. Weisel and R.~A.~S.
  Ari{\"{e}}ns, \emph{Journal of Thrombosis and Haemostasis}, 2012,
  \textbf{10}, 2179--2181\relax
\mciteBstWouldAddEndPuncttrue
\mciteSetBstMidEndSepPunct{\mcitedefaultmidpunct}
{\mcitedefaultendpunct}{\mcitedefaultseppunct}\relax
\EndOfBibitem
\bibitem[Maity \emph{et~al.}(2019)Maity, Li, Chen, and
  Sun]{Maity2019ResponseHydrogels}
D.~Maity, Y.~Li, Y.~Chen and S.~X. Sun, \emph{Soft Matter}, 2019,
  2617--2626\relax
\mciteBstWouldAddEndPuncttrue
\mciteSetBstMidEndSepPunct{\mcitedefaultmidpunct}
{\mcitedefaultendpunct}{\mcitedefaultseppunct}\relax
\EndOfBibitem
\bibitem[Storm \emph{et~al.}(2005)Storm, Pastore, Mackintosh, Lubensky, and
  Janmey]{Storm2005NonlinearGels}
C.~Storm, J.~J. Pastore, F.~C. Mackintosh, T.~C. Lubensky and P.~A. Janmey,
  \emph{Nature}, 2005, \textbf{435}, 191--194\relax
\mciteBstWouldAddEndPuncttrue
\mciteSetBstMidEndSepPunct{\mcitedefaultmidpunct}
{\mcitedefaultendpunct}{\mcitedefaultseppunct}\relax
\EndOfBibitem
\bibitem[Shadwick(1999)]{Shadwick1999MechanicalArteries}
R.~E. Shadwick, \emph{The Journal of experimental biology}, 1999, \textbf{202},
  3305--3313\relax
\mciteBstWouldAddEndPuncttrue
\mciteSetBstMidEndSepPunct{\mcitedefaultmidpunct}
{\mcitedefaultendpunct}{\mcitedefaultseppunct}\relax
\EndOfBibitem
\bibitem[Schultz \emph{et~al.}(2010)Schultz, Ladwig, and
  Wysocki]{Schultz2010ExtracellularWounds}
G.~S. Schultz, G.~Ladwig and A.~Wysocki, \emph{World Wide Wounds}, 2010,
  1--8\relax
\mciteBstWouldAddEndPuncttrue
\mciteSetBstMidEndSepPunct{\mcitedefaultmidpunct}
{\mcitedefaultendpunct}{\mcitedefaultseppunct}\relax
\EndOfBibitem
\bibitem[Burla \emph{et~al.}(2019)Burla, Mulla, Vos, Aufderhorst-Roberts, and
  Koenderink]{Burla2019FromNetworks}
F.~Burla, Y.~Mulla, B.~E. Vos, A.~Aufderhorst-Roberts and G.~H. Koenderink,
  \emph{Nature Reviews Physics}, 2019, \textbf{1}, 249--263\relax
\mciteBstWouldAddEndPuncttrue
\mciteSetBstMidEndSepPunct{\mcitedefaultmidpunct}
{\mcitedefaultendpunct}{\mcitedefaultseppunct}\relax
\EndOfBibitem
\bibitem[Brown \emph{et~al.}(2009)Brown, Litvinov, Discher, Purohit, and
  Weisel]{Brown2009MultiscaleWater}
A.~E.~X. Brown, R.~I. Litvinov, D.~E. Discher, P.~K. Purohit and J.~W. Weisel,
  \emph{Science}, 2009, \textbf{325}, 741--744\relax
\mciteBstWouldAddEndPuncttrue
\mciteSetBstMidEndSepPunct{\mcitedefaultmidpunct}
{\mcitedefaultendpunct}{\mcitedefaultseppunct}\relax
\EndOfBibitem
\bibitem[Roeder(2009)]{Roeder2009FibrilMatrices}
B.~A. Roeder, \emph{Journal of Biomechanical Engineering}, 2009, \textbf{131},
  031004\relax
\mciteBstWouldAddEndPuncttrue
\mciteSetBstMidEndSepPunct{\mcitedefaultmidpunct}
{\mcitedefaultendpunct}{\mcitedefaultseppunct}\relax
\EndOfBibitem
\bibitem[van Oosten \emph{et~al.}(2016)van Oosten, Vahabi, Licup, Sharma,
  Galie, MacKintosh, and Janmey]{vanOosten2016UncouplingStretch-stiffening}
A.~S.~G. van Oosten, M.~Vahabi, A.~J. Licup, A.~Sharma, P.~A. Galie, F.~C.
  MacKintosh and P.~A. Janmey, \emph{Scientific Reports}, 2016, \textbf{6},
  1--9\relax
\mciteBstWouldAddEndPuncttrue
\mciteSetBstMidEndSepPunct{\mcitedefaultmidpunct}
{\mcitedefaultendpunct}{\mcitedefaultseppunct}\relax
\EndOfBibitem
\bibitem[Kim \emph{et~al.}(2014)Kim, Litvinov, Weisel, and
  Alber]{Kim2014StructuralCompression}
O.~V. Kim, R.~I. Litvinov, J.~W. Weisel and M.~S. Alber, \emph{Biomaterials},
  2014, \textbf{35}, 6739--6749\relax
\mciteBstWouldAddEndPuncttrue
\mciteSetBstMidEndSepPunct{\mcitedefaultmidpunct}
{\mcitedefaultendpunct}{\mcitedefaultseppunct}\relax
\EndOfBibitem
\bibitem[Kim \emph{et~al.}(2016)Kim, Liang, Litvinov, Weisel, Alber, and
  Purohit]{Kim2016Foam-likeNetworks}
O.~V. Kim, X.~Liang, R.~I. Litvinov, J.~W. Weisel, M.~S. Alber and P.~K.
  Purohit, \emph{Biomechanics and Modeling in Mechanobiology}, 2016,
  \textbf{15}, 213--228\relax
\mciteBstWouldAddEndPuncttrue
\mciteSetBstMidEndSepPunct{\mcitedefaultmidpunct}
{\mcitedefaultendpunct}{\mcitedefaultseppunct}\relax
\EndOfBibitem
\bibitem[Macminn \emph{et~al.}(2016)Macminn, Dufresne, and
  Wettlaufer]{Macminn2016LargeMaterial}
C.~W. Macminn, E.~R. Dufresne and J.~S. Wettlaufer, \emph{Physical Review
  Applied}, 2016, \textbf{5}, 30\relax
\mciteBstWouldAddEndPuncttrue
\mciteSetBstMidEndSepPunct{\mcitedefaultmidpunct}
{\mcitedefaultendpunct}{\mcitedefaultseppunct}\relax
\EndOfBibitem
\bibitem[Coussy(2004)]{coussy2004poromechanics}
O.~Coussy, \emph{{Poromechanics}}, Wiley, 2004\relax
\mciteBstWouldAddEndPuncttrue
\mciteSetBstMidEndSepPunct{\mcitedefaultmidpunct}
{\mcitedefaultendpunct}{\mcitedefaultseppunct}\relax
\EndOfBibitem
\bibitem[de~Boer(2000)]{deBoer2000TheoryMedia}
R.~de~Boer, \emph{{Theory of Porous Media}}, Springer-Verlag Berlin Heidelberg,
  1st edn, 2000, p. 618\relax
\mciteBstWouldAddEndPuncttrue
\mciteSetBstMidEndSepPunct{\mcitedefaultmidpunct}
{\mcitedefaultendpunct}{\mcitedefaultseppunct}\relax
\EndOfBibitem
\bibitem[Doi(2009)]{Doi2009GelDynamics}
M.~Doi, \emph{Journal of the Physical Society of Japan}, 2009, \textbf{78},
  1--19\relax
\mciteBstWouldAddEndPuncttrue
\mciteSetBstMidEndSepPunct{\mcitedefaultmidpunct}
{\mcitedefaultendpunct}{\mcitedefaultseppunct}\relax
\EndOfBibitem
\bibitem[Brynk \emph{et~al.}(2011)Brynk, Hellmich, Fritsch, Zysset, and
  Eberhardsteiner]{Brynk2011ExperimentalFluctuations}
T.~Brynk, C.~Hellmich, A.~Fritsch, P.~Zysset and J.~Eberhardsteiner,
  \emph{Journal of Biomechanics}, 2011, \textbf{44}, 501--508\relax
\mciteBstWouldAddEndPuncttrue
\mciteSetBstMidEndSepPunct{\mcitedefaultmidpunct}
{\mcitedefaultendpunct}{\mcitedefaultseppunct}\relax
\EndOfBibitem
\bibitem[Cardoso \emph{et~al.}(2013)Cardoso, Fritton, Gailani, Benalla, and
  Cowin]{Cardoso2013AdvancesFlow}
L.~Cardoso, S.~P. Fritton, G.~Gailani, M.~Benalla and S.~C. Cowin,
  \emph{Journal of Biomechanics}, 2013, \textbf{46}, 253--265\relax
\mciteBstWouldAddEndPuncttrue
\mciteSetBstMidEndSepPunct{\mcitedefaultmidpunct}
{\mcitedefaultendpunct}{\mcitedefaultseppunct}\relax
\EndOfBibitem
\bibitem[De~Cagny \emph{et~al.}(2016)De~Cagny, Vos, Vahabi, Kurniawan, Doi,
  Koenderink, MacKintosh, and Bonn]{DeCagny2016PorosityGels}
H.~C. De~Cagny, B.~E. Vos, M.~Vahabi, N.~A. Kurniawan, M.~Doi, G.~H.
  Koenderink, F.~C. MacKintosh and D.~Bonn, \emph{Physical Review Letters},
  2016, \textbf{117}, 1--5\relax
\mciteBstWouldAddEndPuncttrue
\mciteSetBstMidEndSepPunct{\mcitedefaultmidpunct}
{\mcitedefaultendpunct}{\mcitedefaultseppunct}\relax
\EndOfBibitem
\bibitem[Yeromonahos \emph{et~al.}(2010)Yeromonahos, Polack, and
  Caton]{Yeromonahos2010NanostructureClot}
C.~Yeromonahos, B.~Polack and F.~Caton, \emph{Biophysical Journal}, 2010,
  \textbf{99}, 2018--2027\relax
\mciteBstWouldAddEndPuncttrue
\mciteSetBstMidEndSepPunct{\mcitedefaultmidpunct}
{\mcitedefaultendpunct}{\mcitedefaultseppunct}\relax
\EndOfBibitem
\bibitem[Bale \emph{et~al.}(1985)Bale, Mullert, and
  Ferry]{Bale1985EffectsCreep/peptides}
M.~D. Bale, M.~F. Mullert and J.~D. Ferry, \emph{Biochemistry}, 1985,
  \textbf{82}, 1410--1413\relax
\mciteBstWouldAddEndPuncttrue
\mciteSetBstMidEndSepPunct{\mcitedefaultmidpunct}
{\mcitedefaultendpunct}{\mcitedefaultseppunct}\relax
\EndOfBibitem
\bibitem[Macrae \emph{et~al.}(2018)Macrae, Duval, Papareddy, Baker, Yuldasheva,
  Kearney, McPherson, Asquith, Konings, Casini, Degen, Connell, Philippou,
  Wolberg, Herwald, and Ari{\"{e}}ns]{Macrae2018AInvasion}
F.~L. Macrae, C.~Duval, P.~Papareddy, S.~R. Baker, N.~Yuldasheva, K.~J.
  Kearney, H.~R. McPherson, N.~Asquith, J.~Konings, A.~Casini, J.~L. Degen,
  S.~D. Connell, H.~Philippou, A.~S. Wolberg, H.~Herwald and R.~A.~S.
  Ari{\"{e}}ns, \emph{The Journal of Clinical Investigation}, 2018,
  \textbf{128}, 3356--3368\relax
\mciteBstWouldAddEndPuncttrue
\mciteSetBstMidEndSepPunct{\mcitedefaultmidpunct}
{\mcitedefaultendpunct}{\mcitedefaultseppunct}\relax
\EndOfBibitem
\bibitem[Armstrong \emph{et~al.}(1984)Armstrong, Lai, and
  Mow]{Armstrong1984AnCartilage}
C.~G. Armstrong, W.~M. Lai and V.~C. Mow, \emph{Journal of Biomechanical
  Engineering}, 1984, \textbf{106}, 165--173\relax
\mciteBstWouldAddEndPuncttrue
\mciteSetBstMidEndSepPunct{\mcitedefaultmidpunct}
{\mcitedefaultendpunct}{\mcitedefaultseppunct}\relax
\EndOfBibitem
\bibitem[Qiao and Lu(2015)]{Qiao2015AnalyticalLayers}
S.~Qiao and N.~Lu, \emph{International Journal of Solids and Structures}, 2015,
  \textbf{58}, 353--365\relax
\mciteBstWouldAddEndPuncttrue
\mciteSetBstMidEndSepPunct{\mcitedefaultmidpunct}
{\mcitedefaultendpunct}{\mcitedefaultseppunct}\relax
\EndOfBibitem
\bibitem[Piechocka \emph{et~al.}(2010)Piechocka, Bacabac, Potters, MacKintosh,
  and Koenderink]{Piechocka2010StructuralMechanics}
I.~K. Piechocka, R.~G. Bacabac, M.~Potters, F.~C. MacKintosh and G.~H.
  Koenderink, \emph{Biophysical Journal}, 2010, \textbf{98}, 2273--2280\relax
\mciteBstWouldAddEndPuncttrue
\mciteSetBstMidEndSepPunct{\mcitedefaultmidpunct}
{\mcitedefaultendpunct}{\mcitedefaultseppunct}\relax
\EndOfBibitem
\bibitem[Toll(1998)]{Toll1998PackingReinforcements}
S.~Toll, \emph{Polymer Engineering and Science}, 1998, \textbf{38},
  1337--1350\relax
\mciteBstWouldAddEndPuncttrue
\mciteSetBstMidEndSepPunct{\mcitedefaultmidpunct}
{\mcitedefaultendpunct}{\mcitedefaultseppunct}\relax
\EndOfBibitem
\bibitem[Ban \emph{et~al.}(2019)Ban, Wang, Franklin, Liphardt, Janmey, and
  Shenoy]{Ban2019StrongNetworks}
E.~Ban, H.~Wang, J.~M. Franklin, J.~T. Liphardt, P.~A. Janmey and V.~B. Shenoy,
  \emph{Proceedings of the National Academy of Sciences}, 2019, \textbf{116},
  6790--6799\relax
\mciteBstWouldAddEndPuncttrue
\mciteSetBstMidEndSepPunct{\mcitedefaultmidpunct}
{\mcitedefaultendpunct}{\mcitedefaultseppunct}\relax
\EndOfBibitem
\bibitem[Steinwachs \emph{et~al.}(2016)Steinwachs, Metzner, Skodzek, Lang,
  Thievessen, Mark, M{\"{u}}nster, Aifantis, and
  Fabry]{Steinwachs2016Three-dimensionalNetworks}
J.~Steinwachs, C.~Metzner, K.~Skodzek, N.~Lang, I.~Thievessen, C.~Mark,
  S.~M{\"{u}}nster, K.~E. Aifantis and B.~Fabry, \emph{Nature Methods}, 2016,
  \textbf{13}, 171--176\relax
\mciteBstWouldAddEndPuncttrue
\mciteSetBstMidEndSepPunct{\mcitedefaultmidpunct}
{\mcitedefaultendpunct}{\mcitedefaultseppunct}\relax
\EndOfBibitem
\bibitem[Shivers \emph{et~al.}(2019)Shivers, Arzash, and
  MacKintosh]{Shivers2019NonlinearTransition}
J.~L. Shivers, S.~Arzash and F.~C. MacKintosh, \emph{arXiv:1905.09844v1
  [cond-mat.soft]}, 2019,  1--7\relax
\mciteBstWouldAddEndPuncttrue
\mciteSetBstMidEndSepPunct{\mcitedefaultmidpunct}
{\mcitedefaultendpunct}{\mcitedefaultseppunct}\relax
\EndOfBibitem
\bibitem[Shah and Janmey(1997)]{Shah1997StrainClots}
J.~V. Shah and P.~A. Janmey, \emph{Rheologica Acta}, 1997, \textbf{36},
  262--268\relax
\mciteBstWouldAddEndPuncttrue
\mciteSetBstMidEndSepPunct{\mcitedefaultmidpunct}
{\mcitedefaultendpunct}{\mcitedefaultseppunct}\relax
\EndOfBibitem
\bibitem[Yamaue and Doi(2004)]{Yamaue2004TheoryConstraint}
T.~Yamaue and M.~Doi, \emph{Physical Review E - Statistical Physics, Plasmas,
  Fluids, and Related Interdisciplinary Topics}, 2004, \textbf{69}, 5\relax
\mciteBstWouldAddEndPuncttrue
\mciteSetBstMidEndSepPunct{\mcitedefaultmidpunct}
{\mcitedefaultendpunct}{\mcitedefaultseppunct}\relax
\EndOfBibitem
\bibitem[Kurniawan \emph{et~al.}(2016)Kurniawan, Vos, Biebricher, Wuite,
  Peterman, and Koenderink]{Kurniawan2016FibrinScales}
N.~Kurniawan, B.~Vos, A.~Biebricher, G.~Wuite, E.~Peterman and G.~Koenderink,
  \emph{Biophysical Journal}, 2016, \textbf{111}, 1026--1034\relax
\mciteBstWouldAddEndPuncttrue
\mciteSetBstMidEndSepPunct{\mcitedefaultmidpunct}
{\mcitedefaultendpunct}{\mcitedefaultseppunct}\relax
\EndOfBibitem
\bibitem[Weisel and Nagaswami(1992)]{Weisel1992ComputerControlled}
J.~W. Weisel and C.~Nagaswami, \emph{Biophysical Journal}, 1992, \textbf{63},
  111--128\relax
\mciteBstWouldAddEndPuncttrue
\mciteSetBstMidEndSepPunct{\mcitedefaultmidpunct}
{\mcitedefaultendpunct}{\mcitedefaultseppunct}\relax
\EndOfBibitem
\end{mcitethebibliography}


\providecommand*{\mcitethebibliography}{\thebibliography}
\csname @ifundefined\endcsname{endmcitethebibliography}
{\let\endmcitethebibliography\endthebibliography}{}
\begin{mcitethebibliography}{10}
\providecommand*{\natexlab}[1]{#1}
\providecommand*{\mciteSetBstSublistMode}[1]{}
\providecommand*{\mciteSetBstMaxWidthForm}[2]{}
\providecommand*{\mciteBstWouldAddEndPuncttrue}
  {\def\EndOfBibitem{\unskip.}}
\providecommand*{\mciteBstWouldAddEndPunctfalse}
  {\let\EndOfBibitem\relax}
\providecommand*{\mciteSetBstMidEndSepPunct}[3]{}
\providecommand*{\mciteSetBstSublistLabelBeginEnd}[3]{}
\providecommand*{\EndOfBibitem}{}
\mciteSetBstSublistMode{f}
\mciteSetBstMaxWidthForm{subitem}
{(\emph{\alph{mcitesubitemcount}})}
\mciteSetBstSublistLabelBeginEnd{\mcitemaxwidthsubitemform\space}
{\relax}{\relax}

\bibitem[Armstrong \emph{et~al.}(1984)Armstrong, Lai, and
  Mow]{Armstrong1984AnCartilage}
C.~G. Armstrong, W.~M. Lai and V.~C. Mow, \emph{Journal of Biomechanical
  Engineering}, 1984, \textbf{106}, 165--173\relax
\mciteBstWouldAddEndPuncttrue
\mciteSetBstMidEndSepPunct{\mcitedefaultmidpunct}
{\mcitedefaultendpunct}{\mcitedefaultseppunct}\relax
\EndOfBibitem
\bibitem[Qiao and Lu(2015)]{Qiao2015AnalyticalLayers}
S.~Qiao and N.~Lu, \emph{International Journal of Solids and Structures}, 2015,
  \textbf{58}, 353--365\relax
\mciteBstWouldAddEndPuncttrue
\mciteSetBstMidEndSepPunct{\mcitedefaultmidpunct}
{\mcitedefaultendpunct}{\mcitedefaultseppunct}\relax
\EndOfBibitem
\bibitem[Yamaue and Doi(2004)]{Yamaue2004TheoryConstraint}
T.~Yamaue and M.~Doi, \emph{Physical Review E - Statistical Physics, Plasmas,
  Fluids, and Related Interdisciplinary Topics}, 2004, \textbf{69}, 5\relax
\mciteBstWouldAddEndPuncttrue
\mciteSetBstMidEndSepPunct{\mcitedefaultmidpunct}
{\mcitedefaultendpunct}{\mcitedefaultseppunct}\relax
\EndOfBibitem
\bibitem[De~Cagny \emph{et~al.}(2016)De~Cagny, Vos, Vahabi, Kurniawan, Doi,
  Koenderink, MacKintosh, and Bonn]{DeCagny2016PorosityGels}
H.~C. De~Cagny, B.~E. Vos, M.~Vahabi, N.~A. Kurniawan, M.~Doi, G.~H.
  Koenderink, F.~C. MacKintosh and D.~Bonn, \emph{Physical Review Letters},
  2016, \textbf{117}, 1--5\relax
\mciteBstWouldAddEndPuncttrue
\mciteSetBstMidEndSepPunct{\mcitedefaultmidpunct}
{\mcitedefaultendpunct}{\mcitedefaultseppunct}\relax
\EndOfBibitem
\bibitem[Yeromonahos \emph{et~al.}(2010)Yeromonahos, Polack, and
  Caton]{Yeromonahos2010NanostructureClot}
C.~Yeromonahos, B.~Polack and F.~Caton, \emph{Biophysical Journal}, 2010,
  \textbf{99}, 2018--2027\relax
\mciteBstWouldAddEndPuncttrue
\mciteSetBstMidEndSepPunct{\mcitedefaultmidpunct}
{\mcitedefaultendpunct}{\mcitedefaultseppunct}\relax
\EndOfBibitem
\bibitem[Pieters \emph{et~al.}(2012)Pieters, Undas, Marchi, De~Maat, Weisel,
  and Ari{\"{e}}ns]{Pieters2012AnValues}
M.~Pieters, A.~Undas, R.~Marchi, M.~P.~M. De~Maat, J.~W. Weisel and R.~A.~S.
  Ari{\"{e}}ns, \emph{Journal of Thrombosis and Haemostasis}, 2012,
  \textbf{10}, 2179--2181\relax
\mciteBstWouldAddEndPuncttrue
\mciteSetBstMidEndSepPunct{\mcitedefaultmidpunct}
{\mcitedefaultendpunct}{\mcitedefaultseppunct}\relax
\EndOfBibitem
\bibitem[Toll(1998)]{Toll1998PackingReinforcements}
S.~Toll, \emph{Polymer Engineering and Science}, 1998, \textbf{38},
  1337--1350\relax
\mciteBstWouldAddEndPuncttrue
\mciteSetBstMidEndSepPunct{\mcitedefaultmidpunct}
{\mcitedefaultendpunct}{\mcitedefaultseppunct}\relax
\EndOfBibitem
\bibitem[Collet \emph{et~al.}(2005)Collet, Shuman, Ledger, Lee, and
  Weisel]{Collet2005TheClot}
J.-P. Collet, H.~Shuman, R.~E. Ledger, S.~Lee and J.~W. Weisel,
  \emph{Proceedings of the National Academy of Sciences}, 2005, \textbf{102},
  9133--9137\relax
\mciteBstWouldAddEndPuncttrue
\mciteSetBstMidEndSepPunct{\mcitedefaultmidpunct}
{\mcitedefaultendpunct}{\mcitedefaultseppunct}\relax
\EndOfBibitem
\bibitem[Carr and Hermans(1978)]{Carr1978SizeTurbidity}
M.~E. Carr and J.~Hermans, \emph{Macromolecules}, 1978, \textbf{11},
  46--50\relax
\mciteBstWouldAddEndPuncttrue
\mciteSetBstMidEndSepPunct{\mcitedefaultmidpunct}
{\mcitedefaultendpunct}{\mcitedefaultseppunct}\relax
\EndOfBibitem
\bibitem[van Oosten \emph{et~al.}(2016)van Oosten, Vahabi, Licup, Sharma,
  Galie, MacKintosh, and Janmey]{vanOosten2016UncouplingStretch-stiffening}
A.~S.~G. van Oosten, M.~Vahabi, A.~J. Licup, A.~Sharma, P.~A. Galie, F.~C.
  MacKintosh and P.~A. Janmey, \emph{Scientific Reports}, 2016, \textbf{6},
  1--9\relax
\mciteBstWouldAddEndPuncttrue
\mciteSetBstMidEndSepPunct{\mcitedefaultmidpunct}
{\mcitedefaultendpunct}{\mcitedefaultseppunct}\relax
\EndOfBibitem
\end{mcitethebibliography}
\bibliographystyle{rsc} 

\end{document}


\pagestyle{fancy}
\thispagestyle{plain}
\fancypagestyle{plain}{

\fancyhead[C]{\includegraphics[width=18.5cm]{head_foot/header_bar}}
\fancyhead[L]{\hspace{0cm}\vspace{1.5cm}\includegraphics[height=30pt]{head_foot/SM}}
\fancyhead[R]{\hspace{0cm}\vspace{1.7cm}\includegraphics[height=55pt]{head_foot/RSC_LOGO_CMYK}}
\renewcommand{\headrulewidth}{0pt}
}

\makeFNbottom
\makeatletter
\renewcommand\LARGE{\@setfontsize\LARGE{15pt}{17}}
\renewcommand\Large{\@setfontsize\Large{12pt}{14}}
\renewcommand\large{\@setfontsize\large{10pt}{12}}
\renewcommand\footnotesize{\@setfontsize\footnotesize{7pt}{10}}
\makeatother

\renewcommand{\thefootnote}{\fnsymbol{footnote}}
\renewcommand\footnoterule{\vspace*{1pt}%
\color{cream}\hrule width 3.5in height 0.4pt \color{black}\vspace*{5pt}} 
\setcounter{secnumdepth}{5}

\makeatletter 
\renewcommand\@biblabel[1]{#1}            
\renewcommand\@makefntext[1]%
{\noindent\makebox[0pt][r]{\@thefnmark\,}#1}
\makeatother 
\renewcommand{\figurename}{\small{Fig.}~}
\sectionfont{\sffamily\Large}
\subsectionfont{\normalsize}
\subsubsectionfont{\bf}
\setstretch{1.125} 
\setlength{\skip\footins}{0.8cm}
\setlength{\footnotesep}{0.25cm}
\setlength{\jot}{10pt}
\titlespacing*{\section}{0pt}{4pt}{4pt}
\titlespacing*{\subsection}{0pt}{15pt}{1pt}

\fancyfoot{}
\fancyfoot[LO,RE]{\vspace{-7.1pt}\includegraphics[height=9pt]{head_foot/LF}}
\fancyfoot[CO]{\vspace{-7.1pt}\hspace{13.2cm}\includegraphics{head_foot/RF}}
\fancyfoot[CE]{\vspace{-7.2pt}\hspace{-14.2cm}\includegraphics{head_foot/RF}}
\fancyfoot[RO]{\footnotesize{\sffamily{1--\pageref{LastPage} ~\textbar  \hspace{2pt}\thepage}}}
\fancyfoot[LE]{\footnotesize{\sffamily{\thepage~\textbar\hspace{3.45cm} 1--\pageref{LastPage}}}}
\fancyhead{}
\renewcommand{\headrulewidth}{0pt} 
\renewcommand{\footrulewidth}{0pt}
\setlength{\arrayrulewidth}{1pt}
\setlength{\columnsep}{6.5mm}
\setlength\bibsep{1pt}

\makeatletter 
\newlength{\figrulesep} 
\setlength{\figrulesep}{0.5\textfloatsep} 

\newcommand{\topfigrule}{\vspace*{-1pt}%
\noindent{\color{cream}\rule[-\figrulesep]{\columnwidth}{1.5pt}} }

\newcommand{\botfigrule}{\vspace*{-2pt}%
\noindent{\color{cream}\rule[\figrulesep]{\columnwidth}{1.5pt}} }

\newcommand{\dblfigrule}{\vspace*{-1pt}%
\noindent{\color{cream}\rule[-\figrulesep]{\textwidth}{1.5pt}} }

\makeatother

\twocolumn[
  \begin{@twocolumnfalse}
\vspace{3cm}
\sffamily
\begin{tabular}{m{4.5cm} p{13.5cm} }

\includegraphics{head_foot/DOI} & \noindent\LARGE{\textbf{Supplementary Information for: Poroelasticity of (bio)polymer networks during compression: theory and experiment}} \\
\vspace{0.3cm} & \vspace{0.3cm} \\

& \noindent\large{Melle T.J.J.M. Punter,\textit{$^{\ast a}$} Bart E. Vos,\textit{$^{\ast b c}$} Bela M. Mulder,\textit{$^{a}$} and Gijsje H. Koenderink\textit{$^{bd}$}$^{\ddag}$} \\

\end{tabular}

 \end{@twocolumnfalse} \vspace{0.6cm}

  ]

\renewcommand*\rmdefault{bch}\normalfont\upshape
\rmfamily
\section*{}
\vspace{-1cm}

\renewcommand{\thepage}{S\arabic{page}} 
\renewcommand{\thesection}{S\arabic{section}}  
\renewcommand{\thetable}{S\arabic{table}}  
\renewcommand{\thefigure}{S\arabic{figure}}
\renewcommand{\theequation}{S\arabic{equation}}


\footnotetext{$\ast$~These authors contributed equally to this work.}
\footnotetext{\textit{$^{a}$~AMOLF, Theory of Biomolecular Matter, Science Park 104, 1098XG Amsterdam, the Netherlands}}
\footnotetext{\textit{$^{b}$~AMOLF, Biological Soft Matter, Science Park 104, 1098XG Amsterdam, the Netherlands}}
\footnotetext{\textit{$^{c}$}~Current address: \textit{ZMBE, Mechanics of cellular systems Group,
Institute of Cell Biology, Westf{\"a}lische Wilhelms-Universit{\"a}t, Von-Esmarch-Stra{\ss}e 56, 48149 M{\"u}nster, Germany.}}
\footnotetext{\textit{$^{d}$}~Current address: \textit{Department of Bionanoscience, Kavli Institute of Nanoscience Delft, Delft University of Technology, 2629HZ Delft, The Netherlands}}
\footnotetext{$\ddag$~Corresponding Author, email: G.Koenderink@amolf.nl}

In this Supplementary Information we first present in section \ref{sec:app_solution} our approximate solution to the poroelastic equations of motion, equation (1)--(3) in the manuscript. 
Afterwards, we treat the assumptions on which the approximate form of the poroelastic equations is based, as well as the assumptions underlying our solution. 
Next, we consider in section \ref{sec:cube_model} the cubic lattic model used to estimate the dependence of the permeability $k$ of a fibrin network on the concentration of fibrinogen $c$ and the compressive strain $\epsilon$. In section \ref{sec:Comparison-to-experiment}, we present the fit results of the measured normal force in all compression experiments and discuss them systematically. 

\section{Approximate solution\label{sec:app_solution}}

\begin{figure*}[t!]
\includegraphics[scale=0.71]{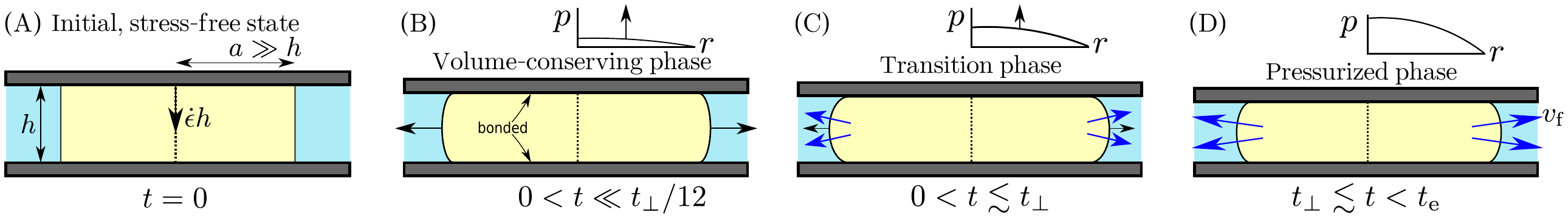}
\caption{The compression phases of (\textbf{A}) an initially stress-free cylindrical fibrin gel (yellow) of radius $a$ and height $h$ with high aspect ratio $S\equiv a/h \gg 1$ (Figure not on scale). 
The gel is ramp compressed in a parallel-plate rheometer with the upper plate (gray) having a constant velocity $\dot{\epsilon} h$, where $\dot{\epsilon}$ is the strain rate. 
\textbf{B}) Compression commences with the volume-conserving phase. The fibrin network starts to bulge out (black arrows) because the gel is bonded to the plates, causing the fluid pressure $p$ to build up. 
\textbf{C} The build-up proceeds, causing a fluid outflow velocity $v_\mathrm{f}$ (blue arrows) due to the fluid pressure gradient, $v_\mathrm{f} \propto - \partial p / \partial r$ with $r$ the radial coordinate in the fibrin gel. After the pressurizing time $t_\perp$, the outward bulging of the gel network, which induces the fluid pressure, stabilizes. 
\textbf{C}) In the pressurized phase, the gel is compressed further at maximal fluid pressure until at time $t_\mathrm{e}$ the compression stops. 
}
\label{fig:compression_phases_SI}
\end{figure*}

The approximate closed-form solution to the poroelastic equations of motion is found as an exact solution to an approximate form of the poroelastic equations. First, we state this exact solution. 
Afterwards, we motivate the approximate form of the poroelastic equations, and detail the underlying assumptions. 

We consider the ramp compression of a cylindrical gel bonded to the plates of a parallel-plate rheometer, where the gel network is treated as a linear elastic solid with shear modulus $G$ and bulk modulus $K$, see Figure \ref{fig:compression_phases_SI} and Figure 2 of the manuscript. 
The gel network has a permeability $k$, the fluid has dynamic viscosity $\eta$, and the gel is compressed at a strain rate $\dot{\epsilon} \equiv v / h$, with $v$ the velocity of the upper plate. 
Before compression, the gel has an axial length $h$, a radius $a$ and a large aspect ratio $S \equiv a / h \gg 1$, see Figure \ref{fig:compression_phases_SI}A. 
The solution of the exact poroelastic equations, equation (1)--(3) of the manuscript, can be specified by the displacement field of the gel network $\boldsymbol{U}$, the velocity field of the fluid $\boldsymbol{v}_\mathrm{f}$ and the fluid pressure $p$. 
For notational convenience, however, we do not use the fluid velocity, but the volume-averaged local velocity of the gel $\boldsymbol{V} \equiv \phi_\mathrm{n} \boldsymbol{v}_\mathrm{n} + \phi_\mathrm{f} \boldsymbol{v}_\mathrm{f}$ instead, with $\phi_\mathrm{f}(\phi_\mathrm{n})$ the volume fraction and $\boldsymbol{v}_\mathrm{f}(\boldsymbol{v}_\mathrm{n} \equiv \partial_t \boldsymbol{U})$ the velocity field of the fluid(gel network).  
In the exact solution of the approximate form of the poroelastic equations,  for any radial position $r$, vertical position $z$ and time $t$ with $(r,z,t) \in ]0,a] \times [0,h] \times [0,\infty]$, the displacement field of the gel network $\boldsymbol{U}(r,z,t) = U(r,z,t)\boldsymbol{\hat{r}} + W(z,t)\boldsymbol{\hat{z}}$, with $U(r,z,t)$ the radial displacement and $W(z,t)$ the vertical displacement, reads the following
\begin{align}
\label{eq:U} U(r,z,t)   &= T(t)\dot{\epsilon} t_\perp \frac{r}{4} \frac{z}{h} \left( 1 - \frac{z}{h} \right),\\
\nonumber W(z,t)     &= -\left( 1 - T(t) \right) \epsilon z \frac{z}{h} \left( 3 - 2 \frac{z}{h} \right) - T(t)\epsilon z + \\
\label{eq:W} & T^2(t) \dot{\epsilon} t_\perp \frac{h}{12} \frac{M - G}{M} \left( \frac{z}{h} - 3 \left( \frac{z}{h} \right)^2 + 2 \left( \frac{z}{h} \right)^3 \right),\\
\label{eq:exp} T(t)     &= 1 - \exp \left( - 12 \frac{t}{t_\perp} \right),
\end{align}
where $t_\perp \equiv \eta h^2 / G k$ is the time for the gel to become pressurized, i.e., $\partial_t p = 0$, during compression, $\epsilon = \dot{\epsilon} t$ is the compressive strain on the gel, $ M = K + 4 G / 3 $ is the oedometric modulus, and $T(t)$ is the transition function which brings the gel from volume-conserving compression into the pressurized phase. 
The fluid pressure $p(r,z,t)$ is given by
\begin{align}
\nonumber p(r,z,t)   &=  T(t) G \frac{S^2}{4} \dot{\epsilon} t_\perp \left( 1 - \left( \frac{r}{a} \right)^2 \right) + 2 \left( 1 - T(t) \right) G \epsilon + p_\mathrm{ext} - \\
\label{eq:p} & 6 \frac{z}{h} \left( 1 - \frac{z}{h} \right) \Bigg[ \left( 1 - T(t) \right) M \epsilon +      \left( T^2(t)-T(t) \right) \frac{M-G}{12} \dot{\epsilon} t_\perp \Bigg],
\end{align}
where $p_\mathrm{ext}$ is the pressure of the fluid in which the gel is immersed, and the volume-averaged local velocity of the gel $\boldsymbol{V}(r,z,t)$ is found as
\begin{align}
\label{eq:Vcm} \boldsymbol{V}(r,z,t)    &= \left( 1 - T(t) \right)\boldsymbol{V}^\mathrm{vc}(r,z) + T(t) \boldsymbol{V}^\mathrm{cp}(r,z),\\
\label{eq:Vcm_inc} \boldsymbol{V}^\mathrm{vc}(r,z)   &= 3 \dot{\epsilon}  r \frac{z}{h} \left( 1 - \frac{z}{h} \right) \boldsymbol{\hat{r}} - z \dot{\epsilon} \frac{z}{h} \left( 3 - 2 \frac{z}{h} \right) \boldsymbol{\hat{z}},\\
\label{eq:Vcm_cp} \boldsymbol{V}^\mathrm{cp}(r,z)     &= \frac{r}{2} \dot{\epsilon} \boldsymbol{\hat{r}} - z \dot{\epsilon} \boldsymbol{\hat{z}},
\end{align}
which shows that the volume averaged local velocity of the gel transitions between that of volume-conserving compression $\boldsymbol{V}^\mathrm{vc}(r,z,t)$ and that in the pressurized phase $\boldsymbol{V}^\mathrm{cp}(r,z,t)$ where the fluid pressure is constant in time. 

Equation \eqref{eq:U}--\eqref{eq:Vcm_cp} exactly solve the following approximate set of poroelastic bulk and boundary equations of motion: equation \eqref{eq:fb_r}--\eqref{eq:bc_fb_r_approx}. First of all, equation \eqref{eq:U}--\eqref{eq:p} solve the exact overall force balance, see equation (1) of the manuscript, which reads
\begin{align}
\label{eq:fb_r} \frac{\partial p}{\partial r}    &= M \frac{\partial}{\partial r} \frac{1}{r} \frac{\partial r U}{\partial r} + (M-G) \frac{\partial^2 W}{ \partial r \partial z} + G \frac{\partial^2 U}{\partial z^2},\\
\label{eq:fb_z} \frac{\partial p}{\partial z}    &= M \frac{\partial^2 W}{\partial  z^2} + (M-G) \frac{1}{r} \frac{\partial}{\partial r} r \frac{\partial U}{\partial z} + G \frac{1}{r} \frac{\partial}{\partial r} r \frac{\partial W}{\partial r}. 
\end{align}
Equation \eqref{eq:U}--\eqref{eq:Vcm_cp} solve an approximate form of Darcy's law, see equation (3) in the manuscript, which has been rewritten using the definition of the volume averaged local velocity of the gel $\boldsymbol{V} \equiv \phi_\mathrm{n} \boldsymbol{v}_\mathrm{n} + \phi_\mathrm{f} \boldsymbol{v}_\mathrm{f}$
\begin{align}
\label{eq:darcy_r} v_{\mathrm{n},r} - V_r &= \frac{k}{\eta} \frac{\partial p}{\partial r},\\
\label{eq:darcy_z_approx} \langle v_{\mathrm{n},z} - V_z \rangle &= \frac{k}{\eta} \Big\langle \frac{\partial p}{\partial z} \Big\rangle,
\end{align}
where $\langle X \rangle \equiv (1/h) \int_0^h dz X$ is the average of $X$ over the axial height of the gel. 
This form of Darcy's law, which stems from the force balance of the fluid, is approximate because it requires Darcy's law in the vertical direction to be obeyed only on average. 
Finally, equation \eqref{eq:Vcm}--\eqref{eq:Vcm_cp} solve the incompressibility condition of the gel which follows from mass conservation
\begin{equation}
\label{eq:mass_cons} \nabla \cdot \boldsymbol{V} = 0.
\end{equation}

The solution in equation \eqref{eq:U}--\eqref{eq:exp} obeys the following essential boundary conditions. 
Ramp compression and the binding of the gel network to the rheometer plates impose
\begin{align}
\label{eq:bc_W_low}    W &= 0\,\mathrm{and}\,V_z = 0,\,\mathrm{at}\,z=0,\\
\label{eq:bc_W_upp}    W &= -\epsilon h\,\mathrm{and}\,V_z = - \dot{\epsilon} h,\,\mathrm{at}\,z=h,\\
\label{eq:bc_U}    U &= 0,\,\mathrm{at\,both}\,z=0\,\mathrm{and}\,z=h.
\end{align}
At the free boundary the solution obeys the average form of the tangential overall stress balance, and an approximate form of the radial overall stress balance combined with the permeability condition
\begin{align}
\label{eq:bc_fb_z_approx} & \langle \sigma'_{rz} \rangle = 0\,\mathrm{at}\,r=a,\\
\label{eq:bc_fb_r_approx} & \langle \sigma'_{rr} \left( 1 - T(t) \right) - p\rangle = - p_\mathrm{ext},\,\mathrm{at}\,r=a,
\end{align}
where $\boldsymbol{\sigma'}$ is the Terzaghi effective stress of the gel network, taken to be that of a linear elastic solid.  

The time-dependent normal force $F(t)$ on the plates which is generated by the gel during compression, can be calculated from the overall stress at the gel-plate interface as
\begin{equation}
    F(t) = - \int_0^a dr 2 \pi r \left(\sigma'_{zz} - p \right)\Big|_{z=0,h},
\end{equation}
which gives with equation \eqref{eq:U}--\eqref{eq:p}
\begin{equation}
\label{eq:F} \frac{F(t)}{\pi a^2} = T(t) \left[ G \left( \frac{S^2}{8} - T(t) \frac{M-G}{12G} \right) \dot{\epsilon} t_\perp + M \epsilon \right] + \left( 1 - T(t) \right) 2 G \epsilon. 
\end{equation}
The condition for the validity of the solution presented in equation \eqref{eq:U}--\eqref{eq:Vcm_cp} is that $ M / G S^2 \ll 1$. 
Using this validity condition, the dominant part of equation \eqref{eq:F} gives equation (4) of the manuscript. 

\subsection{Assumptions\label{sec:assumptions}}

To study the ramp compression of a bonded disk-like gel, we make several assumptions by comparing the bonded gel with a frictionless gel having no friction with the plates of the rheometer. 
Similar to a frictionless gel, we assume the bonded gel to deform in a volume-conserving manner when compression commences, see Figure \ref{fig:compression_phases_SI}B. 
Also similar to a frictionless gel, the fluid pressure is assumed to reach a maximal value during compression, the gel is then pressurized, see Figure \ref{fig:compression_phases_SI}D. 
Finally, the gel transitions in the pressurizing time from the volume-conserving to the pressurized phase in a simple mono-exponential time-dependent fashion, see Figure \ref{fig:compression_phases_SI}C. 

\subsubsection{The pressurized phase\label{subsec:pressurized_phase}}

Consider a cylindrical gel under compression identical to the case we treat in the manuscript, see Figure \ref{fig:compression_phases_SI}, but instead of being bonded to the plates, it experiences no friction with the plates. 
In this case, the exact solution of the poroelastic equations of motion, equation (1)--(3) in the manuscript, is known\cite{Armstrong1984AnCartilage}. 
During sufficiently slow ramp compression, the frictionless gel becomes pressurized, i.e., $\partial_t p = 0$, after a pressurizing time $t_\parallel = a^2 \eta / k M $. 
When pressurized, the exact solution for the displacement field, the volume-averaged velocity field and the fluid pressure $p$ of the frictionless gel reads
\begin{align}
\label{eq:U_frictionless} U(r,t) &= \nu r \epsilon + a \frac{ \eta \dot{\epsilon} a^2 }{ 8 k M } \left( \frac{1}{2} - \nu \right) \left( ( 3 - 2 \nu ) \frac{r}{a} - \left( \frac{r}{a} \right)^3 \right),\\
W(z,t) &= -z \epsilon,\\
\boldsymbol{V}(r,z) &= \frac{r}{2} \dot{\epsilon} \boldsymbol{\hat{r}} - z \dot{\epsilon} \boldsymbol{\hat{z}},\\
p(r) &= \frac{ \eta \dot{\epsilon} }{ 2 k } \left( \frac{1}{2} - \nu \right) \left( a^2 - r^2 \right) + p_\mathrm{ext},
\end{align}
which shows that the gel deformation differs from that of a static frictionless solid by an inhomogeneous radial strain proportional to the compression rate, see equation \eqref{eq:U_frictionless}. 
This inhomogeneous strain causes stress in the gel network, which in turn sources the fluid pressure $p$ required for the outflow of fluid from the gel, see equation \eqref{eq:fb_r} and \eqref{eq:fb_z}. 
The fact that this fluid pressure is constant in time, implies a constant outflow of fluid from the gel, which brings us to the first assumption. 

\textbf{Assumption 1}: similar to a frictionless gel, a bonded gel becomes pressurized, i.e., $\partial_t p = 0$, after some pressurizing time $t_\perp$, see Figure \ref{fig:compression_phases_SI}D. 

Given this assumption, we can write down the following approximate solution for a bonded gel in the pressurized phase with constant pressure, which obeys the exact form of equation \eqref{eq:fb_r}--\eqref{eq:bc_U}
\begin{align}
\label{eq:U_cp}    U^\mathrm{cp}(r,z) &= \dot{\epsilon} t_\perp \frac{r}{4} \frac{z}{h} \left( 1 - \frac{z}{h} \right),\\
\label{eq:W_cp}    W^\mathrm{cp}(z,t) &= -\epsilon  z + \dot{\epsilon} t_\perp \frac{h}{4} \frac{1}{6 \left( 1 - \nu \right)} \left( \frac{z}{h} - 3\left( \frac{z}{h} \right)^2 + 2 \left( \frac{z}{h} \right)^{3} \right),\\
\label{eq:V_cp}    \boldsymbol{V}^\mathrm{cp}(r,z) &= \frac{r}{2} \dot{\epsilon} \boldsymbol{\hat{r}} - z \dot{\epsilon} \boldsymbol{\hat{z}},\\
\label{eq:p_cp}    p(r) &= G\frac{S^{2}}{4}\dot{{\epsilon}}t_{\perp}\left(1-\left(\frac{r}{a}\right)^{2}\right)+p_{\mathrm{{ext}}},
\end{align}
where we note that in the bonded case the vertical strain is inhomogeneous instead of the radial strain in the frictionless case, and we used the permeability condition $p=p_\mathrm{ext}$ at the free boundary $r=a$. 
The volume-averaged gel velocity field $\boldsymbol{V}^\mathrm{cp}$ equals that of a frictionless volume-conserving solid because $\partial_t U = 0$, implying all radial motion is due to fluid flow. 
The velocity field of the fluid equals that of a frictionless volume-conserving solid,
because it is not bonded to the rheometer plates and flows through the radially static, but vertically comoving, gel network. 

To study whether this approximate solution is reasonable, we consider the exact form of the bulk equations of motion, equation \eqref{eq:fb_r}--\eqref{eq:mass_cons}, giving
\begin{align}
\label{eq:fb_r_exact} \frac{\eta}{k} \left( \frac{\partial U}{\partial t} - V_r \right) &= M \frac{\partial}{\partial r} \frac{1}{r} \frac{\partial r U}{\partial r} + (M-G) \frac{\partial^2 W}{ \partial r \partial z} + G \frac{\partial^2 U}{\partial z^2},\\
\label{eq:fb_z_exact} \frac{\eta}{k} \left( \frac{\partial W}{\partial t} - V_z \right) &= M \frac{\partial^2 W}{\partial  z^2} + (M-G) \frac{1}{r} \frac{\partial}{\partial r} r \frac{\partial U}{\partial z} + G \frac{1}{r} \frac{\partial}{\partial r} r \frac{\partial W}{\partial r},\\
\nabla \cdot \boldsymbol{V} &= 0,
\end{align}
where we eliminated the fluid pressure $p$. The boundary conditions (BCs) are equations \eqref{eq:bc_W_low}--\eqref{eq:bc_U}, combined with the exact boundary conditions at the free boundary: the overall force balance and the permeability condition
\begin{align}
\label{eq:bc_fb_r_exact}    \sigma'_{rr} - p &= -p_\mathrm{ext},\,\mathrm{at}\,r=a,\\
\label{eq:bc_fb_z_exact}    \sigma'_{rz} &= 0,\,\mathrm{at}\,r=a,\\
\label{eq:bc_perm_exact}    p &= p_\mathrm{ext},\,\mathrm{at}\,r=a, 
\end{align}
where equation \eqref{eq:bc_fb_r_exact} is the overall balance of forces perpendicular to the free boundary, equation \eqref{eq:bc_fb_z_exact} is the overall balance of forces tangential to the free boundary, and equation \eqref{eq:bc_perm_exact} expresses that the gel network is permeable for fluid. 

The exact solution in the constant pressure phase can be written as $\boldsymbol{U} = \boldsymbol{U}^\mathrm{cp} + \Delta \boldsymbol{U}$ and $\boldsymbol{V} = \boldsymbol{V}^\mathrm{cp} + \Delta \boldsymbol{V} $, with $\Delta \boldsymbol{U}$ and $\Delta \boldsymbol{V}$ the difference solution, i.e., the difference between the exact and the approximate solution in the pressurized phase. 
To show that the difference may be negligible, we scale all quantities to their presumed typical sizes: $\tilde{U} = \Delta U / h$, $\tilde{W} = \Delta W / h$, $\tilde{V}_r = \Delta V_r / (a/t_\mathrm{ext})$, $\tilde{V}_z = \Delta V_z / (h/t_\mathrm{ext}) $, $\xi = z / h $, $ \rho = r / a $, $\tau_\mathrm{ext} = t / t_\mathrm{ext}$ and we define $t_\mathrm{ext}$ as the externally determined time scale of ramp compression, $t_\mathrm{ext} \sim h / v$ with $v$ the velocity of the upper plate. 
The bulk equations of motion for the difference solution read
\begin{align}
\label{eq:fb_r_nodim} \frac{ \partial \tilde{U} }{ \partial \tau_\mathrm{ext} } - S \tilde{V}_r &= \frac{ t_\mathrm{ext} }{ t_\perp } \left( \frac{M}{ G S^2 } \frac{\partial}{\partial\rho} \frac{1}{\rho} \frac{\partial \rho \tilde{U}}{\partial\rho} + \frac{ M - G }{G S} \frac{\partial^2 \tilde{W} }{ \partial \rho \partial \xi} + \frac{ \partial^2 \tilde{U}}{ \partial \xi^2}\right),\\
\label{eq:fb_z_nodim} \frac{ \partial \tilde{W} }{ \partial \tau_\mathrm{ext}} - \tilde{V}_z &= \frac{ t_\mathrm{ext} M }{ t_\perp G } \left( \frac{ \partial^2 \tilde{W} }{ \partial \xi^2 } +  \frac{ 1 - \frac{G}{M} }{S} \frac{1}{\rho} \frac{\partial}{\partial\rho} \rho \frac{ \partial \tilde{U} }{ \partial \xi } + \frac{G}{M S^2} \frac{1}{\rho} \frac{\partial}{\partial\rho} \rho \frac{\partial \tilde{W}}{\partial\rho} \right),\\
\tilde{\nabla} \cdot \boldsymbol{\tilde{V}} &= 0,
\end{align}
where $t_\perp = h^2 \eta / k G$ is the pressurizing time, and the boundary conditions read
\begin{align}
\label{eq:bc_essential_nodim}    \tilde{W} = 0, \tilde{V}_z &= 0\,\mathrm{and}\,\tilde{U}=0,\,\mathrm{at\,both}\,\xi=0\,\mathrm{and}\,\xi=1,\\
\label{eq:bc_fb_r_nodim} \nonumber M \frac{\partial \tilde{U}}{\partial \rho} + \Lambda \frac{\tilde{U}}{\rho} &= - S \Big( \Lambda \frac{\partial \tilde{W}}{\partial \xi} + \Lambda \Big\{ -\epsilon + \frac{\dot{\epsilon} t_\perp}{24(1-\nu)} \Big[ 1 - 6 \left( \frac{z}{h} - \left( \frac{z}{h} \right)^2\right) \Big] \Big\} + \\
&  (M+\Lambda) \frac{\dot{\epsilon} t_\perp}{4} \left\{\frac{z}{h} - \left(\frac{z}{h}\right)^2\right\}  \Big),\,\mathrm{at}\,\rho=1,
\end{align}
\begin{align}
\label{eq:bc_fb_z_nodim} \frac{\partial \tilde{W}}{\partial \rho} &= - S \left( \frac{\partial \tilde{U}}{\partial \xi} + \dot{\epsilon} t_\perp \frac{r}{4 h} \left\{ 1 - 2 \frac{z}{h} \right\} \right),\,\mathrm{at}\,\rho=1.
\end{align}
If we assume that $M/GS^2 \ll 1$, then the $M\partial_r (1/r) \partial_r r U$ term is much smaller than the $G \partial_z^2 U$ term, compare equation \eqref{eq:fb_r_exact} to equation \eqref{eq:fb_r_nodim}, and the $G (1/r) \partial_r r \partial_r W$ term is much smaller than the $M\partial_z^2 W$ term since $G<M$, compare equation \eqref{eq:fb_z_exact} to equation \eqref{eq:fb_z_nodim}. Neglecting the small terms, both equation \eqref{eq:fb_r_nodim} and \eqref{eq:fb_z_nodim} are only first order in $r$ instead of second order. The boundary conditions for $U$ and $W$ at $r=0$ are imposed by smoothness and symmetry: $U=0$ and $\sigma_{rz}=0 \implies \partial_r W = 0$ at $r=0$, and can not be ignored. Therefore, we must neglect the Robin boundary condition for $U$ at $r=a$ in equation \eqref{eq:bc_fb_r_nodim}, and the Neumann boundary condition for $W$ at $r=a$ in equation \eqref{eq:bc_fb_z_nodim}. 
Then, it follows that there are no sources for a nontrivial solution to equation \eqref{eq:fb_r_nodim}--\eqref{eq:bc_essential_nodim}, implying that $\boldsymbol{\tilde{U}} = 0$ and $\boldsymbol{\tilde{V}}=0$ is the solution. 
Therefore, the approximate solution in equation \eqref{eq:U_cp}--\eqref{eq:V_cp} is expected to be accurate in the pressurized phase, provided that $M/GS^2 \ll 1$. 

The neglect of the two small terms is equivalent to treating the dependencies of $\boldsymbol{\tilde{U}}$ and $\boldsymbol{\tilde{V}}$ to the radial coordinate $r$ as static. 
In fact, the relative size of the different network stress terms, provided $M/GS^2 \ll 1$, holds during the whole of compression, which gives the full time-dependent approximate solution, contained in equation \eqref{eq:U},\eqref{eq:W} and \eqref{eq:Vcm}--\eqref{eq:Vcm_cp} a static dependence on $r$: $U(r,z,t) \propto r^1$, $W(z,t) \propto r^0$, $V_r(r,z,t) \propto r^1$ and $V_z(z,t) \propto r^0$ for all times $t$. 

Physically, the condition for the approximate solution to hold, $M/GS^2 \ll 1$, means that the radial force in the gel network due to bending of the gel network, sourced by $G \partial_z^2 U$, is much greater than the radial force due to inhomogeneity in the radial strain, sourced by $M\partial_r (1/r) \partial_r r U$. 
That is, the shear stress induced by friction between the gel and the rheometer plate dominates the network stress in the gel. 
Similarly, the vertical force in the gel network due to inhomogeneity in the vertical strain, sourced by $M \partial_z^2 W$, is much greater than the vertical force induced by vertical out of plane displacements, sourced by $G (1/r) \partial_r r \partial_r W$. 
The term $(M-G) \partial_z \partial_r W$ can be smaller or comparable to the $G \partial_z^2 U$ term, depending on the size of the aspect ratio $S \gg 1$, likewise for the size of $(M-G)(1/r) \partial_r r \partial_z U$ relative to $M\partial_z W$. 
Finally, it should be noted that although the presumed magnitude of $U$, which we take to be $h$, is uncertain, this presumption does not influence the condition to neglect the radial inhomogeneity term, because the shear stress term, relative to which the radial inhomogeneity term is small, also scales with $U$. 

\subsubsection{The volume-conserving phase\label{subsec:vol_cons_phase}}

Next, we consider that in the frictionless case\cite{Armstrong1984AnCartilage} for short times after the commencement of compression, i.e., $\pi^2 t/t_\parallel \ll 1$ with $t_\parallel$ the pressurizing time, the gel deforms to good approximation as a volume-conserving frictionless solid, because the outflow of fluid from the gel network is still small. 
Again, assuming the bonded case to behave analogous to the frictionless case, we arrive at the second assumption. 

\textbf{Assumption 2}: similar to a frictionless gel, for times much shorter than the, yet to be determined, pressurizing time $t_\perp$, a bonded gel deforms like a bonded volume-conserving solid, see Figure \ref{fig:compression_phases_SI}B. 

The (quasi)-static displacement field of a volume-conserving linear elastic solid bonded to the plates is\cite{Qiao2015AnalyticalLayers}
\begin{align}
\label{eq:U_vc}    U^\mathrm{vc}(r,z,t) &= 3 \epsilon r \frac{z}{h} \left( 1 - \frac{z}{h} \right),\\
\label{eq:W_vc}    W^\mathrm{vc}(z,t) &= - z \epsilon \frac{z}{h} \left( 3 - 2 \frac{z}{h} \right),
\end{align}
which gives equation \eqref{eq:fb_r} and \eqref{eq:fb_z} in dimensionless form as
\begin{align}
\label{eq:fb_r_bonded_inc}    \frac{\partial \tilde{p}}{\partial \rho} &= -6S^2 \epsilon \rho,\\
\label{eq:fb_z_bonded_inc}    \frac{\partial \tilde{p}}{\partial \xi} &= -6 \epsilon ( 1 - 2 \xi ),
\end{align}
where we defined $\tilde{p} = p / G$. As $S\gg1$, the majority of the pressure builds up radially, stemming from the $G \partial_z^2 U^\mathrm{vc}$ term in equation \eqref{eq:fb_r}. 
This term quantifies the stress induced by bending of the gel network, which is imposed by the binding of the network to the plates. 
At $z=0$ and $z=h$, we find that $\partial p / \partial z \neq 0$, however, which implies with Darcy's law, see equation (3) of the manuscript, a vertical flow of fluid relative to the gel network through the impermeable plates of the rheometer. 
As this would render the plates permeable, we adopt an approximative form of the vertical part of Darcy's law, see equation \eqref{eq:darcy_z_approx}, by requiring it to hold only when averaged over the axial height of the gel. 

\subsubsection{Transition from the volume-conserving phase to the pressurized phase\label{subsec:transition}}

The frictionless gel transitions from volume-conserving compression to the pressurized phase on a pressurizing time scale $t_\parallel$\cite{Armstrong1984AnCartilage}. Also, previous work in the context of gel compression demonstrated mono-exponential time-dependence, see equation (32) in Yamaue \textit{et al.}\cite{Yamaue2004TheoryConstraint}. This brings us to the final assumption

\textbf{Assumption 3}: we assume the time-dependent dynamics of the radial displacement of the gel network, transitioning from volume-conserving deformation to the pressurized phase, to be proportional to a transition function $T(t)$, see Figure \ref{fig:compression_phases_SI}C. 

We write $U(r,z,t) = T(t) U^\mathrm{cp}$, such that $T(t) \approx 1$ if $t \gtrsim t_\perp $ with $t_\perp$ the pressurizing time, and $T(0)=0$, because at $t=0$ we assume the gel to be stress-free. 
Then, we assume that $\boldsymbol{V}(r,z,t)$, see equation \eqref{eq:Vcm}, interpolates between the volume-conserving velocity field of a bonded solid $\boldsymbol{V}^\mathrm{vc}(r,z,t)$, see the time-derivative of equation \eqref{eq:U_vc} and \eqref{eq:W_vc}, and the volume-conserving velocity field of a frictionless solid $\boldsymbol{V}^\mathrm{cp}(r,z,t)$, see equation \eqref{eq:V_cp}. 
The exact radial force balance, equation \eqref{eq:fb_r} combined with equation \eqref{eq:darcy_r}, can than be solved for $T(t)$, giving the transition function $T(t)$, see equation \eqref{eq:exp}, along with the pressurizing time $t_\perp = h^2 \eta / k G$. 
Then, by requiring also $W(z,t)$ and $p(r,z,t)$ to transition from their volume-conserving form to their form in the pressurized phase, using only $T(t)$, we obtain as a solution to equation \eqref{eq:fb_r}--\eqref{eq:bc_U} the solution contained in equation \eqref{eq:U}--\eqref{eq:Vcm_cp}, but slightly more general with
\begin{align}
\nonumber W(z,t)     &= -\left( 1 - T(t) \right) \epsilon z \frac{z}{h} \left( 3 - 2 \frac{z}{h} \right) - T(t)\epsilon z + \\
\label{eq:W_general} & T^n(t) \dot{\epsilon} t_\perp \frac{h}{12} \frac{M - G}{M} \left( \frac{z}{h} - 3 \left( \frac{z}{h} \right)^2 + 2 \left( \frac{z}{h} \right)^3 \right),\\
\nonumber \label{eq:p_general} p(r,z,t)   &=  T(t) G \frac{S^2}{4} \dot{\epsilon} t_\perp \left( 1 - \left( \frac{r}{a} \right)^2 \right) + C - \\
& 6 \frac{z}{h} \left( 1 - \frac{z}{h} \right) \Bigg[ \left( 1 - T(t) \right) M \epsilon +      \left( T^n(t)-T(t) \right) \frac{M-G}{12} \dot{\epsilon} t_\perp \Bigg],
\end{align}
where $n \geq 2$, and $C$ is an integration constant. The non-uniqueness of this solution, illustrated by the arbitrariness of $n$, may be related to the approximate nature of the equations of motion it solves. 

The terms proportional to $T^n(t)$ are related to the inhomogeneous vertical strain in the pressurized phase, compare equation \eqref{eq:W_cp} to equation \eqref{eq:W_general}. 
This inhomogeneous vertical strain balances the $ ( M - G ) (1/r) \partial_r r \partial_z U $ term in equation \eqref{eq:fb_z_exact}, which measures  the local increase in vertical force on the gel network due to the radial change in the deflection of the gel network. 
If we consider the vertical force balance with the exact form of Darcy's law, and plug in equation \eqref{eq:U}, \eqref{eq:Vcm}--\eqref{eq:Vcm_cp}, and \eqref{eq:W_general}, we find that the left hand side can be grouped in terms proportional to $T(t)^0,T(t)^1,T(t)^{n-1}$ and $T(t)^n$, whereas the right hand side can be grouped in terms proportional to $T(t)^0,T(t)^1$ and $T(t)^n$. 
We then set the value of $n \geq 2$ by requiring the dependencies on the transition function $T(t)$ on both sides of the exact equation to match: for each term on the left hand side proportional to $T(t)^m$, there should be a term on the right hand side also proportional to $T(t)^m$. This requirement uniquely sets $n=2$. 

Finally, to determine the integration constant $C$ for the fluid pressure $p$, we consider the force balance for the overall radial stress at the free boundary. 
The volume-conserving solution, see equation \eqref{eq:U_vc}--\eqref{eq:fb_z_bonded_inc}, can satisfy the weak form of the radial force balance $\langle \sigma'_{rr} - p \rangle = - p_\mathrm{ext}$ at $r=a$. 
On the other hand, our approximate solution, as explained above, ignores the exact boundary conditions at the free boundary in the pressurized phase, see equation \eqref{eq:bc_fb_r_exact}--\eqref{eq:bc_perm_exact}, and thus does not obey the stress-free condition on the gel network $\sigma'_{rr} = 0$ at $r=a$, which is implied by the exact boundary conditions. 
This artifact of our approximate solution comes from ignoring the $M \partial_r (1/r) \partial_r r U$ term, implying we do not take into account the relaxations in the radial strain which are induced by $\sigma'_{rr}=0$ and which are expected to occur, similar to the frictionless case, on a time scale $t_\parallel$. 
Therefore, we assume an approximate condition at the free boundary which interpolates between the weak form of the overall radial force balance in the volume-conserving phase and the weak form of the permeability condition, see equation \eqref{eq:bc_perm_exact}, in the pressurized phase, giving equation \eqref{eq:bc_fb_r_approx}. This approximate condition makes the contribution of the gel network to the radial stress vanish in the pressurized phase, thereby enforcing the weak form of the permeability condition in the pressurized phase. Using equation \eqref{eq:bc_fb_r_approx}, we then find $C = 2 ( 1 - T(t) ) G \epsilon + p_\mathrm{ext}$. Similarly, we assume the force balance of the overall tangential stress at the free boundary, see equation \eqref{eq:bc_fb_z_exact}, to hold in weak form, i.e., $\langle \sigma'_{rz} \rangle = 0$ at $r=a$, which is satisfied trivially due to symmetry. 

\begin{table*}[t!]
\begin{tabular*}{18.4cm}{c|cccccccccc}
\# & $c$(mg/mL) & $a$ (mm) & $\dot{\epsilon}$ ($10^{-3}$/s) & $\epsilon_\mathrm{e}$ (\%) & $k\mathrm{{\,\left(10^{-1}\,\mu m^{2}\right)}}$ & $M$ (kPa) & $G_0$ (kPa) & $\nu$ (-) & $t_{\perp}/12$ (s) & $M/GS^2$\tabularnewline
\hline \hline
1 & 2 & 20 & 1.0 & 10 & 1.26(0.03) & 1.45(0.12) & 0.139 & 0.447(0.005) & 3.3 & 0.026\tabularnewline
2 & 2 & 20 & 1.0 & 10 & 4.97(0.19) & 0.43(0.05) & 0.142 & 0.257(0.039) & 0.8 & 0.008\tabularnewline
3 & 2 & 20 & 1.0 & 10 & 2.79(0.11) & 0.91(0.09) & 0.212 & 0.347(0.019) & 1.0 & 0.011\tabularnewline
4 & 2 & 20 & 1.0 & 10 & 1.73(0.03) & 1.48(0.07) & 0.087 & 0.469(0.002) & 3.8 & 0.042\tabularnewline
5 & 2 & 10 & 1.0 & 10 & 1.54(0.02) & 0.15(0.01) & 0.227 & 1.950(0.235) & 1.6 & 0.007\tabularnewline
6 & 2 & 20 & 1.0 & 5 & 5.00(0.33) & 0.99(0.16) & 0.119 & 0.431(0.013) & 1.0 & 0.021\tabularnewline
7 & 2 & 20 & 1.0 & 5 & 2.29(0.11) & 1.28(0.26) & 0.213 & 0.401(0.024) & 1.2 & 0.015\tabularnewline
8 & 4 & 20 & 1.0 & 10 & 2.17(0.08) & 1.14(0.10) & 0.444 & 0.181(0.048) & 0.6 & 0.006\tabularnewline
9 & 4 & 20 & 1.0 & 10 & 0.63(0.02) & 3.43(0.34) & 1.400 & 0.156(0.058) & 0.7 & 0.006\tabularnewline
10 & 2 & 20 & 5.0 & 10 & 1.20(0.07) & 2.04(1.40) & 0.162 & 0.457(0.032) & 3.0 & 0.032\tabularnewline
11 & 2 & 20 & 5.0 & 10 & 1.05(0.15) &  0(4.10) & 0.139 & - & 3.9 &    -\tabularnewline
12 & 2 & 10 & 0.5 & 5 & 1.98(0.10) & 0.30(0.04) & 0.202 & -0.562(0.418) & 1.4 & 0.015\tabularnewline
13 & 2 & 10 & 0.5 & 5 & 1.97(0.10) & 0.36(0.04) & 0.194 & -0.071(0.133) & 1.5 & 0.019\tabularnewline
14 & 2 & 20 & 0.1 & 10 & 2.16(0.10) & 0.27(0.01) & 0.158 & -0.184(0.075) & 1.7 & 0.004\tabularnewline
15 & 2 & 20 & 0.1 & 10 & 1.83(0.07) & 0.30(0.01) & 0.202 & -0.537(0.132) & 1.5 & 0.004\tabularnewline
\hline \hline
\end{tabular*}
\caption{Experimental conditions and fit results for compression experiments on large-pore fibrin gels at small strain. From the left to the right, the columns provide: the number tag given to each experiment, the concentration of fibrinogen $c$, the initial radius $a$ of the gel before compression, the strain rate at which the gel is compressed $\dot{\epsilon}$, the amount of engineering strain put on the gel after compression $\epsilon_\mathrm{e}$, the fitted permeability $k$, the fitted oedometric modulus $M$, the measured shear modulus just before compression $G_0$ from small-strain rheometry, the calculated Poisson's ratio $\nu = ( M - 2 G_0 ) / ( 2M - 2G_0 )$, the exponential relaxation time $t_{\perp}/12$ with $t_\perp = h^2 \eta / k G$ the pressurizing time, and the validity condition $M / G_0 S^2$. All gels had an initial height of $h = 1\,\mathrm{mm}$ and the estimation uncertainty is in brackets.  \label{tab:large-pore-fibrin-results}}
\end{table*}

\section{Cubic lattice model\label{sec:cube_model}}

\begin{figure}[t!]
\includegraphics[scale=0.35]{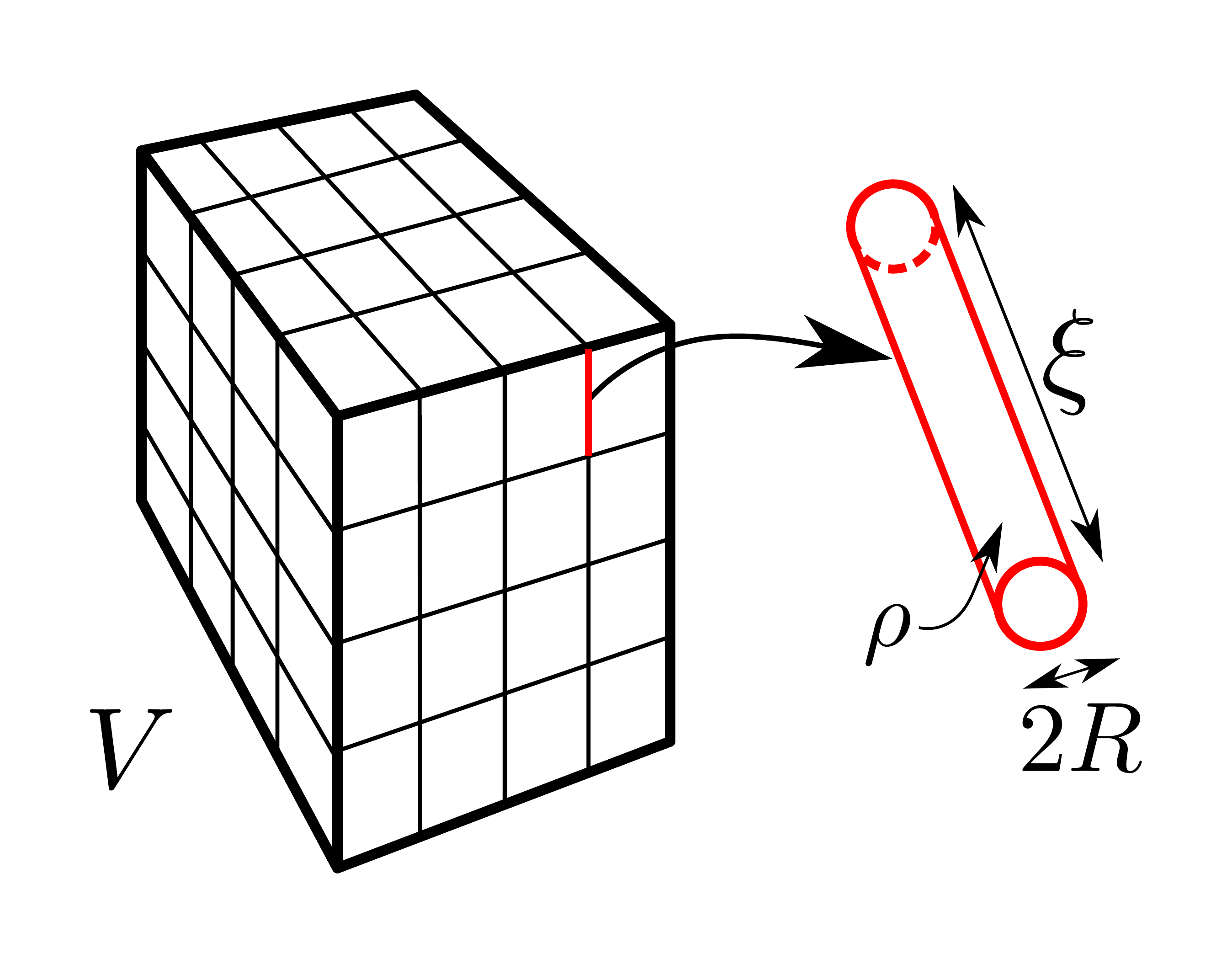}

\caption{In the cubic lattice model a fibrin network of volume $V$ is modelled as a cubic lattice of fibrin fibers, where the edge length equals the mesh size $\xi$. The fibrin fibers have a radius $R$ and mass density $\rho$. 
\label{fig:cubic_lattic_model}}
\end{figure}

To estimate the permeability of a fibrin network from its microscopic properties, we model the network as a cubic lattice. 
Consider a volume $V$ in which a fibrinogen solution with overall mass density $c$ has been polymerized into a large number of $N\gg1$ cubic cells, forming a cubic lattice. 
Because of the large number of cells, the volume $V$ can have any macroscopic shape. 
The cube edges have a length $\xi$, the mesh size, and consist of cylindrical fibrin fibers with radius $R$ and fibrinogen mass density $\rho = \rho_\mathrm{fibrinogen} / f$, see Figure \ref{fig:cubic_lattic_model}, with $\rho_\mathrm{fibrinogen}$ the mass density of pure fibrinogen molecules and $f$ the ratio of the volume which a fibrin fiber encompasses to the volume of fibrinogen molecules in the fiber. 
The total mass $m$ of fibrinogen in the volume is $m=cV$. 
By mass conservation, this mass must be equal to the mass of fibrinogen in the fibrin fibers, implying $m=\rho V_\mathrm{fiber}$ with $V_\mathrm{fiber}$ the total volume of fibrin fibers, where we assume all fibrinogen to be polymerized.  
Assuming the mesh size to be much larger than the radius $R$ of the fibrin fibers, i.e., $\xi / R \gg 1$, the total volume of fibrin fiber can be expressed approximately as $V_\mathrm{fiber}=L\pi R^2$, with $L$ the total axial length of fibrin fiber in the volume. We then obtain
\begin{equation}
c V = \rho L \pi R^{2},\label{eq:Mass_cons_fibrin_network}
\end{equation}
where $V=\xi^{3}N$ and $L=3\xi N$, since the unit cell of a cubic lattice contains three edges of a cube. 
We consider equation \eqref{eq:Mass_cons_fibrin_network} as an implicit function for $\xi$ as a function of $c$, so a given overall fibrinogen mass density $c$ results in a mesh size $\xi$ after polymerization of the fibrinogen.  

If the fibrinogen mass in a fibrin fiber per unit of axial length $\pi R^2 \rho$ is independent of the fibrinogen concentration $c$, we find $\xi \propto c^{-1/2}$. 
Generally, the permeability is expected to scale as $k \propto \xi^2$, implying the permeability to scale inversely with the concentration of fibrinogen. 
On the other hand, if the fibrinogen mass per unit of axial length depends more generally on the fibrinogen concentration as $\pi R^2 \rho \propto c^n$, we obtain $\xi \propto c^{(n-1)/2}$ and $k \propto c^{n-1}$. 
As reported in the manuscript, we find from experiments on small-pore fibrin gels that $k \propto c^m$, with $m=-2.2(0.5)$ and the uncertainty in brackets. 
This result implies that the fibrinogen concentration per unit of axial fiber length $\pi R^2 \rho$ decreases with the overall fibrinogen concentration $c$, suggesting the polymerization kinetics to depend on $c$. 

When a fibrin gel is under large compression, the vertically oriented fibers in the cubic lattic model will buckle. 
Assuming approximate homogeneous deformation, the vertical height of the buckled fibers is $\xi ( 1 - \epsilon )$, where $\epsilon$ is the engineering strain. 
As fluid flows out of the gel radially, and since the permeability $k$ is proportional to the surface area of the pores, we expect the permeability to scale as $k \propto \xi^2 \left( 1 - \epsilon \right)$, giving the compression dependent permeability as $k(\epsilon) = k_0 ( 1 - \epsilon )$ with $k_0$ the permeability at zero strain. This strain-dependent permeability we use in equation (10) of the manuscript. 

\section{Fit results\label{sec:Comparison-to-experiment}}

In section \ref{sec:app_solution} the normal force exerted by a bonded disk-like gel under ramp compression was obtained from an approximate closed-form solution of the poroelastic equations of motion. 
Here, we analyze the compression experiments we performed on fibrin gels.
By varying the experimental conditions, i.e., the amount of strain $\epsilon_\mathrm{e}$, the fibrinogen concentration $c$, the aspect ratio $S=a/h$ with the initial height $h$ and radius $a$ of the gel, and the strain rate $\dot{\epsilon}$, the microscopic response of the fibrin network will be reflected in the fitted elastic constants and the permeability. 
We use the Mathematica function \texttt{NonlinearModelFit} for fitting. 

We conducted experiments at body temperature on fibrin gels having either a large-pore fibrin network or a small-pore network. 
The large-pore gels have a typical mesh size $\xi$ of about 1 micrometer\cite{DeCagny2016PorosityGels,Yeromonahos2010NanostructureClot}. 
The small-pore gels, on the other hand, have a mesh size of about a hundred nanometers. 
For the large-pore fibrin gels, compression experiments with both small and large strain have been conducted. 
In the small strain experiments the engineering strain, i.e., the ratio of the change in gap size over the gap size at polymerization, was maximally 10\%. 
In the large strain experiments the fibrin gels were compressed up to 80\% engineering strain in subsequent steps of 10\% compression. 
For small-pore fibrin gels only small strain compression experiments up to 10\% compression were conducted. 

\subsection{Compression of large-pore fibrin gels\label{subsec:small_comp_large-pore}}

\begin{figure}[t!]
\includegraphics[scale=0.6]{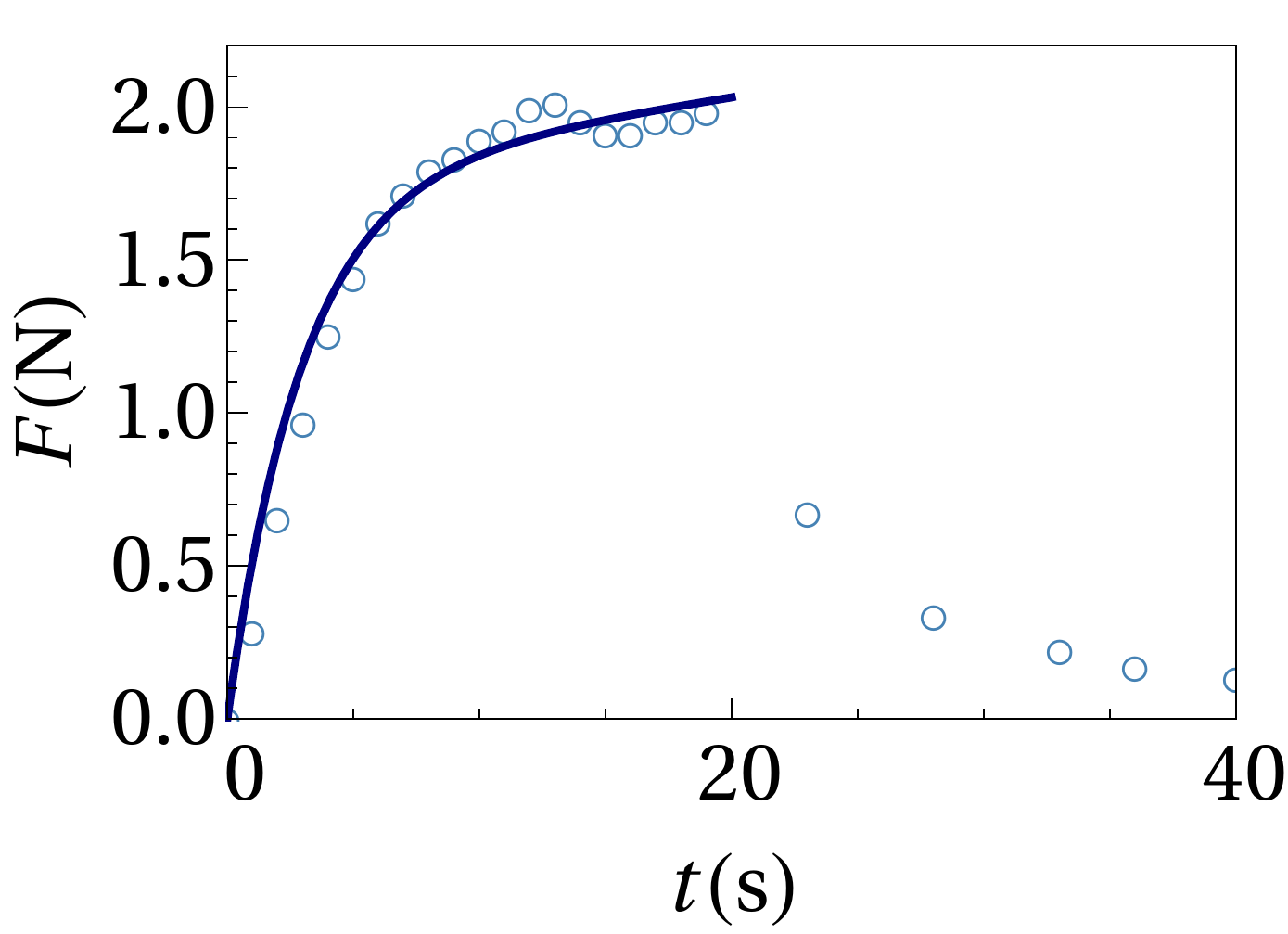}
\caption{
The measured normal force $F$ (blue circles) of a large-pore fibrin gel in response to fast ramp compression, see experiment 10 in Table \ref{tab:large-pore-fibrin-results} for the experimental conditions. 
The pressurizing time for the fluid pressure to build up to its maximal value is $t_\perp = 36\secs$. 
The blue curve is a fit of equation \eqref{eq:F} to the measured normal force during compression, giving the permeability of the fibrin network $k$ and its oedometric modulus $M$ as listed in Table \ref{tab:large-pore-fibrin-results}. 
\label{fig:fast_compression_experiment}}
\end{figure}

We conducted ramp compression experiments on large-pore fibrin gels under variable conditions in which we measured the normal force $F$ exerted by the gel on the rheometer as a function of time $t$. 
See Figure \ref{fig:fast_compression_experiment} for an example of a compression experiment with relatively high strain rate and Figure \ref{fig:slow_compression_experiment} for a compression experiment with relatively low strain rate. 
These experiments were conducted with an initial gap size of $h=\mathrm{{1\,mm}}$, and the standard conditions (SC) were chosen to be $\epsilon_\mathrm{e} = 0.1$, $\dot{{\epsilon}}=10^{-3}\mathrm{{\,/s}}$, $c=2\mathrm{{\,mg/mL}}$ and $S=20$. 
By varying one of these conditions relative to the experiments at standard conditions, the influence of the different conditions could be studied. 
After compression, the gel relaxes while the engineering strain is held constant. 
During and after compression the normal force exerted by the gel on the rheometer is measured. 
Moreover, just before the start of compression, the shear modulus $G_0$ of the gel is measured from small oscillation rheometry. 

Comparing equation \eqref{eq:F} to the measured normal force, we fit the permeability $k$ and the oedometric modulus $M = K + 4 G_0 / 3$. 
Table \ref{tab:large-pore-fibrin-results} gives the results for the fitted material parameters, the experimental conditions, and the exponential relaxation time. Moreover, we calculated $M / G_0 S^2$ which, as shown in section \ref{sec:app_solution}, should be much smaller than unity to validly apply the approximate solution. Indeed, this is the case for all large-pore fibrin compression experiments. 

From the measured shear modulus $G_0$ and the fitted oedometric modulus $M$ follows the Poisson's ratio of the gel network $\nu = (M - 2 G_0 )/( 2 M - 2 G_0 )$, where we assumed the ratio of the standard deviation in the measurement of the shear modulus to its measured value to be 1\%. 
The uncertainties in $G_0$ and $M$ are assumed to be independent because $M$ was fitted against normal force measurements, while $G_0$ was inferred from a torsion measurement. 

Next, we consider the influence of the different experimental conditions on the results for  the permeability $k$ and the Poisson's ratio $\nu$. 

\subsubsection{Compressive strain \label{subsec:Strain}}

The standard amount of engineering strain after compression of the gel is 10\%. 
To probe the influence of strain, two compression experiments have been conducted with 5\% strain instead, experiment 6 and 7. 
The standard condition (SC) experiments, experiment 1 t/m 4, give an average of the best estimates for the permeability of $\bar{k}_{\mathrm{{SC}}}=0.27(0.17)\mathrm{{\,\mu m^{2}}}$, with the standard deviation of the four best estimates in brackets. 
The uncertainty in the best estimate for each of the four individual experiments is smaller than 5\%, see Table \ref{tab:large-pore-fibrin-results}. 
Moreover, the average of the Poisson's ratio best estimates of the SC experiments is $\bar{{\nu}}_{\mathrm{{SC}}}=0.38(0.10)$, with individual uncertainty all less than 15\%. 
We have four experiments at standard conditions, which is the highest number of repeated experiments within a single set of conditions in our data set. 
Since the individual uncertainties are much smaller than the standard deviation of the best estimates, we assume the standard deviation of the permeabilities and the Poisson's ratios to reflect the sample-to-sample variation among the different fibrin gels\cite{Pieters2012AnValues}. 
The variation probably arises from the nature of the polymerization process and the origin and purification process of fibrinogen. 

The permeability estimates of the two 5\% strain experiments are $k_{6}=0.50(0.03)\mathrm{{\,\mu m^{2}}}$ and $k_{7}=0.23(0.01)\mathrm{{\,\mu m^{2}}}$, with the uncertainty of the best estimate in brackets. 
The first of these two estimates is more than one standard deviation from the average of the standard condition experiments. 
Therefore, the 5\% strain permeabilities measurements suggest that possibly the permeability for 5\% strain is significantly different from a gel which is 10\% compressed. 
The best estimates of the Poisson's ratio of the small strain experiments, however, are $\nu_{6}=0.43(0.01)$ and $\nu_{7}=0.40(0.02)$, which both lie within one standard deviation of the standard condition experiments. 
Therefore, the 5\% Poisson's ratios suggest constancy of the Poisson's ratio in the range of 5-10\% strain. 
Due to the small number of experiments, no conclusions can be drawn with respect to the influence of the magnitude of compression on $k$ and $\nu$. 

\subsubsection{Concentration \label{subsec:Concentration}}

\begin{figure}[t!]
\includegraphics[scale=0.6]{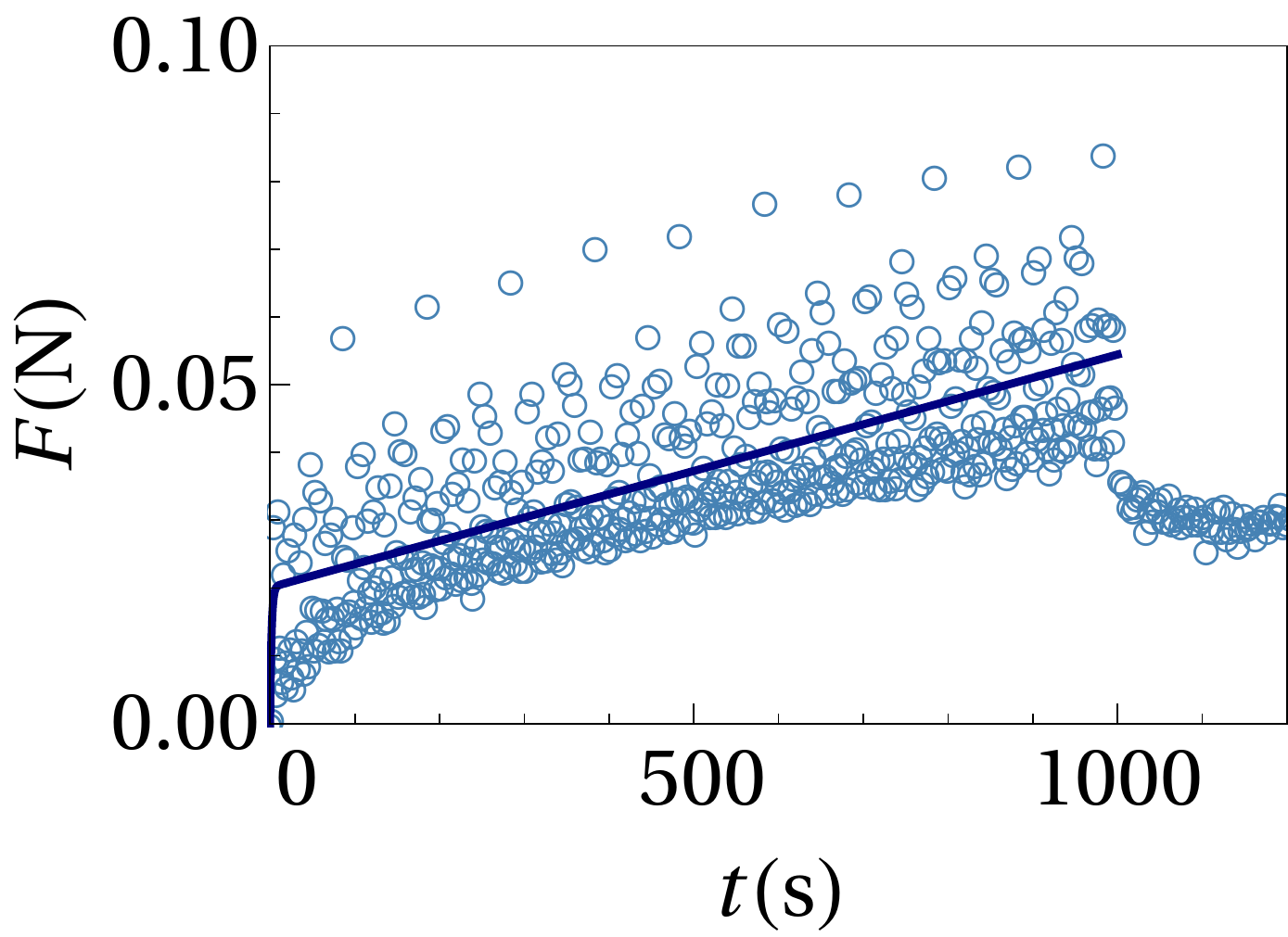}
\caption{
The measured normal force $F$ (blue circles) of a large-pore fibrin gel in response to slow ramp compression, see experiment 14 in Table \ref{tab:large-pore-fibrin-results} for the experimental conditions. 
The pressurizing time for the fluid pressure to build up to its maximal value is $t_\perp = 20\secs$. 
The blue curve is a fit of equation \eqref{eq:F} to the measured normal force during compression, giving the permeability of the fibrin network $k$ and its oedometric modulus $M$ as listed in Table \ref{tab:large-pore-fibrin-results}. 
\label{fig:slow_compression_experiment}}
\end{figure}

The standard condition (SC) in the compression experiments takes a fibrinogen solution with mass concentration $c=2\mathrm{{\,mg/mL}}$, see experiment 1 t/m 4. 
Two experiments were conducted with $c=4\mathrm{{\,mg/mL}}$, experiment 8 and 9, to observe the influence of the concentration of fibrinogen. 
We expect a higher fibrinogen concentration $c$ to give rise to a smaller mesh size $\xi$ and thus a lower permeability $k$. 
Assuming the mass density per unit length of fibrin fiber to be independent of the fibrinogen concentration $c$, the cubic lattice model gives $k \propto c^{-1}$, see section \ref{sec:cube_model}. 

With the scaling relation between the permeability and the mass concentration of the fibrin network, we turn to the fit results. 
Taking the average of the best estimates of the fitted permeability of the standard condition experiments, we find $\bar{k}_{\mathrm{{SC}}}=0.27(0.17)\mathrm{{\,\mu m^{2}}}$, where in brackets is the standard deviation of the estimates, which we assume to measure the sample-to-sample variability of the permeability of different fibrin gels. 
The best estimates of the fitted permeability of the two 4 mg/mL gels are $k_{8}=0.217(0.008)\mathrm{{\,\mu m^{2}}}$ and $k_{9}=0.063(0.002)\mathrm{{\,\mu m^{2}}}$, with the uncertainty in brackets. 
Given the sample-to-sample variation of the permeability, one of the fitted permeabilities of the 4 mg/mL experiments does not lie within one standard deviation, suggesting that possibly the permeabilities of 4 mg/mL experiments differs significantly from the standard condition experiments. 
Given the scaling relation derived above and the permeability estimation from the standard condition experiments, we would expect the permeabilities of the 4 mg/mL gels to equal $\bar{{k}}_{2}/2=0.13(0.08)\mathrm{{\,\mu m^{2}}}$, with the sample-to-sample variation in brackets. The average of the two 4 mg/mL experiments is $\bar{{k}}_{4}=0.14\mathrm{{\,\mu m^{2}}}$, therefore pointing to the validity of the scaling relation. 

Next, the fitted Poisson's ratio from the standard condition experiments is $\bar{{\nu}}_{\mathrm{{SC}}}=0.38(0.10)$. The best estimates for the 4 mg/mL gels are $\nu_{8}=0.18(0.05)$ and $\nu_{9}=0.16(0.06)$, suggesting a significant difference in the Poisson's ratio. Due to the small number of experiments, however, no conclusions can be drawn on the influence of concentration on the permeability $k$ and the Poisson's ratio $\nu$. 

\begin{figure*}[t!]
\includegraphics[scale=0.65]{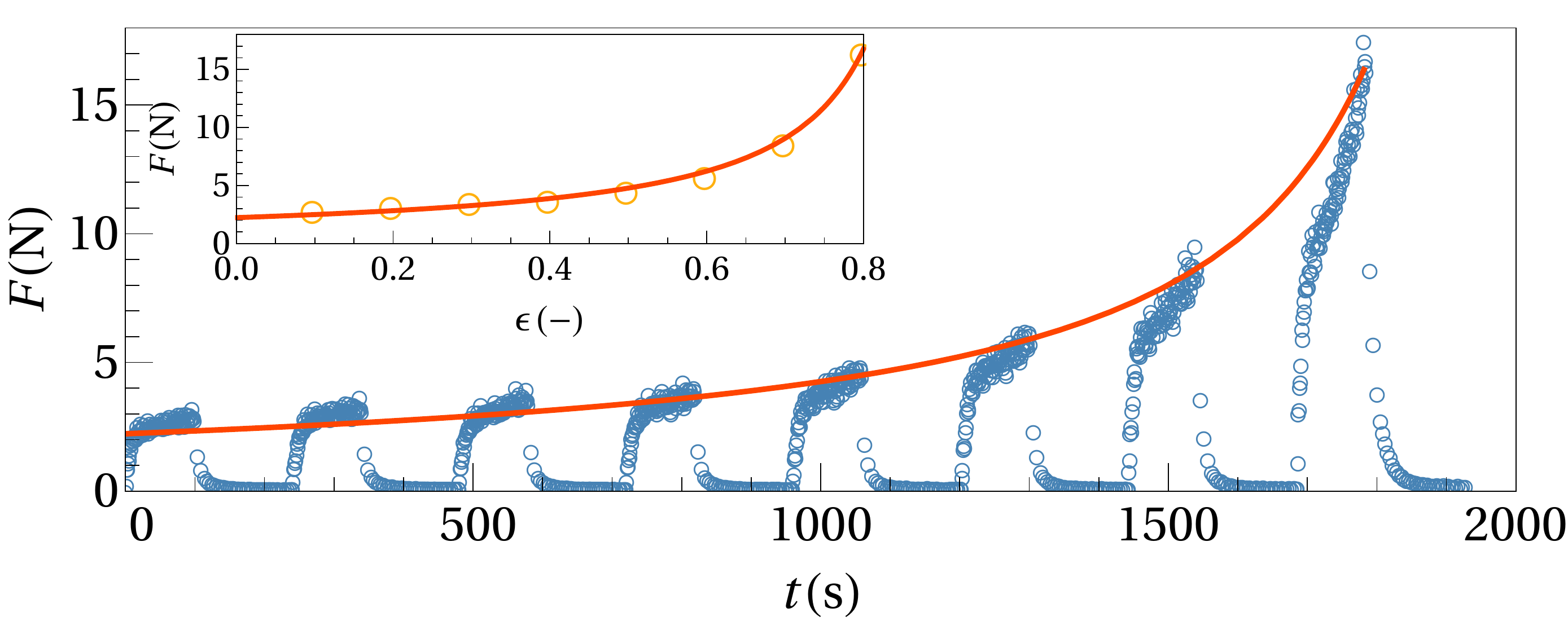}
\caption{
The measured normal force $F$ (blue circles) of a large-pore fibrin gel in response to eight consecutive ramp compression steps of 10\% compressive strain each, see experiment 2 in Table \ref{tab:large-pore-fibrin-results-large-compression} for the experimental conditions. 
The inset displays the maximal normal force during each compression step at the corresponding compressive strain (orange circles), which has been obtained by averaging the last 5\% of data points of the compression step. The orange curve is a fit of equation (10) of the manuscript to the maximal normal force, giving the permeability of the fibrin network $k_0$ in the stress-free initial state and the proportionality constant of the Toll model $b$, as listed in Table \ref{tab:large-pore-fibrin-results-large-compression}. 
\label{fig:large_compression_experiment}}
\end{figure*}

\subsubsection{Geometry \label{subsec:Geometry}}

To observe the influence of geometry, experiment 5 has been conducted, see Table \ref{tab:large-pore-fibrin-results}, where the aspect ratio of the gel was $S=a/h=10$ instead of $S=20$ in the standard condition experiments. 
The fitted permeability of experiment 5 is $k_{5}=0.154(0.002)\mum$, whereas the standard condition experiments give an average permeability of $\bar{{k}}_{\mathrm{{SC}}}=0.27(0.17)\mathrm{{\,\mu m^{2}}}$, with the standard deviation of the best estimates of the permeability of the standard condition experiments in brackets, which is presumed to measure the sample-to-sample variation in permeability. 
Therefore, the permeability of experiment 5 does not seem to be significantly different from the permeability with $S=20$. 
The fitted Poisson's ratio of experiment 5 is $\nu_{10}=2.2(0.3)$, because the oedometric modulus $M$ is fitted to be $160$ Pa while the shear modulus $G$ is 227 Pa, suggesting spontaneous contraction of the fibrin network. 
The reason for this awkward result is probably that the normal force increase during the pressurized phase is approximately $0.005$ N, while the uncertainty in the rheometer measurements is about 0.01 N. 
Experiment 12 and 13 have also been conducted with an aspect ratio of $S=10$, however, also having increases of the normal force during the pressurized phase of less than 0.01 N, but they do give physically reasonable fit values for the Poisson's ratio, $\nu=-0.56(0.42)$ and $\nu=-0.07(0.13)$, albeit with large uncertainties. 
Therefore, the anomalous value of the Poisson's ratio drawn from experiment 5 does not allow for a simple explanation.

\subsubsection{Strain rate \label{subsec:strain_rate}}

The standard strain rate is taken to be $\dot{{\epsilon}}\equiv v/h=\mathrm{{\,10^{-3}/s}}$, with $v=1\mathrm{{\,\mu m/s}}$ the velocity of the upper plate and $h=1\mathrm{{\,mm}}$ the initial height of the sample. 
To observe the influence of strain rate, we conducted two experiments at $\dot{{\epsilon}}=5\cdot\mathrm{{\,10^{-3}/s}}$, see Figure \ref{fig:fast_compression_experiment}, and two at $\dot{{\epsilon}}=0.1\cdot\mathrm{{\,10^{-3}/s}}$, see Figure \ref{fig:slow_compression_experiment}. 
The standard condition (SC) experiments, experiment 1 t/m 4, yield an average permeability $\bar{{k}}_{\mathrm{{SC}}}=0.27(0.17)\mathrm{{\,\mu m^{2}}}$, where the standard deviation due to sample-to-sample variability is between brackets. 
The high strain rate experiments, experiment 10 and 11, yield $k_{10}=0.120(0.007)\mathrm{{\,\mu m^{2}}}$ and $k_{11}=0.11(0.02)\mathrm{{\,\mu m^{2}}}$, which both fall within one standard deviation of the standard condition experiments. 
The low strain rate experiments, experiment 14 and 15, yield $k_{14}=0.22(0.01)\mathrm{{\,\mu m^{2}}}$ and $k_{15}=0.183(0.007)\mathrm{{\,\mu m^{2}}}$, both within the standard deviation of the standard condition experiments. 
Strain rate therefore does not seem to significantly influence the permeability of the fibrin network, as one would expect, because the permeability is expected to depend primarily on the microscopic structure of the fibrin network, see section \ref{sec:cube_model}. 

The Poisson's ratio of the SC experiments yields $\bar{{\nu}}_{1}=0.38(0.10)$, and the high speed experiments, experiment 10 and 11, provide $\nu_{10}=0.46(0.03)$ and $\nu_{11}=1(14)$. 
Experiment 11 clearly gives an unreliable value for the Poisson's ratio. 
In this experiment the normal force decreases significantly during the pressurized phase, causing the fitted value of the oedometric modulus to be fitted to zero, $M=0(4)\mathrm{{\,kPa}}$, though the uncertainty indicates that $M$ is in the order of kPa, similar to experiment 10. 
The Poisson's ratio of experiment 10 falls within the variation of the SC experiments. 
The slow compression experiments, however, yield $\nu_{14}=-0.18(0.07)$ and $\nu_{15}=-0.54(0.13)$, clearly falling out of the sample-to-sample variation implied by the SCEs, see section 3.1 of the manuscript for interpretation of the strain rate dependence of the Poisson's ratio.  

\begin{table}
\begin{tabular}{c|cccc}
\# & $c$(g/L) & $\dot{{\epsilon}}$ ($10^{-3}$/s) & $k\mathrm{{\,\left(10^{-1}\,\mu m^{2}\right)}}$ & $b$ (-)\tabularnewline
\hline \hline
1 & 4 & 1.0 & 0.180(0.030) & 2.6(6.4)\tabularnewline
2 & 4 & 1.0 & 0.195(0.008) & 12.0(1.4)\tabularnewline
3 & 2 & 1.0 & 0.682(0.048) & 3.9(5.7)\tabularnewline
4 & 2 & 1.0 & 1.070(0.050) & 22.0(2.4)\tabularnewline
5 & 2 & 5.0 & 1.510(0.068) & 25.0(8.6)\tabularnewline
6 & 2 & 5.0 & 1.400(0.029) & 76.0(4.2)\tabularnewline
7 & 2 & 0.1 & 1.020(0.031) & 3.4(0.2)\tabularnewline
8 & 2 & 0.1 & 0.943(0.045) & 3.0(0.3)\tabularnewline
\hline \hline
\end{tabular}
\caption{Experimental conditions and fit results for the large compression experiments ($\epsilon \leq 80\%$) on large-pore fibrin. From the left to the right the columns provide: the tag given to each experiment, the concentration of fibrinogen $c$, the strain rate at which the gel is compressed $\dot{{\epsilon}}$, the fitted permeability $k$, and the fitted proportionality constant of the Toll model\cite{Toll1998PackingReinforcements} $b$. All gels have height $h=\mathrm{{1\,mm}}$ and radius $a=\mathrm{{20\,mm}}$ before compression, and the estimation uncertainty is in brackets. \label{tab:large-pore-fibrin-results-large-compression}}
\end{table}

\subsection{Large compression of large-pore fibrin gels \label{subsec:large_comp_large-pore}}

Additional to the small strain experiments on large-pore fibrin gels, see section \ref{subsec:small_comp_large-pore}, large strain compression experiments on large-pore fibrin gels have been conducted to observe the fibrin fiber network response to large compressive strains. 
The standard experimental conditions were equal to that of the small strain experiments: the fibrinogen concentration in the gels was taken to be $c=\mathrm{{2\,mg/mL}}$, the applied strain rate was $\dot{{\epsilon}}=1\cdot10^{-3}\mathrm{{/s}}$, the initial gap size was $h=\mathrm{{1\,mm}}$ and the initial radius of the gel was $a=\mathrm{{20\,mm}}$. 
All gels were compressed up to 80\% engineering strain in a stepwise fashion, see Figure \ref{fig:large_compression_experiment}. 
Each step comprised 10\% strain and after each step the gel was allowed to relax, i.e., the measured normal force decreased after compression to a constant value. 
Figure \ref{fig:large_compression_experiment} shows that the different compression steps have a similar normal force response, although the maximal normal force increases with compression. 
Therefore, we assume that also under large compression the gel enters the pressurized phase during compression. 

In section \ref{sec:app_solution}, we assumed the gel network to be linear elastic, and to have a constant permeability, both rather poor assumptions for a fibrin gel under large compression. 
Therefore, as explained in section 2 of the manuscript, we assume a phenomenological form of the normal force as a function of the compressive strain in the pressurized phase, see equation (10) of the manuscript. 

We assume the Young's modulus of a single fibrin fiber to be in the order of MPa: $E_\mathrm{f} = 1\,\mathrm{MPa}$\cite{Collet2005TheClot}. 
The volume fraction of fibrin fiber before compression is $\phi_0 = f c / \rho_\mathrm{fibrinogen}$, with $\rho_\mathrm{fibrinogen} = 1.4 \cdot 10^3 \mathrm{\,kg/m^3}$ the mass density of pure fibrinogen\cite{Yeromonahos2010NanostructureClot}, $f=5.0$ the volume which a fibrin fiber encompasses relative to the volume of fibrinogen molecules in the fiber\cite{Carr1978SizeTurbidity}, and $c$ the overall concentration of fibrinogen in the gel. As the fibrin fiber network sticks to the plates, one can approximate the average volume fraction as $\phi=\phi_{0}/\left(1-\epsilon\right)$, where we ignore the bulging of the gel since $S = a / h \gg 1$. 

\begin{table*}
\begin{tabular*}{17.1cm}{c|ccccccccccc}
\hline 
\# & $c$(g/L) & $h$ (mm) & $a$ (mm) & $\epsilon_\mathrm{e}$ (\%) & $k\mathrm{{\,\left(10^{-3}\,\mu m^{2}\right)}}$ & $M$ (kPa) & $G_{0}$ (kPa) & $G_\mathrm{c}$ (kPa) & $t_{\mathrm{{c}}}$ (s) & $t_{\perp}/12$ (s)\tabularnewline
\hline \hline 
18 & 2 & 0.5 & 20 & 10 & 6.64(0.24) & 6.1(2.7) & 0.04 & 0.19(0.02) & 6.4(0.6) & 11\tabularnewline
19 & 2 & 1.0 & 20 & 10 & 3.86(0.15) & 0(4.6) & 0.05 & 0.59(0.02) & 7.9(0.3) & 25\tabularnewline
20 & 2 & 1.0 & 20 & 10 & 4.75(0.20) & 0(4.1) & 0.04 & 0.50(0.02) & 7.9(0.3) & 24\tabularnewline
21 & 2 & 1.0 & 20 & 5 & 3.37(0.11) & - & 0.05 & 0.49(0.01) & 6.2(0.3) & 35\tabularnewline
22 & 2 & 1.0 & 20 & 5 & 3.40(0.13) & - & 0.04 & 0.50(0.02) & 6.7(0.3) & 34\tabularnewline
23 & 4 & 1.0 & 20 & 5 & 0.73(0.05) & - & 0.24 & 1.03(0.02) & 5.8(0.3) & 77\tabularnewline
24 & 4 & 1.0 & 20 & 5 & 0.73(0.06) & - & 0.19 & 0.90(0.02) & 9.3(0.2) & 88\tabularnewline
25 & 6 & 1.0 & 20 & 5 & 0.22(0.04) & - & 0.47 & 1.22(0.03) & 5.6(0.3) & 211\tabularnewline
26 & 6 & 1.0 & 20 & 5 & 0.30(0.04) & - & 0.49 & 1.39(0.04) & 5.9(0.3) & 137\tabularnewline
27 & 6 & 1.0 & 20 & 5 & 0.48(0.05) & - & 0.36 & 1.08(0.03) & 5.7(0.3) & 110\tabularnewline
28 & 6 & 1.0 & 10 & 5 & 0.80(0.07) & - & 0.36 & 1.17(0.05) & 10.8(0.4) & 62\tabularnewline
29 & 6 & 1.0 & 10 & 5 & 0.43(0.04) & - & 0.40 & 1.99(0.07) & 9.9(0.4) & 68\tabularnewline
30 & 6 & 1.0 & 10 & 5 & 0.74(0.06) & - & 0.49 & 1.36(0.05) & 9.9(0.4) & 57\tabularnewline
31 & 8 & 1.0 & 20 & 5 & 0.22(0.03) & - & 0.65 & 1.69(0.04) & 7.2(0.3) & 152\tabularnewline
32 & 8 & 1.0 & 20 & 5 & 0.16(0.04) & - & 0.70 & 1.24(0.03) & 9.8(0.4) & 295\tabularnewline
33 & 10 & 1.0 & 20 & 10 & 0.17(0.05) & - & 0.97 & 1.32(0.05) & 14.4(1.0) & 264\tabularnewline
\hline \hline
\end{tabular*}
\caption{Experimental conditions and fit results for compression experiments on small-pore fibrin gels. From left to right, the columns provide: the tag given to each experiment, the concentration of fibrinogen $c$, the initial height $h$ and radius $a$ of the gel before compression, the strain rate at which the gel is compressed $\dot{{\epsilon}}$, the amount of engineering strain put on the gel after compression $\epsilon_\mathrm{e}$, the fitted permeability $k$, the fitted oedometric modulus $M$, the measured shear modulus just before compression $G_{0}$ from small-strain rheometry, the augmented shear modulus during compression $G_\mathrm{c}$, and the exponential relaxation time $t_{\perp}/12$ with $t_\perp$ the pressurizing time. The strain rate is $\dot{\epsilon} = 10^{-3}\,\mathrm{/s}$ in all experiments, and the uncertainty is given in brackets.  \label{tab:small-pore-fibrin-results}}
\end{table*}

To fit equation (10) of the manuscript to the measured normal force in the pressurized phase, in principle one needs to know when the pressurized phase sets in. 
This is unknown outside of the linear regime. 

As the exponential relaxation time in the first compression step is in the order of seconds, however, see Table \ref{tab:large-pore-fibrin-results}, we assume that at the end of each compression step, with $t_\mathrm{e} = 20\-100\,\mathrm{s}$, the gel has entered the pressurized phase, and we fit equation (10) of the manuscript to the maximal normal forces. 
Due to the uncertainty in the measured normal force, we take the last five percent of data points of a given compression step and average both the normal force and the strain of the different data points, to obtain a single data point per compression step, see the inset of Figure \ref{fig:large_compression_experiment}. 

Two of the large compression experiments were performed at standard conditions, two at high strain rate $\dot{{\epsilon}}=5\cdot10^{-3}\mathrm{{\,/s}}$, two at low strain rate $\dot{{\epsilon}}=0.1\cdot10^{-3}\mathrm{{\,/s}}$, and two with a doubled concentration of fibrinogen $c=\mathrm{{4\,mg/mL}}$, see Table \ref{tab:large-pore-fibrin-results-large-compression}. 
Comparing experiment 1 and 2 to experiment 3 and 4, the fitted prefactors $b$ do not appear to be influenced by the fibrinogen concentration $c$. This result confirms the assumed form of the stress response of the fibrin fiber network as that of a randomly structured fiber network in the Toll model\cite{Toll1998PackingReinforcements}, where the concentration of fibrinogen is taken into account by the volume fraction of fibrin fibers $\phi$. 
Comparison of experiment 3-8, where the influence of strain rate is probed, suggests a monotonous increase of $b$ with the strain rate, see Figure 4 of the manuscript. 
This increase is analogous with the findings described in section \ref{subsec:strain_rate}, which showed that for higher strain rate the gel responds more like a volume-conserving solid. 
The increase in strain rate suggests that for the strain rate going to zero the proportionality constant $b$ of the fiber network response could be of order 1 or less. 
The dependence of the normal force to strain rate in large compression experiments shows that the strain rate dependence of the mechanical response of the fibrin fiber network holds both for small and large compressive strain. 

Next, as noted in section \ref{subsec:Concentration}, we expect the permeability to scale as $k\propto c^{-1}$, and we do not expect it to depend on the strain rate, which determines the fluid velocity through the fiber network, as the permeability is supposed to be determined by the architecture of the network. 
The network architecture would depend on the amount of compressive strain, but not on the strain rate. 
Therefore, we expect the fitted permeability at zero compression $k_{0}$ to be independent of the strain rate, although we do expect sample-to-sample variation, see section \ref{subsec:Strain}. 
Experiment 3-8 provide permeabilities which indeed fall within approximately one standard deviation of the permeability of experiment 1-4 of the small strain experiments under standard conditions, see Table \ref{tab:large-pore-fibrin-results} and \ref{tab:large-pore-fibrin-results-large-compression}. 
All fitted permeabilities are at the lower end, however, which seems somewhat unexpected. 
Moreover, given the scaling relation $k\propto c^{-1}$, the permeability of experiment 1 and 2 with $c=\mathrm{{4\,mg/mL}}$ would be expected to be approximately half of that of experiment 3-8. 
This is not confirmed by the fitted values, however. 
Due to the low number of experiments no conclusions can be drawn. 

\subsection{Compression of small-pore fibrin gels\label{subsec:small_comp_small-pore}}

Next to the compression experiments on large-pore fibrin gels, see section \ref{subsec:small_comp_large-pore}, gels of small-pore fibrin networks have been compressed while measuring the normal force. 
The standard conditions were, similar to those in section \ref{subsec:small_comp_large-pore} and \ref{subsec:large_comp_large-pore}, a fibrinogen concentration of $c=\mathrm{{2\,mg/mL}}$, a strain rate of $\dot{{\epsilon}}=10^{-3}\mathrm{{\,/s}}$, an initial gel height of $h=\mathrm{{1\,mm}}$, and an initial radius of $20\mathrm{{\,mm}}$. 
We increased the concentration of fibrinogen up to 10 mg/mL to observe its influence on the permeability. 
Moreover, we varied the geometry of the gels to test the validity of the approximate solution presented in section \ref{sec:app_solution}. 
Below, we first discuss the numerical details of the fitting procedure. 
Afterwards, we consider the influence of fibrinogen concentration and geometry on the permeability and shear modulus of small-pore fibrin networks. 

\subsubsection{Strain stiffening}

As noted in section 2 of the manuscript, we accommodate for strain stiffening by replacing  $Gt \rightarrow \int_0^t G(t') dt'$ where $G$ enters in equation \eqref{eq:F}, with $G(t)$ as given in equation (7) of the manuscript. For the numerical fit routine, we use a sigmoidal function with very high power $n=33$ to interpolate approximately stepwise and analytically from $G_0$ to the augmented shear modulus $G_\mathrm{c}$. This gives for the time-dependent shear modulus
\begin{equation}
G(t) = G_0 + \left( G_\mathrm{c} - G_{0} \right) \frac{ \left( t / t_\mathrm{c} \right)^n }{ 1 + \left( t / t_\mathrm{c} \right)^n}. \label{eq:Gvariable}
\end{equation}

All small-pore fibrin experiments can be described well by fitting the permeability $k$, the augmented shear modulus $G_\mathrm{c}$, the onset time for strain stiffening $t_\mathrm{c}$, and, if possible, the oedometric modulus $M$, see Table \ref{tab:small-pore-fibrin-results}. 

In experiment 18, the fit routine was able to converge with a meaningful value of the oedometric modulus $M=6.2(2.7)\mathrm{{\,kPa}}$. 
The reason for convergence is that in this experiment the initial height of the gel was $h=\mathrm{{0.5\,mm}}$, causing a relatively short exponential relaxation time $t_\perp / 12 =  h^2 \eta / 12 k G_\mathrm{c} $ of 11 s. 
Therefore, during a significant portion of compression the gel is to good approximation in the pressurized phase, in which the increase in normal force is solely due to the oedometric modulus, see equation \eqref{eq:F}. 
For experiment 21-33, we put $M=0$ by hand, because otherwise the fit routine does not converge. This presumption is also justified, however, for the following reasons. 
Arguably, the non-convergence is due to the negligible effect of the mechanical response, $\pi a^2 M \epsilon_\mathrm{e}$ with $a$ the initial radius of the gel and $\epsilon_\mathrm{e}$ the amount of strain after compression, in the pressurized phase. 
Moreover, these experiments all have $h=1\mathrm{{\,mm}}$, implying a longer pressurizing time than in experiment 18. 
In part of the experiments, the gel is compressed with only 5\% engineering strain $\epsilon=0.05$, reducing the mechanical response contribution even further. 
Finally, with increasing concentration the relaxation time $t_{\perp}$ seems to increase, thereby increasing the time needed to enter the pressurized phase and decreasing the influence of $M$, see Figure \ref{fig:ExampleFineHighConc} for a high fibrinogen concentration experiment with $M=0$. 

\subsubsection{Concentration and geometry}

\begin{figure}[t!]
\includegraphics[scale=0.65]{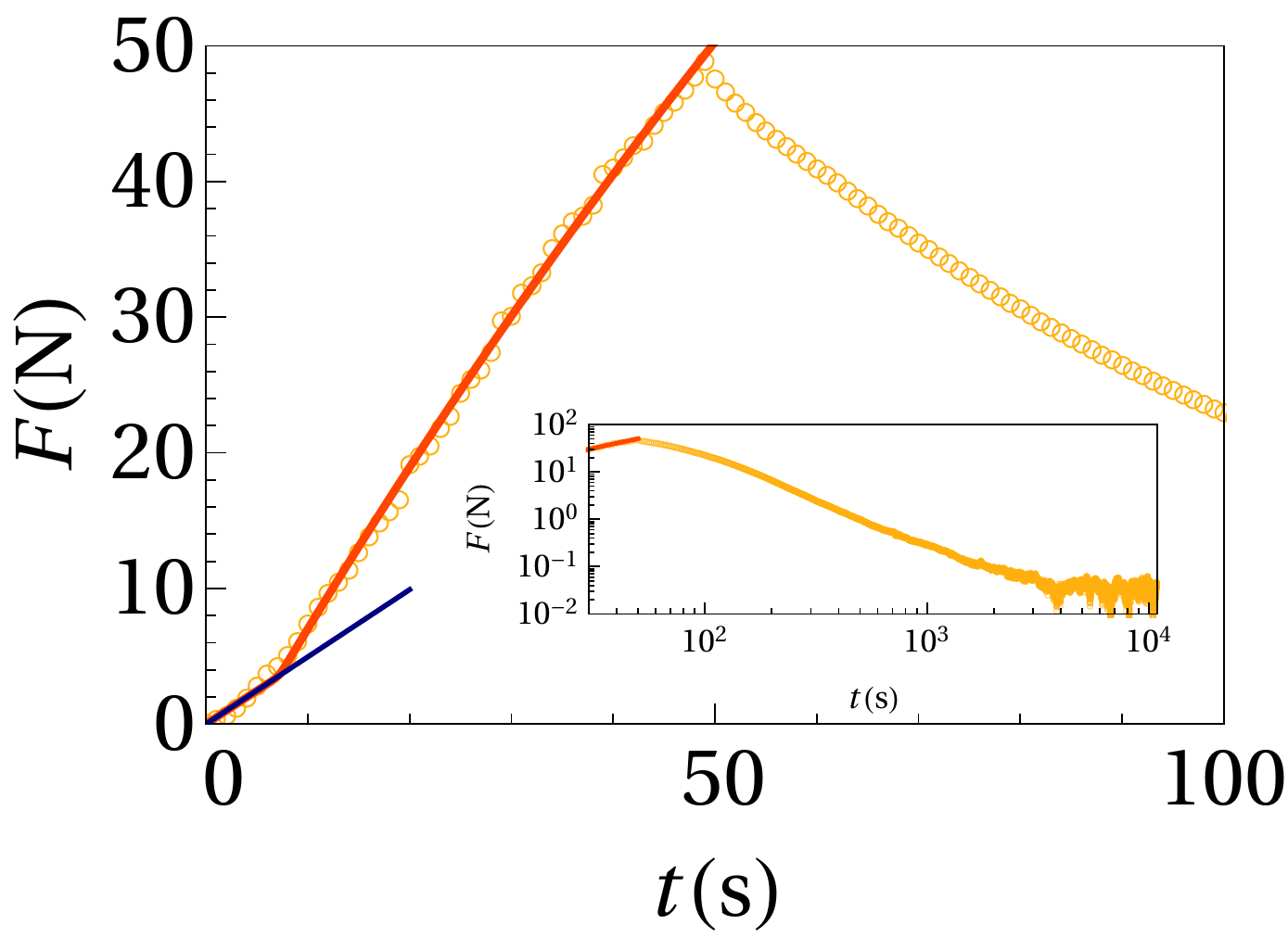}
\caption{
The measured normal force $F$ (orange circles) of a small-pore fibrin gel with high fibrinogen concentration, see experiment 31 in Table \ref{tab:small-pore-fibrin-results} for the experimental conditions. 
The pressurizing time for the fluid pressure to build up to its maximal value is $t_\perp = 1824\secs$. 
The orange curve is a fit of equation \eqref{eq:F} to the measured normal force during compression, giving the permeability of the fibrin network $k$, the onset time of strain stiffening $t_\mathrm{c}$ and the augmented shear modulus $G_\mathrm{c}$ as listed in Table \ref{tab:small-pore-fibrin-results}. 
Before the onset time of strain stiffening, the normal force follows the time dependence expected for a volume-conserving solid with the measured initial shear modulus $G_0$. 
The inset shows an extended process of gel relaxation after compression stops. 
\label{fig:ExampleFineHighConc}}
\end{figure}

To observe the influence of the concentration of fibrinogen $c$ we conducted experiments with $c=2,4,6,8,10\mathrm{{\,mg/mL}}$, see Table \ref{tab:small-pore-fibrin-results} for the results and Figure \ref{fig:ExampleFineHighConc} for a high fibrinogen gel. The permeability $k$ decreases with increasing fibrinogen concentration $c$, see Figure (7) of the manuscript. 

Up to the increase of the shear modulus at $t_\mathrm{c}$, the fibrin gels deform in an approximately volume-conserving manner. 
Therefore, the compression up to $t=t_\mathrm{c}$ can be considered as a shear deformation of the sample. 
The tangential radial stress at the sample-plate interface therefore shears the sample. Within the gel the shear stress is lower because the gel can bulge out. 
The tangential stress at the sample-plate interface before stiffening is given as $\sigma_{\mathrm{{n}},rz}=G_{0}\left(\partial_{z}U+\partial_{r}W\right)=G_{0}\partial_{z}U$, since $W$ is independent of $r$, as we presumed in this entire analysis. 
The radial deformation $U$ grows linearly with the radial coordinate $r$, see equation \eqref{eq:U}. 
Therefore, we average the tangential stress over the gel plate interface to obtain the critical stress: $\sigma_\mathrm{c} \propto \left( 1 / \pi  a^2 \right) \int_0^a dr 2 \pi r \sigma'_{rz}|_{z=h,t=t_\mathrm{c}}$, giving equation (9) of the manuscript, and the dependence of $\sigma_\mathrm{c}$ on the concentration of fibrinogen $c$ is given in Figure 6 of the manuscript. 

Experiments 25-30 have a fibrinogen concentration of $c=\mathrm{{6\,mg/mL}}$, whereas experiments 25-27 have an initial radius of $a=20\mathrm{{\,mm}}$ and experiments 28-30 have $a=10\mathrm{\mathrm{{\,mm}}}$. 
According to equation \eqref{eq:F}, a difference in the magnitude of the normal force $F_{\mathrm{{N}}}\propto a^{4}$ is to be expected, but no difference in the fitted permeability $k$, the shear modulus $G_\mathrm{c}$ and the critcial time $t_{\mathrm{{c}}}$. 
The three fit parameters are all of the same order, their differences are probably due to sample-to-sample variation, and the coefficient of determination exceeds 0.9992 for every experiment. 
Therefore, our model seems to account correctly for the influence of geometry. 

\begin{figure}[t!]
\includegraphics[scale=0.65]{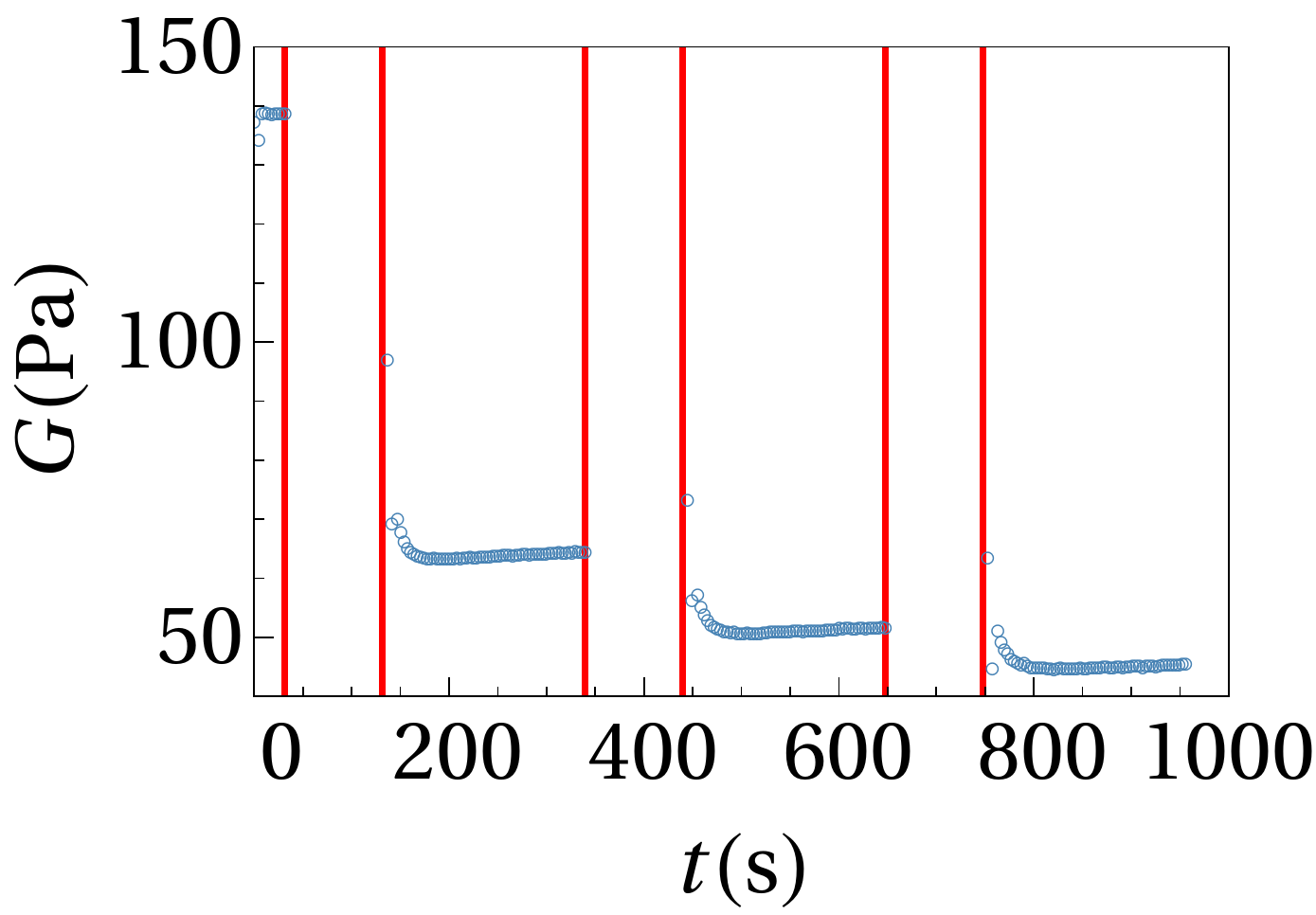}
\caption{The measured shear modulus $G$ (blue dots) before and after three compression steps of 10\% compressive strain for a large-pore fibrin sample. The beginning and end of a compression step are denoted by a red line. Before each compression step the shear modulus is (approximately) constant. After each step the shear modulus decreases rapidly to a smaller value than before compression.  \label{fig:Large_comp_measured_G}}
\end{figure}

\subsection{Shear modulus}

In section \ref{subsec:small_comp_large-pore} we found the measured normal force in all large-pore fibrin compression experiments during compression to be explained by the approximate solution of section \ref{sec:app_solution}, while using the measured value of the shear modulus $G_{0}$ just before compression. 
For small-pore fibrin gels, however, we found the shear modulus to increase at a critical stress and to remain constant afterwards. 
In this section, we also consider the evolution of the shear modulus after compression. 

For large-pore fibrin gels undergoing large compression, see section \ref{subsec:large_comp_large-pore}, the shear modulus has been measured before and after compression, see for example Figure \ref{fig:Large_comp_measured_G}. 
Before compression, the shear modulus has a constant value $G_{0}$. 
After compression, it first decreases rapidly. 
Subsequently, it increases very slowly and can be regarded as approximately constant, i.e., as if the gel is relaxed. 
Just after compression, the shear modulus seems to be at approximately the same value as before compression, suggesting the shear modulus in large-pore fibrin gels to be approximately constant during compression, as we assumed in section \ref{subsec:small_comp_large-pore}. 
With each compression step, the magnitude of the relaxed shear modulus is decreased compared to its value before compression, in accordance with literature\cite{vanOosten2016UncouplingStretch-stiffening}. 

\begin{figure}[t!]
\includegraphics[scale=0.37]{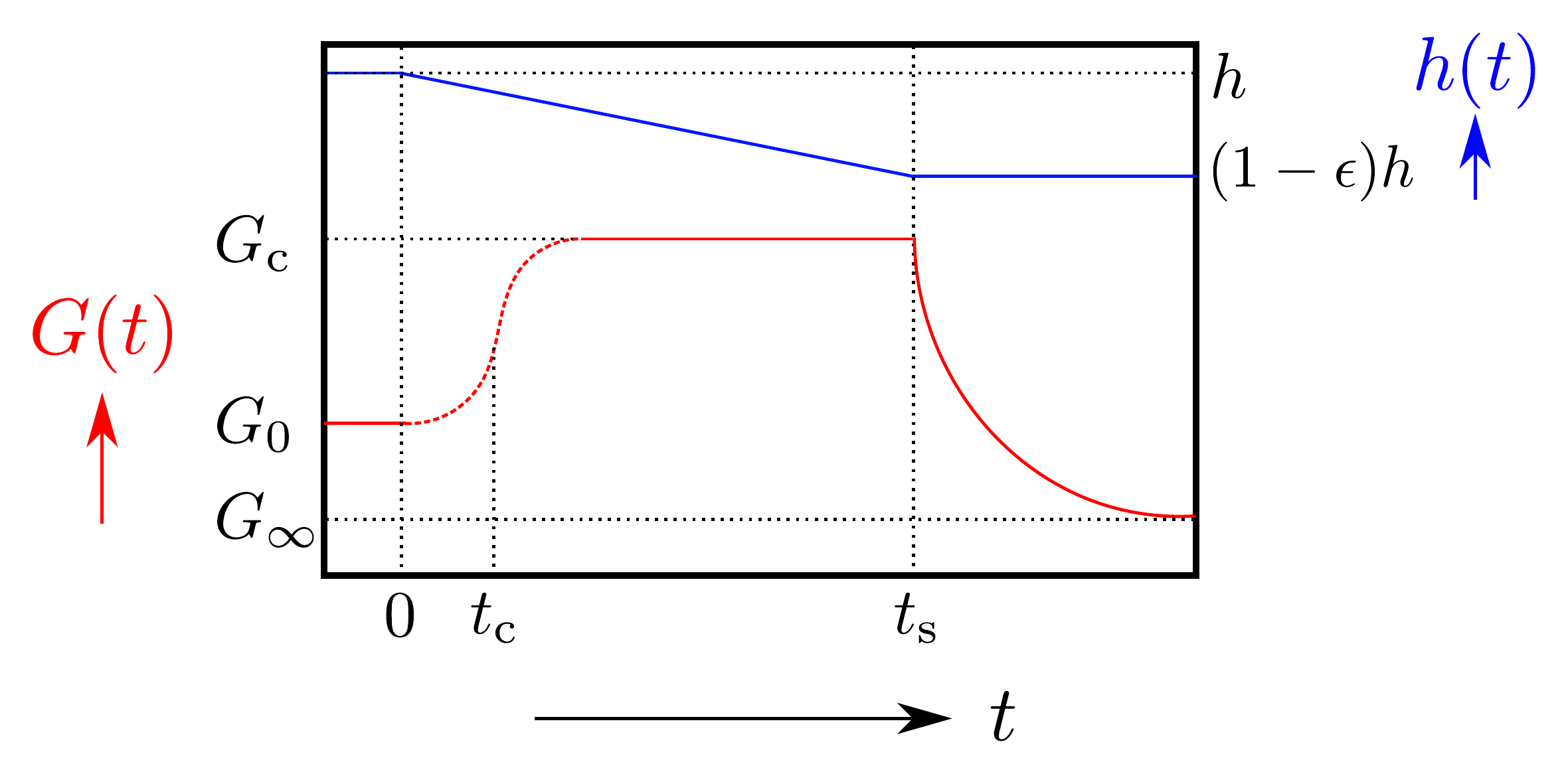}
\caption{Schematic of the development of the shear modulus $G(t)$ in response to changes in the height $h(t)$ of a ramp-compressed fibrin gel during and after compression. 
Initially, for $t<0$ the shear modulus has magnitude $G_0$. 
As compression commences, the shear modulus increases to $G_\mathrm{c}$, with the point of maximum increase at the onset time $t_\mathrm{c}$. 
The difference between $G_\mathrm{c}$ and $G_{0}$ is negligibly small for large-pore fibrin gels but may be significant for small-pore fibrin gels. 
This difference is probably due to the much shorter volume-conserving phase of large-pore fibrin gels. 
After compression stops at $t_\mathrm{e}$, the shear modulus decreases to its static value $G_{\infty}$, which is determined by the compressive strain. \label{fig:G_summary}}
\end{figure}

Combining the results from section \ref{subsec:small_comp_large-pore}, \ref{subsec:small_comp_small-pore} and the shear modulus measurements just after compression, we can form a coherent picture of the evolution of the shear modulus during and after the compression of fibrin gels, see Figure \ref{fig:G_summary}. 
Before compression, the gel has a constant shear modulus which reflects that the gel is in equilibrium. 
As soon as compression starts it is first compressed in a volume-conserving manner because the (low) permeability of the gel prevents fluid to be squized out instantaneously. 
In this phase the shear modulus may increase around an onset stress $\sigma_\mathrm{c}$ in the network, see equation (9) of the manuscript. 
If fluid starts to flow out before the gel is stressed to $\sigma_\mathrm{c}$, the shear modulus remains constant throughout the whole of the ramp compression, as with the large-pore compression experiments in section \ref{subsec:small_comp_large-pore}. 
If $\sigma_\mathrm{c}$ is reached while being in the volume-conserving phase, the shear modulus can increase significantly and remains so during the rest of compression, see section \ref{subsec:small_comp_small-pore}. 
After compression, the shear modulus relaxes to its static value. 

The evolution of the shear modulus of a fibrin gel of Figure \ref{fig:G_summary}, and the difference in stiffening depending on whether the gel reaches the onset stress $\sigma_\mathrm{c}$ in the volume-conserving phase or not, suggests that the occurrence of strain stiffening has a strong connection with the flow of fluid through the fibrin network. 
In the volume-conserving phase, when no fluid flows relative to the fibrin network, the shear modulus shows compressive stiffening, as most materials do when compressed. 
As the gel transitions into the pressurized phase and the fluid velocity relative to the network increases to a nonzero value, however, the modulus becomes fixed. 
Therefore, it seems as if flow of fluid through the network prevents further stiffening of the shear modulus. 
After compression stops, however, and the relative velocity starts to decrease to zero, the shear modulus also decreases to its new equilibrium value. 

\balance

\bibliography{references}
\bibliographystyle{rsc} 